\documentclass[11pt,letterpaper,oneside]{article}
\usepackage[utf8]{inputenc}
\usepackage{amsmath}
\usepackage{amsfonts}
\usepackage{amssymb}
\usepackage{enumerate}
\usepackage{siunitx}
\usepackage{makeidx}
\usepackage[toc,page]{appendix} 
\usepackage{pdflscape} 
\usepackage[margin=0.75in]{geometry}
\usepackage{float}
\usepackage[table]{xcolor}
\definecolor{prp}{RGB}{108,43,145}
\usepackage{makecell}
\usepackage{caption}

\usepackage{verbatim}
\usepackage{listing}
\usepackage{parskip} 
\usepackage{setspace} 
\usepackage{multirow}

\usepackage{graphicx}
\usepackage{booktabs}
\usepackage{longtable}
\usepackage{tabu}
\usepackage{etoolbox}

\usepackage{threeparttablex} 
\usepackage{adjustbox}
\usepackage{mathtools}
\DeclarePairedDelimiter\abs{\lvert}{\rvert}%
\usepackage[super]{nth}
\usepackage{colortbl}
\usepackage{subcaption}
\usepackage{caption}
\usepackage{natbib}
\usepackage{everypage}
\def\PageTopMargin{1in}
\def\PageLeftMargin{1in}
\newcommand\atxy[3]{%
	\AddThispageHook{\smash{\hspace*{\dimexpr-\PageLeftMargin-\hoffset+#1\relax}%
			\raisebox{\dimexpr\PageTopMargin+\voffset-#2\relax}{#3}}}}

\usepackage{afterpage}
\usepackage{capt-of}

\usepackage{perpage}
\MakePerPage{footnote}

\usepackage{natbib}
\usepackage{hyperref}
\usepackage{etoolbox}
\DeclareMathSymbol{\mh}{\mathord}{operators}{`\-}

\DeclareCaptionFormat{myformat}{%
  \begin{varwidth}{\linewidth}%
    \centering
    #1#2#3%
  \end{varwidth}%
}

\usepackage{tikz}
\usepackage{graphicx}
\usetikzlibrary{positioning}

\makeatletter

\pretocmd{\NAT@citex}{%
	\let\NAT@hyper@\NAT@hyper@citex
	\def\NAT@postnote{#2}%
	\setcounter{NAT@total@cites}{0}%
	\setcounter{NAT@count@cites}{0}%
	\forcsvlist{\stepcounter{NAT@total@cites}\@gobble}{#3}}{}{}
\newcounter{NAT@total@cites}
\newcounter{NAT@count@cites}
\def\NAT@postnote{}

\def\NAT@hyper@citex#1{%
	\stepcounter{NAT@count@cites}%
	\hyper@natlinkstart{\@citeb\@extra@b@citeb}#1%
	\ifnumequal{\value{NAT@count@cites}}{\value{NAT@total@cites}}
	{\ifNAT@swa\else\if*\NAT@postnote*\else%
		\NAT@cmt\NAT@postnote\global\def\NAT@postnote{}\fi\fi}{}%
	\ifNAT@swa\else\if\relax\NAT@date\relax
	\else\NAT@@close\global\let\NAT@nm\@empty\fi\fi
	\hyper@natlinkend}
\renewcommand\hyper@natlinkbreak[2]{#1}

\patchcmd{\NAT@citex}
{\ifNAT@swa\else\if*#2*\else\NAT@cmt#2\fi
	\if\relax\NAT@date\relax\else\NAT@@close\fi\fi}{}{}{}
\patchcmd{\NAT@citex}
{\if\relax\NAT@date\relax\NAT@def@citea\else\NAT@def@citea@close\fi}
{\if\relax\NAT@date\relax\NAT@def@citea\else\NAT@def@citea@space\fi}{}{}

\newcommand*{\rom}[1]{\expandafter\@slowromancap\romannumeral #1@}

\hypersetup{
    colorlinks=true,     
    linkcolor=blue,      
    urlcolor=blue,       
    citecolor=blue,      
    hyperfootnotes=false , 
}

\usepackage[bottom]{footmisc}

\newcounter{customfootnote} 
\renewcommand{\footnote}[1]{%
    \stepcounter{customfootnote}%
    \textsuperscript{\thecustomfootnote}%
    \footnotetext[\value{customfootnote}]{#1}%
}

\usepackage{hyperref}
\usepackage[nameinlink]{cleveref}

\crefname{section}{section}{sections}
\crefname{subsection}{section}{sections}
\crefname{subsubsection}{section}{sections}

\title{Realised Volatility Forecasting: Machine Learning \\ 
via Financial Word Embedding}

\bigskip

\author{Eghbal Rahimikia, Stefan Zohren, and Ser-Huang Poon\thanks{Eghbal Rahimikia (corresponding author) (\href{mailto:eghbal.rahimikia@manchester.ac.uk}{eghbal.rahimikia@manchester.ac.uk}) is affiliated with the Alliance Manchester Business School at the University of Manchester and the Alan Turing Institute; Stefan Zohren (\href{mailto:stefan.zohren@eng.ox.ac.uk}{stefan.zohren@eng.ox.ac.uk}) is with the Department of Engineering Science, University of Oxford; and Ser-Huang Poon (\href{mailto:ser-huang.poon@manchester.ac.uk}{ser-huang.poon@manchester.ac.uk}) is also at the Alliance Manchester Business School, University of Manchester. We are grateful to participants and discussants at the 2021 FMA Conference on Derivatives and Volatility, 2022 British Accounting and Finance Association (BAFA) Annual Conference, Economics of Financial Technology Conference, \nth{12} Financial Markets and Corporate Governance Conference, 2022 FMA European Conference, \nth{14} Annual SoFiE Conference, Advances in Data Science Conference, Greater China Area Finance Conference, 2022 FMA Annual Meeting, \nth{3} Frontiers of Factor Investing Conference, London-Oxford-Warwick Mathematical Finance Workshop, 2022 Cardiff FinTech Conference, 2023 International Finance and Banking Society (IFABS) Conference, Wolfe Research \nth{6} Annual Quantitative and Macro Investment Conference, 2023 Market Volatility: NLP Trends in Research, Dow Jones, IFZ FinTech Colloquium at Lucerne University of Applied Sciences and Arts, Switzerland, and the Workshop on Financial Management and Technology for a Sustainable Future at Sheffield University Management School. We express sincere appreciation to the IT services of the University of Manchester for their continuous support and for providing the computational infrastructure utilised in this study. We extend our gratitude to Google for awarding an academic research grant in support of this study.}}

\date{\parbox{\linewidth}{\centering%
  \endgraf\medskip
  \small First version: July 29, 2021 $\bullet$ This revision: \today}}

\begin{document}
\renewcommand{\baselinestretch}{1.29}
\normalsize

\clearpage\maketitle
\thispagestyle{empty}

\linespread{1.5}

\begin{abstract}
We examine whether financial news can improve realised volatility forecasting using a parsimonious NLP-based framework that incorporates specialised financial word embeddings alongside general-purpose alternatives. News-only forecasts contain useful predictive information but generally do not outperform strong volatility-history benchmarks. Crucially, combining stock-related news forecasts with a strong volatility-history benchmark lowers forecast losses for several specifications and increases realised utility, providing evidence consistent with forecast complementarity. Performance varies across news types, embedding representations, and volatility regimes. SHAP attributions associate forecast variation with economically interpretable firm-specific and macroeconomic news themes.
\end{abstract}



\vspace{0.5cm}
\doublespacing
\clearpage
\section{Introduction} \label{intro_section}
Realised volatility forecasting in empirical finance has traditionally been dominated by econometric specifications that exploit the strong persistence of volatility while treating information arrivals only indirectly \citep{engle1993measuring}, despite substantial evidence that the timing and incorporation of information are closely linked to volatility dynamics \citep{french1986stock}. Media content and tone have also been shown to predict market activity and price pressure, consistent with news conveying economically relevant information beyond standard market variables \citep{tetlock2007giving}. Related evidence links firm-specific public information directly to volatility. In particular, \citet{engle2021news} show that indicators of firm-specific public information arrival explain substantial variation in idiosyncratic volatility, reinforcing the view that observable news flow contains information relevant for volatility dynamics. However, despite the success of modern natural language processing (NLP) models in extracting signals from text in other fields, empirical finance has historically relied on relatively simple textual representations rather than modern language models, even as text-based methods have expanded rapidly across the broader economics literature \citep{gentzkow2019text}. Motivated by these advances, this study asks whether a forecast constructed from financial news contains predictive information for next-day realised volatility that is not captured by standard volatility-history forecasts.

Most studies on realised volatility forecasting use historical realised volatility as the primary source of data for prediction within a linear framework. Heterogeneous autoregressive (HAR) models \citep{corsi2009simple} are well-known, straightforward linear models for realised volatility forecasting that parsimoniously capture volatility persistence across multiple horizons. Further developments expanded the HAR-family with variations such as HAR-J (HAR with jumps) and CHAR (continuous HAR) from \citet{andersen2007roughing}, SHAR (semivariance-HAR) from \citet{patton2015good}, and the HARQ model from \citet{bollerslev2016exploiting}. \citet{bollerslev2018risk} extend these models by demonstrating that exploiting cross-asset volatility similarities improves out-of-sample forecasting performance, while \citet{patton2022bespoke} show that further gains can be achieved by customising realised volatility measures. An additional strand of the literature emphasises measurement error in realised volatility and its implications for modelling and forecasting. For instance, \citet{cipollini2021realized} show that such errors are heteroskedastic and can bias standard models, proposing robust specifications to address this. Moreover, \citet{gallo2015forecasting} show that modelling time-varying, regime-dependent average volatility improves forecast performance. Over the last decade, research on the theory and application of machine learning (ML) in finance has grown substantially. Recent evidence shows that ML models can outperform traditional financial models in asset pricing and return prediction \citep{gu2020empirical}, short-term price forecasting using high-frequency data \citep{sirignano2019universal}, and text-based financial forecasting using news articles \citep{adammer2020forecasting}. \citet{medeiros2021forecasting} show that ML methods can deliver substantial out-of-sample forecasting gains in a data-rich macroeconomic setting, with random forests benefiting from both nonlinearities and variable selection. \citet{bybee2024business} employ a latent Dirichlet allocation (LDA) topic model to analyse the textual content of Wall Street Journal articles, demonstrating that business news attention closely tracks economic activity and improves forecasts of macroeconomic dynamics. More recently, \citet{hong2025forecasting} extract interpretable economic narratives from a large corpus of Wall Street Journal articles using topic modelling and employ ML methods to exploit these textual signals for forecasting. They show that narrative information improves out-of-sample inflation forecasts and provides incremental predictive information beyond a high-dimensional macroeconomic information set.

In the context of realised volatility forecasting, \citet{bucci2020realized}, \citet{christensen2023machine}, and \citet{rahimikia2020machine} show that machine-learning and neural-network methods can provide competitive improvements over traditional econometric approaches to volatility prediction. \citet{ma2023stock} show that combining macroeconomic and financial predictors with variable selection significantly improves volatility forecasting performance. Relatedly, \citet{zhang2024volatility} show that ML methods can exploit intraday volatility commonality and market-level information to improve realised volatility forecasts, highlighting the importance of expanding the predictor information set beyond an asset's own volatility history. \citet{moreno2024deepvol} introduce a realised volatility forecasting model using dilated causal convolutions and find that it improves prediction accuracy compared to traditional methods. \citet{li2025automated} introduce an automated volatility forecasting system that leverages a broad array of features and multiple algorithms, achieving superior out-of-sample forecasting performance across different horizons. Earlier evidence from \citet{hillebrand2010benefits} also shows that bootstrap aggregation can improve realised volatility forecasts from HAR-type and nonlinear specifications, highlighting the potential value of forecast aggregation in the presence of model and parameter uncertainty. \citet{10.1093/jjfinec/nbag019} show that neural-network volatility forecasting depends importantly on the estimation regime and data scale, with global estimation and larger, more diverse training samples generating substantially greater gains than increases in model size or changes in network architecture. A closely related strand of research shows that textual news can be processed automatically to quantify market reactions, with \citet{gross2011machines} demonstrating that machine-read news contains economically meaningful information for high-frequency market activity and volatility. Particularly related to our setting, \citet{zhang2016distillation} extract sentiment from financial text using alternative lexical representations and show that the resulting textual measures contain incremental information for subsequent stock volatility, trading volume, and returns. Notwithstanding these advances, other studies caution that the empirical performance of ML models can be unstable and context-dependent \citep{audrino2016lassoing, branco2024forecasting, audrino2025hard}. Beyond realised volatility, there are a handful of studies that specifically focus on NLP and large language models (LLMs), such as \citet{van2024almost}, who employ ML techniques using word embeddings to construct a 170-year-long measure of economic sentiment from 200 million pages of US local newspapers, demonstrating its predictive power for GDP growth, employment, and monetary policy decisions. More recently, studies such as \citet{chen2022expected}, \citet{huang2023finbert}, and \citet{rahimikia2024r} have incorporated information extracted from advanced LLMs for various financial tasks.

Taken together, prior research has improved realised volatility forecasting primarily by refining the dynamics extracted from historical volatility and by expanding the predictor set with additional market information, while the text-as-data literature shows that financial news contains economically relevant information that is not necessarily summarised by standard market variables. This motivates the central question of this study: whether the information contained in financial news expands the predictive information set for next-day realised volatility beyond that captured by the HAR-family volatility-history benchmarks considered here. In addressing this question, the distinction between substitution and complementarity is important, as the usefulness of news does not require a news-only model to dominate established HAR-family models. Instead, the news-only experiment provides a deliberately restrictive assessment of how much predictive information is contained in text without conditioning on lagged realised volatility or other market variables, while the economically more important question is whether the resulting textual signal contains incremental information that improves forecasts produced by strong volatility-history models.\footnote{Throughout the paper, `incremental' refers to predictive content beyond the HAR-family volatility-history benchmarks considered here, rather than beyond every potentially observable financial predictor or forecasting architecture. Option-implied volatility, market-level predictors, regularised high-dimensional specifications, and cross-asset information may contain additional predictive information and are outside the scope of the present comparison.} Accordingly, the empirical analysis addresses four related questions: how informative financial news is when used as the sole forecasting information set; whether its predictive content varies across market conditions and between firm-specific and general news; how the forecasting performance of richer linguistic representations compares with simpler textual proxies such as news intensity and dictionary-based sentiment; and, most importantly, whether combining the textual forecast with an established realised volatility benchmark improves statistical forecasting performance and economic value. To answer these questions, we develop a parsimonious NLP forecasting framework that maps news headlines into next-day realised volatility forecasts using fixed word embeddings and a convolutional neural network, evaluate the textual model both in isolation and as a complementary information source to standard HAR-family forecasts, and use explainability analysis to identify the textual features associated with variation in the resulting forecasts.

We conduct a set of benchmarks to assess the quality and relevance of alternative word embeddings. At the language level, specialised financial word embeddings outperform general-purpose word embeddings in capturing economically meaningful relationships, particularly for stock-related financial analogies, even though they do not dominate on generic linguistic benchmarks. Turning to realised volatility forecasting, first, when we restrict the information set to news only, the NLP forecaster contains useful predictive information but generally does not outperform the best volatility-history model. For stock-related news, full out-of-sample average loss ratios range from 1.106 to 1.187 under MSE and from 1.476 to 1.838 under QLIKE, relative to CHAR, our common HAR-family reference benchmark, while realised utility ranges from 1.6280\% to 2.1047\% compared with 2.7540\% for the benchmark model. Second, the contribution of news is strongly regime-dependent, with the relative performance of stock-related news across volatility states depending partly on the loss function. Third, the informativeness of news depends on its scope, the volatility regime, and the loss function: stock-related news performs more favourably over the full sample and is more competitive during high volatility days under MSE but during normal volatility days under QLIKE, whereas general news is more competitive during normal volatility days. Specialised representations tend to be more competitive for stock-related inputs, while general-purpose representations remain competitive for broader market-wide news. Fourth, combining the NLP news signal with the CHAR reference benchmark via a simple ensemble provides evidence consistent with complementary predictive information, with full-sample loss ratios for the stock-related ensembles ranging from 0.961 to 1.020 under MSE and from 0.937 to 1.096 under QLIKE, with full-sample gains concentrated in stock-related news and regime-specific performance differing across loss functions, and with the best-performing ensemble increasing realised utility to approximately $2.9321\%$. Finally, explainability analysis links the forecasts to economically intuitive phrases and themes, revealing distinct classes of news such as earnings announcements, corporate actions, macroeconomic releases, and policy-related developments that contribute to volatility forecasts and help clarify which forms of information arrival are associated with predictive performance.

The contribution of this study is therefore not that an embedding-based model replaces established volatility dynamics, nor that classical word embeddings represent the frontier of language modelling. Instead, we use a transparent, reproducible, and computationally tractable text representation to study an econometric information-set question: whether news observed before the forecast origin contains incremental predictive information for next-day realised volatility beyond that already contained in strong volatility-history forecasts. This perspective shifts the emphasis from the novelty of the underlying NLP architecture to the economic and econometric value of incorporating unstructured information into volatility forecasting. The empirical design links ML-based measurement of textual information to conventional out-of-sample forecast evaluation, comparisons with established econometric benchmarks, economic assessment, and model interpretability. In particular, the news-only specification should be viewed primarily as a diagnostic experiment that isolates the predictive content of textual information in the absence of lagged realised volatility and other market variables, rather than as an attempt to replace established volatility models. The principal empirical object is instead forecast complementarity: whether information extracted from news can improve forecasts that already incorporate the persistent dynamics of realised volatility. This distinction is important because a textual signal may be economically useful even when it is not sufficiently informative to dominate a strong volatility-history model when used in isolation. Finally, the analysis is explicitly predictive rather than causal. The results concern incremental out-of-sample information content and the extent to which textual and volatility-history information sets complement one another; they should not be interpreted as identifying a structural causal effect of news arrivals or particular news content on subsequent volatility.

The remainder of the paper is organised as follows. \Cref{we_section} outlines the theory and construction of word embeddings. \Cref{fwe_section} presents the specialised word embeddings developed in this study and reports visualisation and evaluation results at multiple language levels. \Cref{RV_section} reviews realised volatility and the HAR-family of models and describes the proposed NLP framework for realised volatility forecasting. \Cref{results_section} reports the out-of-sample forecasting results, including both statistical performance and economic evaluation, and summarises the accompanying explainability evidence. \Cref{conclusions_section} concludes by summarising the main findings and discussing directions for future research.

\section{Word Embedding} \label{we_section}
Text data are inherently high-dimensional and unstructured, which makes them difficult to incorporate directly into statistical models. A central challenge is therefore to construct a numerical representation of text that is both low-dimensional and informative. Word embeddings address this challenge by mapping words to vectors in a continuous space, such that words used in similar contexts have similar representations. Earlier approaches relied on one-hot encoding, in which each word is represented by a binary vector with a single nonzero entry; under this representation, vectors corresponding to distinct words are orthogonal, which precludes any notion of similarity and limits the ability to capture structure in language. For example, while one-hot encoding treats words such as \emph{stock} and \emph{equity}, or \emph{fell} and \emph{declined}, as unrelated despite their similar economic meaning, word embeddings assign similar vectors to these words because they appear in comparable local contexts near words related to prices, markets, or volatility. This context-based representation captures economically meaningful similarities that are lost under one-hot encoding.

Let the text corpus be represented as an ordered sequence of tokens,
\begin{equation}
(w_1, w_2, \ldots, w_T),
\label{eq:token_sequence}
\end{equation}
where a token is a sequence of characters corresponding to a unit of meaning, such as a word, number, or punctuation mark. Each token $w_t$ takes a value in a finite set
\begin{equation}
\mathcal{V} = \{1,2,\ldots,N\},
\label{eq:vocabulary}
\end{equation}
referred to as the vocabulary. Each token $w \in \mathcal{V}$ is associated with a vector
\begin{equation}
\mathbf{e}_w \in \mathbb{R}^d,
\label{eq:word_vector}
\end{equation}
where $d$ is the embedding dimension. Collecting these vectors yields an embedding matrix
\begin{equation}
E = (\mathbf{e}_1, \mathbf{e}_2, \ldots, \mathbf{e}_N)',
\label{eq:embedding_matrix}
\end{equation}
with dimension $N \times d$. The vectors in $E$ are estimated from the data and are not specified ex ante. In empirical applications, $d$ is typically much smaller than $N$. The embedding matrix $E$ is estimated using information on local co-occurrence patterns in the text. Tokens that tend to appear in similar textual environments are assigned similar vectors, so that distances or inner products between vectors provide a quantitative measure of similarity.

In this study, we focus on models that are explicitly designed to estimate word embeddings from text data. Although modern LLMs rely on word embedding layers, such representations form only one component of a substantially more complex architecture and are optimised jointly with many downstream layers for predictive performance rather than for standalone interpretability, making them difficult to isolate from an econometric perspective. By contrast, embedding-based models provide an explicit and reproducible mapping from raw text to numerical variables that is well suited for empirical finance. The resulting representations are low-dimensional, computationally tractable for large corpora, and easily aggregated within standard econometric frameworks. Efficient procedures for estimating such representations include Word2Vec and FastText, which are summarised in \Cref{word2vec_subsection} and \Cref{fasttext_subsection}.

\subsection{Word2Vec} \label{word2vec_subsection} 
Word2Vec is an unsupervised method for estimating the embedding matrix $E$ in \Cref{eq:embedding_matrix} using local co-occurrence patterns in text \citep{mikolov2013efficient}. This method operates on the ordered sequence of tokens $(w_1,\ldots,w_T)$ defined in \Cref{eq:token_sequence} and is based on the idea that tokens appearing in similar local contexts should have similar embedding vectors.

Fix a context window of size $k$. For each token $w_t$, define its local context as the set of surrounding tokens within $k$ positions of $t$. For exposition, let $\mathbf{v}_w$ and $\mathbf{u}_w$ denote the input and output vectors associated with token $w$, respectively. The canonical skip-gram objective can then be written using the full-softmax log-likelihood as follows:
\begin{equation}
L(V,U)
=
\frac{1}{T}
\sum_{t=1}^{T}
\sum_{\substack{j=-k \\ j \neq 0}}^{k}
\log p\!\left(w_{t+j} \mid w_t\right),
\label{eq:word2vec_objective}
\end{equation}
where the conditional probability is defined as
\begin{equation}
p(w_c \mid w_t)
=
\frac{\exp\!\left(\mathbf{u}_{w_c}' \mathbf{v}_{w_t}\right)}
{\sum_{l \in \mathcal{V}} \exp\!\left(\mathbf{u}_{l}' \mathbf{v}_{w_t}\right)}.
\label{eq:word2vec_prob}
\end{equation}

The objective in \Cref{eq:word2vec_objective} corresponds to the skip-gram specification, in which the centre token $w_t$ is used to predict each surrounding context token $w_{t+j}$. The continuous bag-of-words (CBOW) specification reverses the prediction task by predicting the centre token $w_t$ from the average representation of its surrounding context. Let

\begin{equation}
\bar{\mathbf{v}}_{\mathcal{C}_t}
=
\frac{1}{|\mathcal{C}_t|}
\sum_{w_c \in \mathcal{C}_t}
\mathbf{v}_{w_c},
\label{eq:cbow_context}
\end{equation}

where $\mathcal{C}_t$ denotes the set of context tokens surrounding $w_t$. The corresponding conditional probability is

\begin{equation}
p(w_t \mid \mathcal{C}_t)
=
\frac{
\exp\!\left(
\mathbf{u}_{w_t}'\bar{\mathbf{v}}_{\mathcal{C}_t}
\right)
}{
\sum_{l \in \mathcal{V}}
\exp\!\left(
\mathbf{u}_{l}'\bar{\mathbf{v}}_{\mathcal{C}_t}
\right)
}.
\end{equation}

Under both specifications, the embedding vectors are estimated solely from local co-occurrence information. Tokens that appear in similar textual environments are assigned similar vectors, while tokens that appear in different contexts are placed further apart in the embedding space. When the vocabulary size $N$ is large, evaluating the softmax probability in \Cref{eq:word2vec_prob} becomes computationally expensive. One approach to address this issue is hierarchical softmax, which replaces the flat softmax with a tree-based decomposition to reduce computational cost \citep{morin2005hierarchical}. Another approach is negative sampling, which reformulates the estimation problem as a set of binary classification tasks that distinguish observed word--context pairs from randomly sampled noise pairs \citep{mikolov2013distributed}. In the empirical implementation, the FinText embeddings are estimated using negative sampling rather than the full-softmax objective shown above; specifically, five negative samples are used for each observed word--context pair, as detailed in \Cref{Appendix_word_embedding_details}.

\subsection{FastText} \label{fasttext_subsection} 
FastText extends the Word2Vec framework by modifying how word representations are constructed, while retaining a similar estimation strategy \citep{bojanowski2017enriching}. As in Word2Vec, FastText estimates word embeddings using local co-occurrence patterns in the text and optimises a prediction-based objective function, such as the skip-gram or CBOW objective described in \Cref{word2vec_subsection}. The key difference lies in how each token is represented in the model.

In Word2Vec, each token is treated as an atomic unit and is associated with a single embedding vector. In contrast, FastText represents each token as a collection of smaller units, referred to as character n-grams. A character n-gram is a contiguous sequence of $n$ characters extracted from a word. For example, when $n=3$, the token \emph{profit} is decomposed into overlapping character trigrams such as \emph{pro}, \emph{rof}, \emph{ofi}, and \emph{fit}. FastText also includes boundary symbols at the beginning and end of each token to distinguish complete words from character sequences that may appear as parts of other words. This ensures, for instance, that the standalone token \emph{fit} is treated differently from the sequence \emph{fit} appearing inside a longer word.

Formally, let a token $w$ be decomposed into $m_w$ character n-grams together with a whole-word token. FastText associates an embedding vector with each n-gram as well as with the token itself. The representation of token $w$ is then constructed as the sum of its component vectors,
\begin{equation}
\mathbf{e}_w = \sum_{i=0}^{m_w} \mathbf{u}_i,
\label{eq:fasttext_embedding}
\end{equation}
where $\mathbf{u}_0$ denotes the vector associated with the token itself and $\mathbf{u}_i$, for $i=1,\ldots,m_w$, denotes the vector associated with the $i$th character n-gram. This composite vector $\mathbf{e}_w$ replaces the single-token embedding used in Word2Vec. The estimation procedure in FastText is otherwise similar to that of Word2Vec. In both models, the embedding vectors are estimated by maximising a prediction-based objective defined on local co-occurrence patterns in the text. When the vocabulary size is large, direct evaluation of the softmax probability becomes computationally expensive. To address this issue, FastText adopts the same approximation techniques commonly used in Word2Vec.

This representation has important implications for rare and out-of-vocabulary (OOV) tokens. Because embeddings are constructed from character n-grams, FastText can generate meaningful vectors for tokens that appear infrequently or were not observed during training, provided their character components have been seen elsewhere in the corpus. This feature is particularly useful in settings with specialised terminology, abbreviations, or morphological variation. In contrast, Word2Vec can only assign embeddings to tokens that appear explicitly in the training data, which limits its ability to handle rare or unseen words. For comparability across the downstream forecasting specifications, however, we apply a common treatment for out-of-vocabulary tokens when constructing the CNN inputs: a token not contained in the vocabulary of the corresponding pre-trained embedding is mapped to the common padding index. Consequently, FastText's ability to construct subword-based vectors for previously unseen tokens is not exploited in the forecasting exercise.
\section{Financial Word Embedding} \label{fwe_section}
At the time of writing, several well-known general-purpose word embeddings are available. A key characteristic used to compare such embeddings is the size of the corpus on which they are trained, commonly measured by the total number of processed tokens. Corpus size captures how much text the model is exposed to during estimation and is widely used as a proxy for the scale and generality of a word embedding, since larger corpora tend to encompass a broader range of language usage and contextual relationships. Prominent examples include the word embedding of \citet{mikolov2013efficient}, trained using the Word2Vec algorithm described in \Cref{word2vec_subsection} on the Google News dataset with approximately 100 billion tokens, and the WikiNews word embedding, trained using the FastText algorithm described in \Cref{fasttext_subsection} on datasets such as Wikipedia, the UMBC web-based corpus, and the StatMT news dataset, comprising approximately 16 billion tokens.

Although general-purpose word embeddings have demonstrated strong performance in broad language tasks, it is important to consider the potential benefits of developing specialised word embeddings and to conduct a comparative evaluation with general-purpose alternatives at the language level. Financial language exhibits distinctive usage patterns, institutional references, and contextual relationships that are often absent from general text corpora. As a result, many tokens may carry meanings, associations, or economic implications that differ substantially from their usage in non-financial contexts. Using general-purpose word embeddings without prior validation may therefore introduce measurement error into text-based variables and weaken their economic interpretability. This consideration motivates the development and evaluation of specialised word embeddings tailored to financial language before employing them in downstream financial applications. \Cref{design_subsection} describes the steps from data collection to the training of the financial word embedding, together with the associated design considerations, and \Cref{eval_subsection} subsequently presents the evaluation of both general-purpose and financial word embeddings using representation analysis and multiple language-level benchmarks.

\subsection{Design} \label{design_subsection}
To construct specialised word embeddings, the text used for training must be domain-specific, large in scale, and linguistically consistent, with minimal noise and redundancy. Prior research shows that the quality of word representations depends critically on corpus size, contextual richness, and stable domain usage \citep{mikolov2013efficient, pennington2014glove}. Moreover, the use of professionally produced financial text is essential for capturing economically meaningful language patterns and avoiding semantic distortions arising from generic corpora \citep{loughran2011liability, gentzkow2019text}. To satisfy these requirements, this study uses all news stories from the Dow Jones Newswires Text News Feed for the period from 1 January 2000 to 11 September 2015 to develop specialised word embeddings. The Dow Jones Newswires provide extensive, professionally edited coverage of financial markets and corporate events, making them particularly well suited for this purpose. A detailed description of the data cleaning and pre-processing steps applied to this corpus is provided in \Cref{Appendix_data_cleaning}.

		

\begin{figure}
	\begin{subfigure}{0.48\linewidth}
	\centering
	\includegraphics[scale=0.42, trim = {0cm 0.3cm -1.1cm 0cm}]{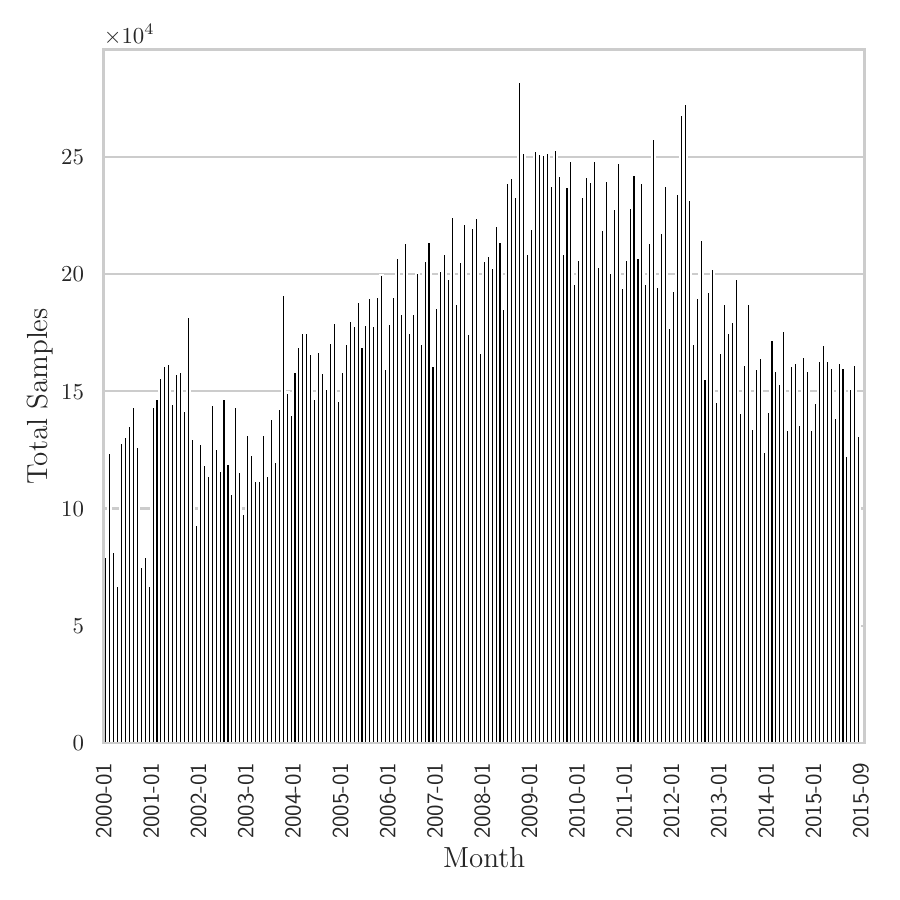}
	\caption{Total Number of Samples}
	\label{word_cloud_stock_relatedxx}
	\end{subfigure}
	\begin{subfigure}{0.48\linewidth}
	\centering
	\includegraphics[scale=0.42, trim = {0cm 0.3cm -1.1cm 0cm}]{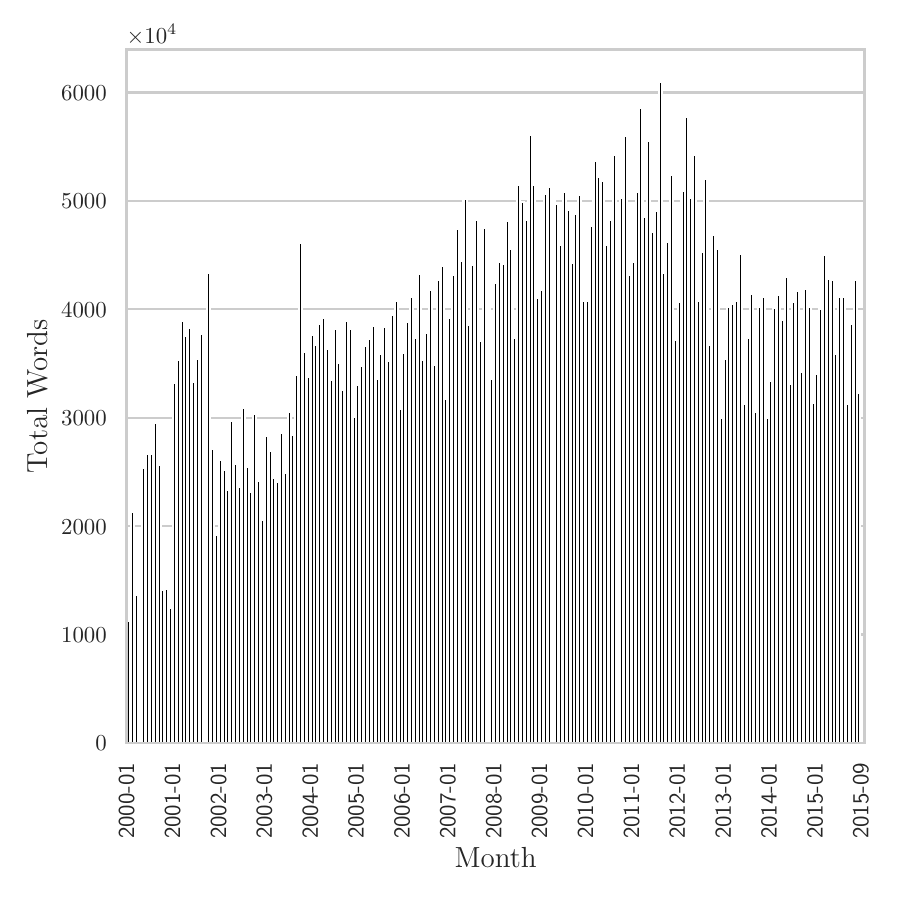}
	\caption{Total Number of Words}
	\label{word_cloud_general_hotxx}
	\end{subfigure}
	\caption{Monthly Corpus Sample and Word Count}
	\begin{minipage}{0.97\linewidth} 
		\rule{\linewidth}{0.03em} \vspace{.15mm} \footnotesize
		{\emph{Notes:} The monthly total sample count and word count used for training the FinText word embedding, covering data from 1 January 2000 to 11 September 2015, are presented in figures (a) and (b), respectively.}
	\end{minipage}  
	\label{WE_world_sample_plot}  
\end{figure}

Turning to corpus size, the Google Word2Vec, WikiNews, and FinText corpora contain approximately 100~billion, 16~billion, and 4.32~billion words, respectively, implying that FinText relies on a substantially smaller corpus than these general-purpose word embeddings. Nevertheless, the FinText dataset expands substantially over time, with the number of monthly samples and the total word count exhibiting clear long-run growth over the 15-year period covered by the news archive, albeit with noticeable time variation, as illustrated in \Cref{WE_world_sample_plot}. Following all pre-processing steps, the resulting FinText corpus comprises 2,733,035 unique tokens.

The details of the word embedding construction, including model specifications and associated hyperparameter settings, are provided in \Cref{Appendix_word_embedding_details}. All pre-processing procedures and hyperparameter configurations closely follow the standard choices commonly adopted in the word embedding literature for general-purpose language models, ensuring methodological consistency across implementations. These common implementation choices facilitate comparisons across the FinText specifications. Comparisons with the publicly available Google and WikiNews embeddings, however, may also reflect differences in training corpora, vocabulary composition, embedding algorithms, and implementation choices. Accordingly, the results should be interpreted as comparisons of the resulting embedding representations rather than as a causal decomposition of the effect of domain-specific training alone.

In total, we trained four versions of the FinText word embedding: Word2Vec (skip-gram), Word2Vec (CBOW), FastText (skip-gram), and FastText (CBOW). In this notation, the first term indicates the underlying embedding algorithm, while the term in parentheses specifies the training architecture used to estimate the word representations. We consider all possible combinations of algorithms and training architectures in order to systematically assess their relative performance and to obtain a comprehensive understanding of how these design choices affect the resulting word embeddings. This approach is also consistent with the construction of widely used general-purpose word embeddings, which are themselves developed under different algorithm–architecture combinations.\footnote{Word embeddings trained from large text corpora may reflect biases inherent in the underlying data. In the present setting, this risk is partially mitigated by training on a long-span corpus of professionally edited news covering a large set of international news agencies, which reduces sensitivity to transient language, short-lived narratives, and idiosyncratic editorial perspectives.}

\subsection{Language-Level Evaluation} \label{eval_subsection}
We evaluate both general-purpose and specialised word embeddings using a combination of standard language benchmarks, visualisation techniques, and domain-specific tasks. The analysis begins with widely used general-purpose benchmarks to assess how different embedding algorithms and training architectures capture generic semantic relationships. We then examine the structure of the embedding spaces through low-dimensional visualisations to gain insight into how linguistic and sectoral information is organised. Finally, we turn to finance-oriented evaluations, including analogy tasks and a purpose-built gold-standard financial benchmark, to assess whether word embeddings trained on specialised financial corpora capture economically meaningful relationships that are not adequately represented by general-purpose models.

\begin{table}
	\centering
	\begin{adjustbox}{max width={0.8\textwidth}}
		\begin{threeparttable}
  \caption{Word Embedding Comparison (Google Analogy)}
    \begin{tabular}{ccccccc}
    \toprule
    
    &  \multicolumn{3}{c}{\textbf{Word2Vec}\tnote{a}} & \multicolumn{3}{c}{\textbf{FastText}}\\

    \cmidrule(lr){2-4}\cmidrule(lr){5-7}
    
    \textbf{Section} & \begin{tabular}{@{}c@{}}\textbf{FinText\tnote{b}} \\ \scalebox{0.8}{(CBOW)\tnote{c}}\end{tabular} & \begin{tabular}{@{}c@{}}\textbf{FinText} \\ \scalebox{0.8}{(skip-gram)}\end{tabular} & \begin{tabular}{@{}c@{}}\textbf{Google} \\ \scalebox{0.8}{(skip-gram)}\end{tabular} & \begin{tabular}{@{}c@{}}\textbf{WikiNews} \\ \scalebox{0.8}{(skip-gram)}\end{tabular} & \begin{tabular}{@{}c@{}}\textbf{FinText} \\ \scalebox{0.8}{(skip-gram)}\end{tabular} & \begin{tabular}{@{}c@{}}\textbf{FinText} \\ \scalebox{0.8}{(CBOW)}\end{tabular} \\   
    \midrule 
    
    capital-common-countries & 77.27 & 85.50 & 83.60 & 100 & 85.93 & 47.40\\
  
    capital-world & 63.60 & 75.87 & 82.72 & 98.78 & 71.06 & 35.79\\
  
    currency & 22.49 & 36.69 & 39.84 & 25.00 & 32.54 & 10.65\\

    city-in-state & 19.93 & 60.48 & 74.64 & 81.41 & 58.20 & 15.83\\
 
    family & 63.46 & 70.51 & 90.06 & 98.69 & 58.97 & 59.62\\

    gram1-adjective-to-adverb & 27.47 & 33.00 & 32.27 & 70.46 & 50.59 & 79.45\\
  
    gram2-opposite & 33.33 & 32.50 & 50.53 & 73.91 & 50.83 & 71.67\\

    gram3-comparative & 77.65 & 75.04 & 91.89 & 97.15 & 77.06 & 87.39\\
  
    gram4-superlative & 61.67 & 55.00 & 88.03 & 98.68 & 62.14 & 90.71\\

    gram5-present-participle & 62.30 & 61.24 & 79.77 & 97.53 & 70.63 & 76.06\\

    gram6-nationality-adjective & 88.11 & 93.23 & 97.07 & 99.12 & 94.05 & 79.05\\

    gram7-past-tense & 42.02 & 39.92 & 66.53 & 87.25 & 37.98 & 31.09\\
 
    gram8-plural & 59.23 & 62.46 & 85.58 & 98.69 & 70.92 & 79.54\\
   
    gram9-plural-verbs & 53.26 & 54.53 & 68.95 & 97.38 & 61.59 & 79.17\\
   \midrule
    Overall & 53.65 & 62.86 & 77.08 & 91.44 & 65.00 & 55.74\\
    \bottomrule
    \end{tabular}%
			\begin{tablenotes}[para,flushleft]
				\footnotesize
				\emph{Notes:} \item[a] To learn word embeddings from textual datasets, \textbf{Word2Vec} was developed by \citet{mikolov2013efficient}, and \textbf{FastText}, an extension of the Word2Vec algorithm, was developed by \citet{bojanowski2017enriching}. \item[b] The following word embeddings are utilised in this study: a financial word embedding developed specifically for this study (\textbf{FinText}); a publicly available word embedding trained on a portion of the Google news dataset (\textbf{Google}); and a publicly available word embedding trained on the Wikipedia dataset, UMBC web-based corpus, and StatMT news dataset (\textbf{WikiNews}). \item[c] CBOW and skip-gram are the unsupervised learning models proposed by \citet{mikolov2013efficient} for learning distributed representations of tokens.

			\end{tablenotes}
			\label{word_embedding_compare_google_analogy}
		\end{threeparttable}
	\end{adjustbox}
	
\end{table}

\begin{table}
	\centering
	\begin{adjustbox}{max width={0.73\textwidth}}
		\begin{threeparttable}
  \caption{Word Embedding Comparison (Gold-Standard Collections)}
    \begin{tabular}{ccccccc}
    \toprule
    
    &  \multicolumn{3}{c}{\textbf{Word2Vec}\tnote{a}} & \multicolumn{3}{c}{\textbf{FastText}}\\

    \cmidrule(lr){2-4}\cmidrule(lr){5-7}
    
    \textbf{Benchmark} & \begin{tabular}{@{}c@{}}\textbf{FinText\tnote{b}} \\ \scalebox{0.8}{(CBOW)\tnote{c}}\end{tabular} & \begin{tabular}{@{}c@{}}\textbf{FinText} \\ \scalebox{0.8}{(skip-gram)}\end{tabular} & \begin{tabular}{@{}c@{}}\textbf{Google} \\ \scalebox{0.8}{(skip-gram)}\end{tabular} & \begin{tabular}{@{}c@{}}\textbf{WikiNews} \\ \scalebox{0.8}{(skip-gram)}\end{tabular} & \begin{tabular}{@{}c@{}}\textbf{FinText} \\ \scalebox{0.8}{(skip-gram)}\end{tabular} & \begin{tabular}{@{}c@{}}\textbf{FinText} \\ \scalebox{0.8}{(CBOW)}\end{tabular} \\  
    \midrule
    \begin{tabular}{@{}c@{}}WordSim-353\tnote{d} \\ (relatedness)\end{tabular} & \begin{tabular}{@{}c@{}}0.3821 \end{tabular} & \begin{tabular}{@{}c@{}}0.4993 \end{tabular} & \begin{tabular}{@{}c@{}}0.6096 \\ \end{tabular} & \begin{tabular}{@{}c@{}}0.6018 \\ \end{tabular} & \begin{tabular}{@{}c@{}}0.4425 \\ \end{tabular} & \begin{tabular}{@{}c@{}}0.1677 \\\end{tabular}  \\
    \midrule
    \begin{tabular}{@{}c@{}}WordSim-353 \\ (similarity)\end{tabular} & \begin{tabular}{@{}c@{}}0.6126 \\ \end{tabular} & \begin{tabular}{@{}c@{}}0.6436 \\ \end{tabular} & \begin{tabular}{@{}c@{}}0.7407 \\ \end{tabular} & \begin{tabular}{@{}c@{}}0.6713 \\ \end{tabular} & \begin{tabular}{@{}c@{}}0.6393 \\ \end{tabular} & \begin{tabular}{@{}c@{}}0.4722 \\ \end{tabular} \\
    \midrule
    SimLex & \begin{tabular}{@{}c@{}}0.2657 \\ \end{tabular} & \begin{tabular}{@{}c@{}}0.2650 \\ \end{tabular} & \begin{tabular}{@{}c@{}}0.3638 \\ \end{tabular} & \begin{tabular}{@{}c@{}}0.3985 \\ \end{tabular} & \begin{tabular}{@{}c@{}}0.2772
    \\ \end{tabular} & \begin{tabular}{@{}c@{}}0.2574
    \\ \end{tabular}\\ 

    \bottomrule
    \end{tabular}%
			\begin{tablenotes}[para,flushleft]
				\footnotesize
				\emph{Notes:} \item[a] To learn word embeddings from textual datasets, \textbf{Word2Vec} was developed by \citet{mikolov2013efficient}, and \textbf{FastText}, an extension of the Word2Vec algorithm, was developed by \citet{bojanowski2017enriching}. \item[b] The following word embeddings are utilised in this study: a financial word embedding developed specifically for this study (\textbf{FinText}); a publicly available word embedding trained on a portion of the Google news dataset (\textbf{Google}); and a publicly available word embedding trained on the Wikipedia dataset, UMBC web-based corpus, and StatMT news dataset (\textbf{WikiNews}). \item[c] CBOW and skip-gram are the unsupervised learning models proposed by \citet{mikolov2013efficient} for learning distributed representations of tokens. \item[d] WordSim-353 refers to the similarity and relatedness splits of the original WordSim-353 dataset, as re-annotated by \citet{agirre2009study}, while SimLex \citep{hill2015simlex} serves as another gold-standard dataset that focuses specifically on similarity.

			\end{tablenotes}
			\label{word_embedding_compare_gold_standard_collections353}
		\end{threeparttable}
	\end{adjustbox}
\end{table}

\Cref{word_embedding_compare_google_analogy} compares four versions of FinText against general-purpose word embeddings based on the Google Analogy benchmark. The Google Analogy benchmark is a widely used dataset for evaluating the quality of word embeddings. Each section in the Google Analogy benchmark contains a set of analogies. For example, under the `capital-common-countries' section, the model is evaluated on analogies such as `London to England is like Paris to ?'. From \Cref{word_embedding_compare_google_analogy}, it is apparent that, except for `currency' and `gram1-adjective-to-adverb', WikiNews achieves the highest predictive accuracy. The overall evaluation score supports this observation. For this general-purpose benchmark, FinText is outperformed by Google under Word2Vec and by WikiNews under FastText. The individual and overall scores for FinText indicate that the skip-gram model performs better than CBOW.

\Cref{word_embedding_compare_gold_standard_collections353} presents the predictive accuracy based on the gold-standard collections, namely \mbox{WordSim-353} \citep{agirre2009study} for measuring word relatedness and similarity, and SimLex \citep{hill2015simlex}, which focuses on similarity. Relatedness and similarity capture distinct types of relationships between words. Relatedness refers to an associative connection; words may be conceptually linked even if they do not share a direct meaning. For example, \textit{doctor} and \textit{hospital} are related because they appear in similar contexts but are not synonyms. Similarity, on the other hand, requires words to have a close or nearly identical meaning. For instance, \textit{doctor} and \textit{physician} are similar, as they essentially refer to the same concept. Accounting for these differences is crucial for language models, as it enables them to capture a broader range of semantic relationships between words. All these collections contain human-assigned judgements about the relatedness and similarity of word pairs. Performance is measured by Spearman's rank correlation coefficient. It is evident from \Cref{word_embedding_compare_gold_standard_collections353} that Google’s word embeddings outperform under WordSim-353, while WikiNews embeddings perform better under SimLex. As before, FinText is outperformed by Google under Word2Vec and outperformed by WikiNews under FastText. Additionally, for both Word2Vec and FastText algorithms, the skip-gram model is generally superior to the CBOW model, with only one exception.

\afterpage{%
	\clearpage
	\thispagestyle{empty}
	\atxy{\dimexpr\paperwidth-0.45in}{.5\paperheight}{\rotatebox[origin=center]{90}{\thepage}}
\begin{landscape}
\begin{figure}
	\centering
	\includegraphics[scale=0.55, trim = {0cm 1.2cm 0cm 0cm}]{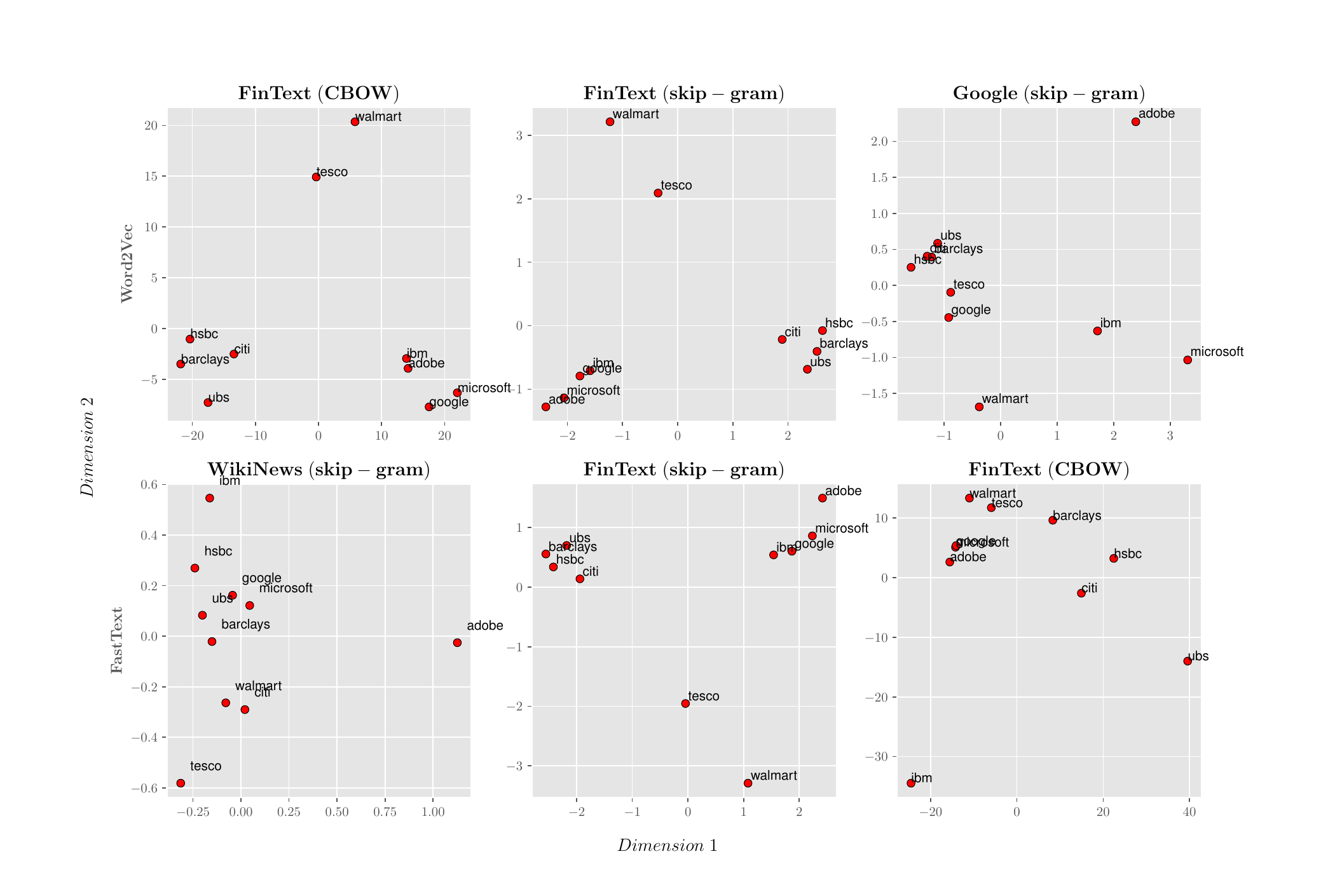}
	\caption{2D Visualisation of Word Embeddings}
	\label{2D_visualisation_word_embedding}
		\begin{minipage}{0.8\linewidth}
		\rule{\linewidth}{0.02em}
		\footnotesize
		{\emph{Notes:} This figure presents a two-dimensional visualisation of selected word embeddings obtained via PCA, which reduces the original 300-dimensional vectors to two dimensions. The selected tokens comprise technology firms (\textit{microsoft}, \textit{ibm}, \textit{google}, \textit{adobe}), financial services and investment banking firms (\textit{barclays}, \textit{citi}, \textit{ubs}, \textit{hsbc}), and retail firms (\textit{tesco}, \textit{walmart}). The first and second rows correspond to Word2Vec and FastText embeddings, respectively, estimated using FinText as well as publicly available Google Word2Vec and WikiNews embeddings. As the sign of each principal component is arbitrary, the direction of each axis may be reversed without affecting the PCA representation. The visualisation is therefore intended to illustrate relative clustering and separation patterns among embeddings rather than to support a literal interpretation of axis directions or distances.}  
	\end{minipage}
\end{figure}
\end{landscape}
}

To obtain a clearer understanding of how different word embeddings handle financial terminology, we illustrate their behaviour through a visualisation-based example that examines sectoral relationships among firm names. In this setting, tokens representing firms operating in similar sectors are expected to be embedded closer together, as their names tend to appear in comparable financial contexts. As discussed in \Cref{we_section}, each token is represented by a 300-dimensional embedding vector, which we project onto a two-dimensional space using principal component analysis (PCA) for visual inspection. \Cref{2D_visualisation_word_embedding} presents the resulting projection for selected company tokens. Word2Vec results are shown in the top row, while FastText results appear in the bottom row. The projection suggests that, under FinText, technology firms such as \textit{microsoft}, \textit{ibm}, \textit{google}, and \textit{adobe}, financial institutions such as \textit{barclays}, \textit{citi}, \textit{ubs}, and \textit{hsbc}, and retail firms such as \textit{tesco} and \textit{walmart} appear more coherently grouped relative to the general-purpose word embeddings. This pattern suggests that domain-specific training enables FinText to capture sector-related usage of firm names.

\begin{table}
\centering
\caption{Financial Analogy Examples}
	\begin{adjustbox}{max width={0.72\textwidth}}
		\begin{threeparttable}
\label{financial_analogy_examples_table}
\begin{tabular}{cccc}
\toprule
 & \multicolumn{3}{c}{\textbf{Word Embedding}} \\
\cmidrule(lr){2-4}
 \textbf{Analogy} & \textbf{Google} & \textbf{WikiNews} & \textbf{FinText\tnote{a}} \\
 \cmidrule(lr){1-4}
debit:credit :: positive:X & positive & negative & negative \\
bullish:bearish :: rise:X & rises & rises & fall \\
apple:iphone :: microsoft:X & windows\_xp & iphone & windows \\
us:uk :: djia:X & NONE\tnote{b} & NONE & ftse\_100 \\
microsoft:msft :: amazon:X & aapl & hmv & amzn \\
bid:ask :: buy:X & tell & ask- & sell \\
creditor:lend :: debtor:X & lends & lends & borrow \\
rent:short\_term :: lease:X & NONE & NONE & long\_term \\
growth\_stock:overvalued :: value\_stock:X & NONE & NONE & undervalued \\
us:uk :: nyse:X & nasdaq & hsbc & lse \\
call\_option:put\_option :: buy:X & NONE & NONE & sell \\

\bottomrule
\end{tabular}
\begin{tablenotes}[para,flushleft]
				\footnotesize
                \emph{Notes:} \item[a] Among the four variants considered, this table reports the FinText specification developed using the Word2Vec algorithm with a skip-gram model. \item[b] Not included in the vocabulary list.
			\end{tablenotes}
		\end{threeparttable}
	\end{adjustbox}
\end{table}

\Cref{financial_analogy_examples_table} also reports a set of simple financial analogy tasks designed to assess whether different word embeddings capture financially meaningful relationships. Each row follows the structure `A is to B as C is to X', where the intended answer reflects a financially meaningful association. Analogy tests provide a transparent and intuitive way to evaluate word embeddings, as they directly reveal whether economically meaningful relationships are encoded in the embedding space without relying on complex downstream models. For example, just as a \emph{debit} is offset by a \emph{credit}, a \emph{positive} position should be offset by a \emph{negative} one; similarly, \emph{bullish} corresponds to rising prices, while \emph{bearish} corresponds to falling prices. A word embedding performs well if it returns the financially intuitive counterpart rather than a token that is merely lexically similar. The results show that the general-purpose word embeddings frequently fail to return meaningful financial answers. In contrast, FinText more frequently recovers financially intuitive relationships, including exchange identifiers, trading concepts, and valuation terminology. Overall, these examples demonstrate that FinText encodes financial concepts and institutional knowledge that are not captured by general-purpose word embeddings.

While general-purpose language benchmarks provide useful information about overall linguistic competence, they are not sufficient in finance, where language encodes institutional facts and economically meaningful relationships. Also, visualisations and illustrative examples are helpful for intuition but remain qualitative and cannot provide a systematic assessment of whether word embeddings capture financial knowledge. This motivates the introduction of a gold-standard financial benchmark to rigorously evaluate finance-specific relationships. The benchmark is based on firm-level information from Bureau van Dijk’s Orbis database and consists of seven groups of financial analogy tasks covering different settings, with full construction details provided in \Cref{Appendix_gold_benchmak}. Similar to the previous benchmarks, this benchmark evaluates word embeddings using simple financial analogies. For example, if \textit{Apple} is associated with the ticker \textit{AAPL}, then \textit{Amazon} should be associated with \textit{AMZN}; if \textit{Microsoft} is listed on \textit{NASDAQ}, then \textit{IBM} should be listed on \textit{NYSE}; and if \textit{HSBC} is headquartered in the \textit{UK}, then \textit{JPMorgan} should be headquartered in the \textit{US}. A word embedding performs well if it consistently recovers these known financial relationships.

\begin{table}
	\centering
	\begin{adjustbox}{max width=0.94\textwidth}
		\begin{threeparttable}
  \caption{Gold-Standard Financial Benchmark}
    \begin{tabular}{lcccccc}
    \toprule
    
    & \multicolumn{3}{c}{\textbf{Word2Vec}\tnote{a}} & \multicolumn{3}{c}{\textbf{FastText}}\\

    \cmidrule(lr){2-4}\cmidrule(lr){5-7}
    
    \textbf{Group} & \begin{tabular}{@{}c@{}}\textbf{FinText\tnote{b}} \\ \scalebox{0.8}{(CBOW)\tnote{c}}\end{tabular} & \begin{tabular}{@{}c@{}}\textbf{FinText} \\ \scalebox{0.8}{(skip-gram)}\end{tabular} & \begin{tabular}{@{}c@{}}\textbf{Google} \\ \scalebox{0.8}{(skip-gram)}\end{tabular} & \begin{tabular}{@{}c@{}}\textbf{WikiNews} \\ \scalebox{0.8}{(skip-gram)}\end{tabular} & \begin{tabular}{@{}c@{}}\textbf{FinText} \\ \scalebox{0.8}{(skip-gram)}\end{tabular} & \begin{tabular}{@{}c@{}}\textbf{FinText} \\ \scalebox{0.8}{(CBOW)}\end{tabular} \\   
    \midrule 
    
   Ticker to City (US) & 14.74 & 23.68 & 0.26 & 0.00 & 15.00 & 1.05 \\
  
   Name to Ticker (US) & 62.37 & 62.89 & 0.00 & 0.00 & 54.21 & 37.11 \\
  
   Name to Incorporation year (US) & 7.37 & 5.79 & 0.00 & 2.63 & 10.00 & 5.79 \\

   Name to Exchange (US) & 29.74 & 5.53 & 0.00 & 0.26 & 0.26 & 0.00 \\
 
   Name to State (US) & 6.58 & 3.42 & 0.00 & 0.26 & 1.32 & 0.00 \\

   Name to Country (US \& UK) & 10.79 & 8.16 & 0.00 & 0.00 & 1.58 & 0.00 \\

   Name to Country (US, UK, China, \& Japan) & 5.53 & 4.47 & 0.00 & 0.00 & 5.26 & 0.00 \\

   \midrule
    Overall & 19.59 & 16.28 & 0.04 & 0.45 & 12.52 & 6.28 \\
    \bottomrule
    \end{tabular}%
			\begin{tablenotes}[para,flushleft]
				\footnotesize
				{\emph{Notes:} \item[a] To learn word embeddings from textual datasets, \textbf{Word2Vec} was developed by \citet{mikolov2013efficient}, and \textbf{FastText}, an extension of the Word2Vec algorithm, was developed by \citet{bojanowski2017enriching}. \item[b] The following word embeddings are utilised in this study: a financial word embedding developed specifically for this study (\textbf{FinText}); a publicly available word embedding trained on a portion of the Google news dataset (\textbf{Google}); and a publicly available word embedding trained on the Wikipedia dataset, UMBC web-based corpus, and StatMT news dataset (\textbf{WikiNews}). \item[c] CBOW and skip-gram are the unsupervised learning models proposed by \citet{mikolov2013efficient} for learning distributed representations of tokens.}
			\end{tablenotes}
			\label{word_embedding_fintext_benchmark}
		\end{threeparttable}
	\end{adjustbox}
\end{table}

The results are presented in \Cref{word_embedding_fintext_benchmark}. Each benchmark group contains 380 analogy questions, yielding 2,660 financial analogies across all seven groups, and performance is measured using top-5 accuracy. General-purpose embeddings perform poorly across these finance-specific relationships, while a FinText specification achieves the highest accuracy in every benchmark group. Among the FinText variants, Word2Vec with the CBOW architecture achieves the highest overall top-5 accuracy of 19.59\%. The benchmark results should primarily be interpreted in relative rather than absolute terms. Although FinText considerably outperforms the general-purpose alternatives, an overall top-5 accuracy of approximately 20\% indicates that the benchmark remains demanding and that domain-specific training does not amount to comprehensive financial-language understanding. The evidence therefore supports the narrower conclusion that specialised financial embeddings encode the finance-specific relationships considered here more effectively than the general-purpose alternatives.\footnote{The FinText word embeddings and the gold-standard financial benchmark are available for download at \href{https://fintext.ai}{FinText.ai}.}

\section{Realised Volatility Forecasting} \label{RV_section}
A substantial body of literature establishes that financial market volatility is fundamentally driven by the arrival and interpretation of information. Seminal contributions such as \citet{engle1993measuring} and \citet{andersen1998answering}, together with subsequent intraday evidence, show that information arrival and return innovations are closely linked to volatility clustering and sudden volatility episodes \citep{gallo2000effects, kalev2004public}. More broadly, this line of research emphasises that volatility dynamics reflect heterogeneous information flows operating across different horizons, which in turn generate the persistent behaviour commonly observed in financial markets \citep{andersen2007roughing}. However, most empirical volatility forecasting models incorporate this information only indirectly, relying on past volatility measures as reduced-form summaries of how markets have responded to information arrivals. Although this approach has proven empirically successful, it abstracts from the underlying informational content and treats volatility persistence primarily as a time-series phenomenon. Recent evidence further shows that public-information arrival, sentiment, and attention measures are closely related to volatility dynamics and can provide predictive information beyond conventional market variables \citep{audrino2020impact,engle2021news}. This motivates testing textual news as a direct proxy for information arrival.

\subsection{Realised Volatility Forecasting and HAR Models} \label{review_rv_subsection}
Assume that \( P_t \) denotes the stock price process, whose dynamics are given by
\begin{equation}
d\log(P_t) = \mu_t\,dt + \sigma_t\,dW_t + \kappa_t\,dQ_t,
\label{eq_pricedynamics}
\end{equation}
where \( \mu_t \) is the drift component, assumed to be continuous, \( \sigma_t \) is the càdlàg volatility process, \( W_t \) is a standard Brownian motion, \( \kappa_t \) denotes the jump size, and \( Q_t \) is a Poisson process. Over the interval \( [t-1,t] \), the integrated variance (IV) is defined as
\begin{equation}
IV_t = \int_{t-1}^{t} \sigma_s^2 \, ds.
\label{eq_iv}
\end{equation}
A corresponding high-frequency measure of return variation is the realised variance,
\begin{equation}
RV_t \equiv \sum_{i=1}^{M} r_{t,i}^2,
\label{main_RV_def_eq2}
\end{equation}
where \( M = 1/\delta \) and the \( \delta \)-period intraday return, expressed in percentage points, is defined as
\begin{equation}
r_{t,i}
\equiv
100\left[
\log\!\left(P_{t-1+i\delta}\right)
-
\log\!\left(P_{t-1+(i-1)\delta}\right)
\right].
\end{equation}
Accordingly, \(RV_t\), \(BPV_t\), and the realised semivariances are expressed in squared percentage-point units, while \(RQ_t\) is expressed in fourth-power percentage-point units.

Given that intraday returns are expressed in percentage points, \(RV_t/10^4\) is a consistent estimator of \(IV_t\) as \(\delta \to 0\) in the absence of jumps \citep{barndorff2002estimating}. Building on this framework, the HAR-family of models is employed to forecast realised variance (RV), which constitutes the main forecasting task of this study. To evaluate NLP models for RV forecasting, it is necessary to establish a strong and well-defined set of benchmark models. In general, these models can be written as
\begin{equation}
RV_{t+1} = f\!\left(\overline{RV}_{t-i}, J_t, \overline{BPV}_{t-i}, RV_t^{+}, RV_t^{-}, \overline{RQ}_{t-i}\right),
\label{har_general_spec}
\end{equation}
where \( \overline{RV}_{t-i} \) denotes the average RV over the previous \( i \) days. The jump component is defined as \( J_t = \max(RV_t - BPV_t, 0) \), with bipower variation given by
\begin{equation}
BPV_t = \frac{\pi}{2} \sum_{i=1}^{M-1} |r_{t,i}|\,|r_{t,i+1}|,
\label{bpv}
\end{equation}
where \(M\) denotes the number of intraday returns. The positive and negative realised semivariances are defined as
\begin{equation}
RV_t^{+} \equiv \sum_{i=1}^{M} r_{t,i}^2 \mathbb{I}(r_{t,i} > 0),
\qquad
RV_t^{-} \equiv \sum_{i=1}^{M} r_{t,i}^2 \mathbb{I}(r_{t,i} < 0),
\label{har_SHAR}
\end{equation}
where \( \mathbb{I}(\cdot) \) is an indicator function. Finally, realised quarticity (RQ) is defined as
\begin{equation}
RQ_t \equiv \left(\frac{M}{3}\right)\sum_{i=1}^{M} r_{t,i}^4,
\label{har_HARQ}
\end{equation}
where \( \overline{RQ}_{t-i} \) denotes its average over the previous \( i \) days. 

Within this setting, all models considered can be viewed as restricted versions of the general specification in \Cref{har_general_spec}. The AR model restricts the information set to the daily realised variance \( RV_t \) only. The benchmark HAR model of \citet{corsi2009simple} extends this specification by including RV measured at heterogeneous horizons, namely the daily component \( RV_t \) and its weekly and monthly averages \( \overline{RV}^w_t \) and \( \overline{RV}^m_t \), where the weekly component is defined as the average RV over the preceding five trading days and the monthly component as the average over the preceding twenty-one trading days. The SHAR model of \citet{patton2015good} replaces the daily realised variance component with the positive and negative realised semivariances \(RV_t^{+}\) and \(RV_t^{-}\) defined in \Cref{har_SHAR}, while retaining the weekly and monthly RV components. The HAR-J model extends the HAR framework by adding the jump component \( J_t \), while the CHAR model replaces RV with the continuous volatility measure based on bipower variation as defined in \Cref{bpv} \citep{andersen2007roughing}. To address measurement error in RV, the ARQ and HARQ models of \citet{bollerslev2016exploiting} allow the coefficient on lagged realised variance to vary with realised quarticity. Specifically, the HARQ specification augments HAR with the interaction $\sqrt{RQ_t}RV_t$, so that the persistence of daily RV varies with the magnitude of measurement error. Following \citet{bollerslev2016exploiting}, the square-root quarticity term is demeaned using its mean over the corresponding rolling estimation window for ease of interpretation; for notational simplicity, \(RQ_t^{1/2}\) continues to denote the demeaned quantity. The ARQ model is obtained by excluding the weekly and monthly HAR components. We additionally consider a further HARQ extension, denoted HARQ-F in this study, which incorporates realised quarticity information over weekly and monthly horizons.\footnote{As noted by \citet{andersen2007roughing}, one possible strategy for handling heavy-tailed distributions is to model the logarithm of RV. However, as discussed by \citet{patton2015good}, this approach effectively targets forecasts of log-volatility rather than volatility in levels, even though the latter is generally of greater relevance for most economic and financial applications. Accordingly, and consistent with the established literature on the HAR-family of models, this study focuses on forecasting RV in levels.}

To formally implement the forecasting experiments, the sample periods, estimation procedure, and data construction are specified as follows. The in-sample period spans from 27 July 2007 to 11 September 2015, yielding 2,046 estimation observations, while the out-of-sample period extends from 14 September 2015 to 27 January 2022, yielding 1,604 out-of-sample forecasts. Importantly, September 2015 marks the end of the embedding-estimation and in-sample periods rather than the end of the empirical dataset; forecast evaluation continues through January 2022.\footnote{The rolling estimation window comprises 2,046 daily observations, corresponding to approximately eight years of data. This choice reflects a standard bias–variance trade-off in volatility forecasting, providing sufficient observations for stable estimation while remaining responsive to gradual structural change. As an additional robustness check, we also consider an estimation window equal to half this size; the corresponding results are reported in \Cref{har_outofsample_table_base_short}.} RV is constructed from 5-minute intraday returns based on transaction prices, consistent with \citet{liu2015does}, who highlight the difficulty of significantly beating the 5-minute RV benchmark. The HAR-family forecasts are generated using a daily rolling-window estimation scheme with a fixed window length of 2,046 days.\footnote{As documented in \Cref{design_subsection}, the FinText word embedding dataset covers the period from 1 January 2000 to 11 September 2015. For the FinText-based specifications, the out-of-sample forecasting exercise is therefore free from look-ahead bias in the estimation of the word embeddings and forecasting models, with the FinText embeddings treated as fixed representations trained only on the historical corpus ending on 11 September 2015. The publicly available general-purpose embeddings are treated as external representation benchmarks; accordingly, results based on these embeddings should not be interpreted as strictly real-time evaluations of the embedding construction process.} More generally, at each forecast origin, all textual and market information entering the forecasting models is available before the intraday returns used to construct the corresponding next-day RV are realised. Accordingly, the empirical design evaluates the incremental out-of-sample predictive information contained in news beyond volatility-history variables and should be interpreted as a predictive rather than causal exercise. RV is computed over NASDAQ trading hours from 9:30~AM to 4:00~PM Eastern Time using limit order book (LOB) data obtained from the LOBSTER database \citep{huang2011lobster}. The empirical analysis is conducted on a cross-section of 23 stocks selected based on liquidity and data availability over the sample period.\footnote{This study focuses on a sample of 23 large, liquid U.S.-listed securities. These stocks are selected to ensure high-quality and continuous LOB data, reliable RV measurement, and sufficiently rich and regular firm-specific news coverage over the sample period. The number of stocks included in the analysis is also consistent with the range commonly adopted in the empirical RV forecasting literature. Moreover, expanding the cross-sectional dimension beyond this level would impose substantially higher computational requirements that are beyond the scope of this study. Accordingly, the external validity of the results is primarily limited to large, liquid, and heavily covered equities, and the findings need not extend unchanged to smaller, less liquid, or sparsely covered firms.} All data-cleaning procedures follow the guidelines of \citet{barndorff2009realized}, with a detailed description provided in \Cref{appendix_RV_cleaning}. Descriptive statistics for RV across these 23 stocks are presented in \Cref{descriptive_stat_table}.

\begin{table}
	\centering
	\begin{adjustbox}{max width=0.83\textwidth}
		\begin{threeparttable}
			\caption{RV Descriptive Statistics}
			\begin{tabular}{cccccccccc} \toprule
				{Ticker} & {Min} & {Max} & {\nth{1} Quartile} & {Median} & {\nth{3} Quartile} & {Mean} & {STD} & {Kurtosis} & {Skewness}\\ \midrule
				{AAPL} & 0.102 & 229.420 & 0.899 & 1.733 & 3.680 & 4.623 & 12.596 & 111.012 & 9.124  \\
				{MSFT} & 0.067 & 216.181 & 0.829 & 1.449 & 2.814 & 3.237 & 8.125 & 194.004 & 11.275 \\
				{INTC} & 0.030 & 318.697 & 1.103 & 1.873 & 3.577 & 4.299 & 11.628 & 294.963 & 13.982 \\
				{CMCSA} & 0.004 & 237.387 & 0.910 & 1.632 & 3.320 & 3.821 & 9.697 & 192.169 & 11.462 \\
				{QCOM} & 0.122 & 373.543 & 1.024 & 1.975 & 4.129 & 5.073 & 15.380 & 200.609 & 12.100 \\
				{CSCO} &0.047 & 343.946 & 0.886 & 1.561 & 3.028 & 4.115 & 13.160 & 212.453 & 12.258 \\
				{EBAY} & 0.205 & 252.608 & 1.319 & 2.271 & 4.356 & 5.082 & 12.592 & 142.684 & 10.009 \\
				{GILD} & 0.064 & 259.489 & 1.167 & 1.892 & 3.379 & 4.304 & 12.930 & 182.820 & 12.063 \\        
				{TXN} & 0.177 & 287.897 & 1.047 & 1.905 & 3.748 & 4.014 & 9.820 & 311.666 & 14.242 \\
				{AMZN} & 0.065 & 547.030 & 1.305 & 2.336 & 4.808 & 6.200 & 19.359 & 242.205 & 12.735 \\
				{SBUX} & 0.052 & 265.094 & 0.864 & 1.594 & 3.423 & 4.201 & 11.237 & 161.435 & 10.626 \\
				{NVDA} & 0.159 & 1104.351 & 2.282 & 4.358 & 9.084 & 9.756 & 30.117 & 586.612 & 20.058 \\        
				{MU} & 0.292 & 484.388 & 3.570 & 6.246 & 11.912 & 12.818 & 25.734 & 89.141 & 7.960 \\
				{AMAT} & 0.292 & 531.579 & 1.783 & 3.028 & 5.712 & 6.005 & 14.632 & 532.194 & 18.338 \\
				{NTAP} & 0.119 & 462.821 & 1.503 & 2.587 & 5.154 & 6.289 & 18.008 & 201.510 & 11.934 \\        
				{ADBE} & 0.119 & 569.720 & 1.099 & 2.020 & 3.908 & 4.947 & 15.003 & 588.095 & 18.867 \\
				{XLNX} & 0.229 & 265.374 & 1.296 & 2.363 & 4.787 & 5.005 & 11.941 & 194.718 & 11.764 \\
				{AMGN} & 0.032 & 214.156 & 0.969 & 1.593 & 2.872 & 3.398 & 9.612 & 183.759 & 11.898 \\
				{VOD} & 0.055 & 219.033 & 0.687 & 1.342 & 3.137 & 3.933 & 10.869 & 122.252 & 9.601 \\
				{CTSH} & 0.189 & 485.894 & 0.984 & 1.764 & 4.161 & 5.288 & 15.757 & 325.214 & 14.287 \\
				{KLAC} & 0.154 & 499.808 & 1.456 & 2.710 & 5.416 & 5.919 & 16.878 & 354.626 & 16.033 \\
				{PCAR} & 0.039 & 389.930 & 1.157 & 2.162 & 4.633 & 5.125 & 12.108 & 313.338 & 13.010 \\
				{ADSK} & 0.268 & 693.772 & 1.644 & 2.765 & 5.167 & 6.644 & 22.377 & 388.131 & 16.554 \\
				\bottomrule
			\end{tabular}
			\begin{tablenotes}[para,flushleft]
			\footnotesize
			\emph{Notes:} The period for the descriptive statistics spans from 27 July 2007 to 27 January 2022.
			\end{tablenotes}
			\label{descriptive_stat_table}
		\end{threeparttable}
	\end{adjustbox}
\end{table}

\subsection{NLP for RV Forecasting} \label{nlp_ml_structure_subsection}
An NLP-based RV forecasting model is introduced in which textual information from news serves as the sole input and the next-day RV, \( RV_{t+1} \), is the output. The news input consists of stock-related or general headlines released during the open-to-open interval associated with trading day \( t \), reflecting information available to the market prior to the realisation of \( RV_{t+1} \). Importantly, the model does not incorporate any past RV or other market-based variables, allowing the predictive content of text to be examined in isolation. While HAR-type models summarise the market’s reaction to information arrivals through lagged volatility measures in \Cref{review_rv_subsection}, the proposed NLP structure allows news content to enter the forecasting process explicitly and in a flexible manner. The model architecture is intentionally simple, computationally tractable, and easy to train, while remaining sufficiently expressive to capture nonlinear relationships between textual information and future RV.

To make the timing of the forecasting exercise explicit, the forecast of \(RV_{t+1}\) is formed at 9:30~AM Eastern Time on trading day \(t+1\), immediately before the regular trading session from which \(RV_{t+1}\) is constructed. At this forecast origin, all volatility-history variables used by the HAR-family models have been observed through the close of trading day \(t\), while the textual information set contains news headlines released during the open-to-open interval from 9:30~AM on day \(t\) to immediately before 9:30~AM on day \(t+1\). Consequently, both the volatility-history and news information sets are observable when the forecast is formed, whereas none of the intraday returns used to construct \(RV_{t+1}\) has yet been realised. This timing convention ensures that the forecasting exercise does not use future information and allows the empirical analysis to assess whether news available at the forecast origin contains incremental predictive information about next-day realised volatility beyond that already captured by lagged realised-volatility measures.

Formally, let $\mathcal{X}_t = \{X_{(t,1)}, X_{(t,2)}, \ldots, X_{(t,k_t)}\}$ denote the sequence of $k_t$ tokens extracted from news headlines released during the open-to-open information window associated with day $t$, where each token $X_{(t,i)}$ belongs to a finite vocabulary and is mapped to a fixed-dimensional word embedding vector $\mathbf{e}_{(t,i)} \in \mathbb{R}^{d}$ using a word embedding matrix. After padding the sequence to a fixed length $K$, the embedded tokens form a sentence matrix $\mathbf{S}_t \in \mathbb{R}^{K \times d}$, which provides a numerical representation of the information set available to the market prior to the realisation of $RV_{t+1}$. A nonlinear feature extraction operator $\Psi(\cdot)$, implemented via convolutional filters of varying window sizes followed by pooling operations, maps the sentence matrix $\mathbf{S}_t$ into a low-dimensional latent feature vector $\mathbf{z}_t = \Psi(\mathbf{S}_t)$. The forecast of next-day RV is then obtained as
\begin{equation}
\widehat{RV}_{t+1}
= f_{\text{NLP}}(\mathcal{X}_t)
= \phi(\mathbf{z}_t;\boldsymbol{\theta}),
\end{equation}
where $\phi(\cdot)$ denotes a nonlinear mapping with parameters $\boldsymbol{\theta}$. This formulation replaces the traditional HAR-family information set based on lagged components in \Cref{har_general_spec} with a direct, text-driven representation of information arrival, allowing semantic patterns in news to enter the RV forecasting process explicitly and flexibly.

\begin{figure}
	\centering
	\includegraphics[scale=0.88, trim = {0cm 0.4cm 0cm 0cm}]{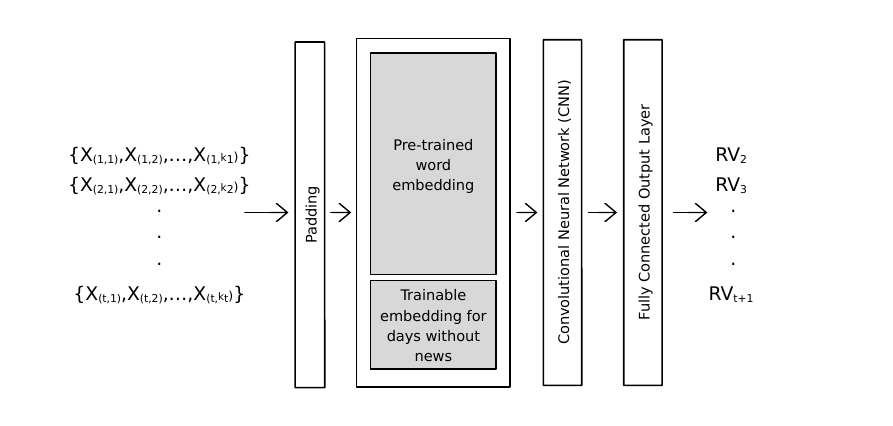}
	\caption{An Abstract Representation of the NLP Model}
	\label{abstract_rep_nlp_ml}
\begin{minipage}{0.8\linewidth}
		\rule{\linewidth}{0.02em}
		\footnotesize
		{\emph{Notes:} The set $X_t = \{X(t,1), X(t,2), \dots, X(t,k_t)\}$ consists of the sequence of tokens extracted from news headlines released during the open-to-open information window associated with day $t$, where $X(t,k)$ represents the $k$-th token in the aggregated daily token sequence. Additionally, $RV_{t+1}$ denotes the RV for day $t+1$ (i.e., the next day’s RV). Padding up to a maximum length of 500 tokens (for stock-related news) is applied to ensure that all inputs to the model have a consistent length. The pre-trained word embeddings are fixed and non-trainable. Index zero is used for both post-padding and OOV tokens, while a dedicated trainable no-news representation is repeated across the full input sequence on days with no qualifying news.}  
	\end{minipage}	
\end{figure}

\Cref{abstract_rep_nlp_ml} provides an abstract representation of the NLP forecasting structure. In implementation, the token sequence $\mathcal{X}_t$ is post-padded or post-truncated to a fixed maximum length $K$, where $K=500$ for stock-related news and $K=2{,}000$ for general hot news. If a sequence contains more than $K$ tokens, the $K$ most recent tokens are retained while preserving their chronological order, and the remaining earlier tokens are discarded; if it contains fewer than $K$ tokens, padding is appended to the end of the sequence. For each word embedding, tokens that are not contained in the corresponding vocabulary are mapped to index zero, which is also the index used for padding. On days with no qualifying news, all $K$ positions are assigned a dedicated no-news index associated with a trainable embedding. We use only news headlines rather than full news bodies, as aggregating complete articles generates extremely long token sequences even for a single day. Such sequences substantially increase computational demands and may lead to overfitting, particularly given the relatively limited size of the training sample. Moreover, news headlines are specifically designed to convey the most salient information contained in the full article.

As shown in \Cref{abstract_rep_nlp_ml}, the NLP model consists of three main components: a word embedding block,
a convolutional neural network (CNN) block, and a fully connected output block. The word embedding block transforms each input token \( X_{(t,k_{t})} \) into a numerical vector representation. For days with news, each token is mapped to a fixed \( 1 \times 300 \) word embedding vector. This word embedding is fixed and non-trainable to reduce model complexity and computational cost. The resulting \( K \times 300 \) sentence matrix provides a structured numerical representation of the daily news input. On days without news, a dedicated trainable $1\times300$ no-news representation is repeated across all $K$ input positions, allowing the model to learn a baseline representation in the absence of textual information. The CNN block processes the sentence matrix by extracting informative local patterns and higher-level features from the embedded text. CNNs are well suited for this task because they efficiently capture local dependencies and hierarchical structures in sequential data through convolution and pooling operations \mbox{\citep{lecun2015deep}}. The fully connected output block then maps the pooled features directly to a single scalar output, producing the forecast of next-day RV. This layered architecture enables the model to capture nonlinear relationships between news content and future RV while maintaining a transparent and parsimonious structure.\footnote{Apart from the reasons discussed in \Cref{we_section}, the embedding-based CNN architecture is chosen to provide a transparent, reproducible, and computationally tractable mapping from news text to RV forecasts in a rolling out-of-sample setting. This design allows for regular retraining while retaining interpretability, which is central to the empirical objectives of this study. The use of a parsimonious architecture is also consistent with recent evidence that, in financial volatility forecasting, training regime and data scale can have substantially larger effects on neural-network performance than increasing model size or changing network architecture \citep{10.1093/jjfinec/nbag019}.}

\begin{figure}
	\centering
	\includegraphics[scale=0.72, trim = {3cm 4.3cm 4.5cm 10cm}]{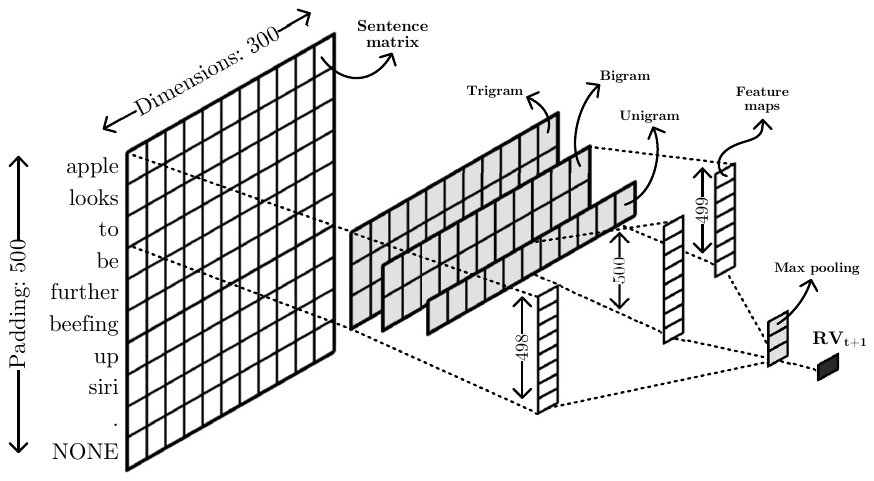}
	\caption{A Detailed Representation of the NLP Model}
\begin{minipage}{0.76\linewidth}
		\rule{\linewidth}{0.02em}
		\footnotesize
		{\emph{Notes:} The sentence matrix is a $500 \times 300$ matrix with a maximum padding length of 500 (for stock-related news) and word embedding dimensions of 300. In this matrix, each token is represented by a vector of 300 values. This structure employs convolutional filters with three window sizes, 1, 2, and 3. For each window size, $q$ filters are applied, where $q\in\{25,50,75,100\}$ across the model specifications. For an input length of 500, the corresponding feature maps have lengths 500, 499, and 498 for window sizes 1, 2, and 3, respectively. Global max pooling reduces each feature map to a single value, yielding a $3q$-dimensional feature vector. A dropout layer is applied before a fully connected output layer produces the forecast of the following day's RV ($RV_{t+1}$).}  
	\end{minipage}
	\label{NN_detailed_rep}
\end{figure}

\Cref{NN_detailed_rep} illustrates the detailed architecture of the NLP model using an example input. The input is a numerical sentence matrix constructed from the news headline `apple looks to be further beefing up siri,' where each word is mapped to a fixed-length numerical vector using a word embedding. The sentence matrix is processed by a CNN. Three sets of one-dimensional convolutional filters with window sizes ${1,2,3}$ are applied simultaneously, corresponding to unigram, bigram, and trigram representations based on one word, two consecutive words, and three consecutive words, respectively. The filters move across the sentence with a stride of one word, meaning that the filter window shifts forward by one token at a time and evaluates all overlapping word sequences. The convolution is applied using valid padding, so filters are evaluated only where they fully overlap with the input sentence, resulting in output feature maps that are shorter than the original sentence representation. For each window size, 25, 50, 75, and 100 distinct filters are considered in separate model specifications, allowing the CNN to learn multiple types of unigram, bigram, and trigram patterns. For a specification with $q$ filters per window size, the three convolutional branches produce $q$ feature maps of lengths 500, 499, and 498 for window sizes 1, 2, and 3, respectively. Global max pooling reduces each feature map to its strongest activation, yielding a $3q$-dimensional feature vector that captures the most informative word or phrase patterns in the daily news input.\footnote{When multiple headlines occur within the same daily information window, their tokenised representations are concatenated in their existing order to form a single input sequence. No additional separator token is inserted between headlines; terminal punctuation retained during pre-processing provides the boundary between consecutive headlines.} A dropout layer is applied to the pooled feature vector before a fully connected output layer maps the extracted features to a scalar forecast of next-day RV. The NLP models are estimated by minimising MSE, while QLIKE is used as an additional out-of-sample evaluation criterion; the implications of this distinction are discussed in \Cref{appendix_RV_model}.

\begin{figure}
	\begin{subfigure}{0.5\linewidth}
	\centering
	\includegraphics[scale=0.30, trim = {3cm 0cm 1cm 0cm}]{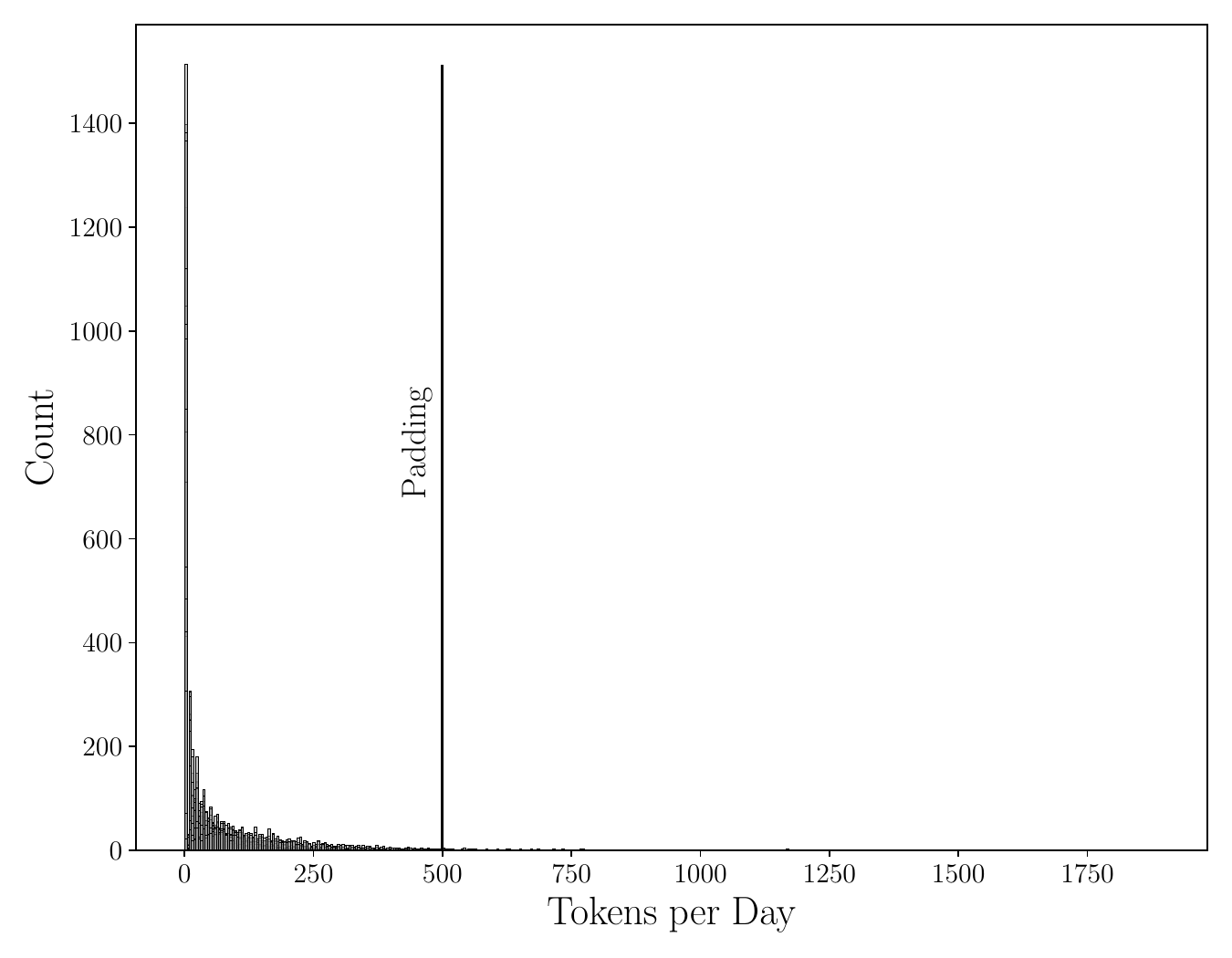}
	\caption{Stock-Related Headlines}
	\label{token_distribution_ticker_related}
	\end{subfigure}
	\begin{subfigure}{0.5\linewidth}
	\centering
	\includegraphics[scale=0.30, trim = {0cm 0cm -1cm 0cm}]{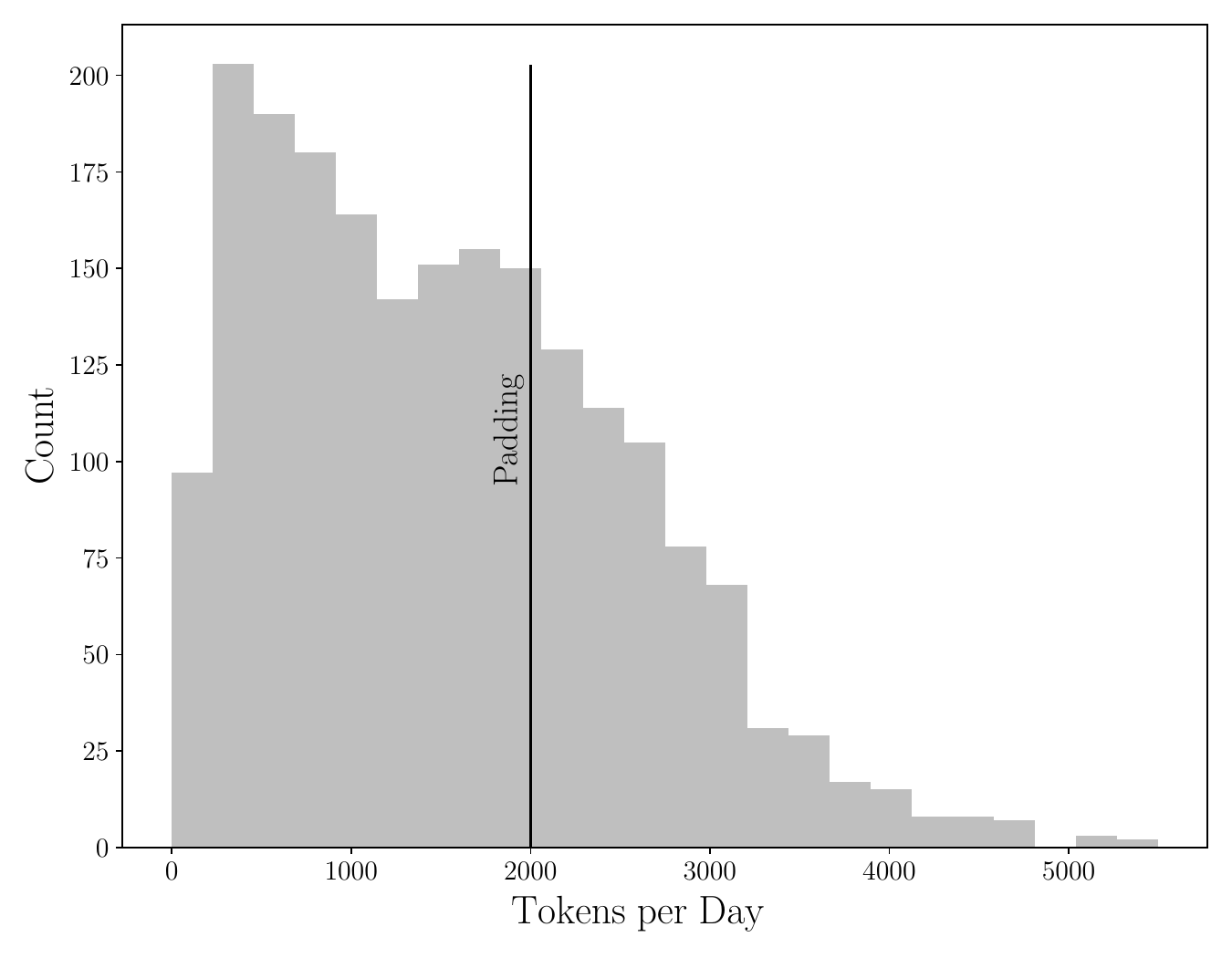}
	\caption{General Hot Headlines}
	\label{token_distribution_ticker_related_hot_political}
	\end{subfigure}
	\caption{Distribution of Daily Tokens}
	\begin{minipage}{1\linewidth} 
		\rule{\linewidth}{0.03em} \vspace{.15mm} \footnotesize
		{\emph{Notes:} The number of daily tokens is calculated, and their histograms are plotted for the stock-related news (left plot) and general hot news (right plot) (training data spanning 2,046 days). The vertical line represents the chosen maximum padding length. For clarity, ticker names are excluded from the left plot.}
		
	\end{minipage}  
	\label{token_distribution}
\end{figure}

News headlines are aggregated over an open-to-open rolling window from 9:30~AM Eastern Time on day $t$ to 9:30~AM Eastern Time on day $t+1$, so that the forecast of $RV_{t+1}$ is formed at 9:30~AM on day $t+1$, before the intraday returns used to construct $RV_{t+1}$ are realised. Because daily re-estimation of the NLP model is computationally intensive, the model is retrained every thirty trading days and used for forecasting in the subsequent period. At each re-estimation date, a separate NLP forecasting model is estimated for each stock using a fixed rolling window of the most recent 2,046 daily observations. The estimation sample ends before the first forecast generated by the newly estimated model, so that no observations from the subsequent forecasting block enter model estimation.\footnote{The pre-trained word embeddings are held fixed during the out-of-sample forecasting period. At each forecast origin, newly arriving headlines therefore require only the pre-processing and tokenisation described above, mapping into the fixed embedding space, and generation of the forecast using the most recently estimated NLP model. We do not report wall-clock processing times because these depend on the specific hardware and implementation environment.} The Dow Jones Newswires Text News Feed assigns tags to news stories to indicate their relevance to specific stocks; for stock-related news, we use the `about' tag, which denotes news stories in which the firm is a main subject. To examine the impact of broader information arrivals on RV, we additionally consider general hot news. According to their definition, `hot' denotes news stories deemed important or timely under Dow Jones editorial standards. Additionally, we define a news story as `general' when it is not associated with any specific stock. In summary, stock-related news comprises headlines in which the firm is the primary subject, such as earnings announcements, management changes, and firm-specific corporate events. General hot news consists of major market-moving headlines that are not attributable to any single firm but instead convey broader economic, financial, or political information, and are classified as important or timely for the market. Based on these definitions, stock-related news and general hot news are mutually exclusive. To limit the volume of textual input and maintain economic relevance, general hot news is restricted to U.S.\ market news only.\footnote{Dow Jones Newswires is used because it provides a long and consistently timestamped archive of professionally edited financial news together with structured stock-relevance metadata, and has been widely used in the empirical finance and financial-news literature. These characteristics allow the textual information set to be defined consistently at each forecast origin and avoid the need for ex-post keyword matching to identify firm-specific stories. The use of a single provider also ensures consistency in editorial standards, timestamp conventions, and stock-level tagging throughout the sample, while facilitating comparability with the existing literature. A corresponding limitation is that the findings are conditional on the Dow Jones information environment and need not generalise unchanged to alternative news providers with different coverage, editorial selection, or tagging practices.}

\begin{figure}
	\begin{subfigure}{0.5\linewidth}
	\centering
	\includegraphics[scale=0.23, trim = {0cm 3cm 0cm 0cm}]{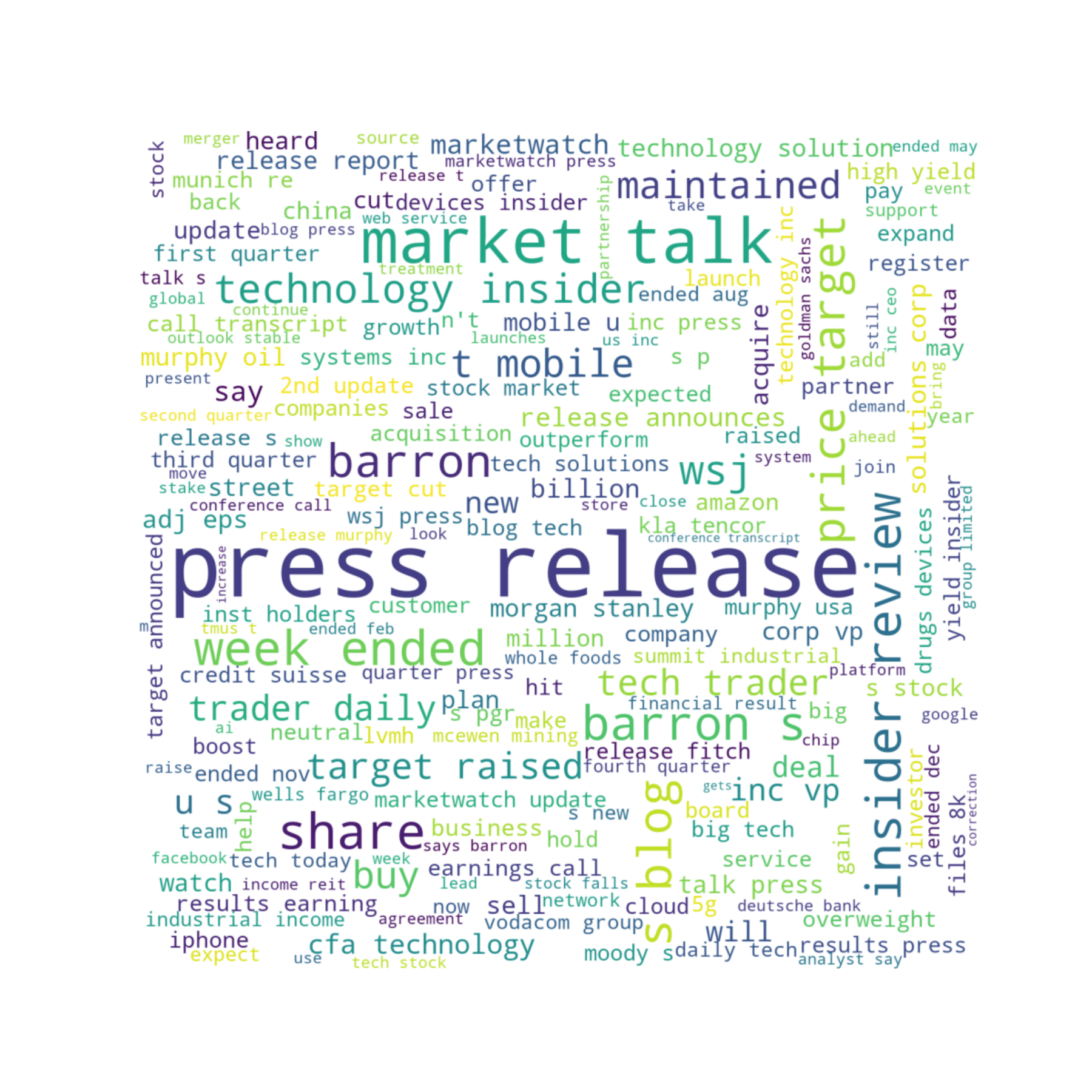}
	\caption{Stock-Related Headlines}
	\label{word_cloud_stock_related}
	\end{subfigure}
	\begin{subfigure}{0.5\linewidth}
	\centering
	\includegraphics[scale=0.23, trim = {0cm 3cm 0cm 0cm}]{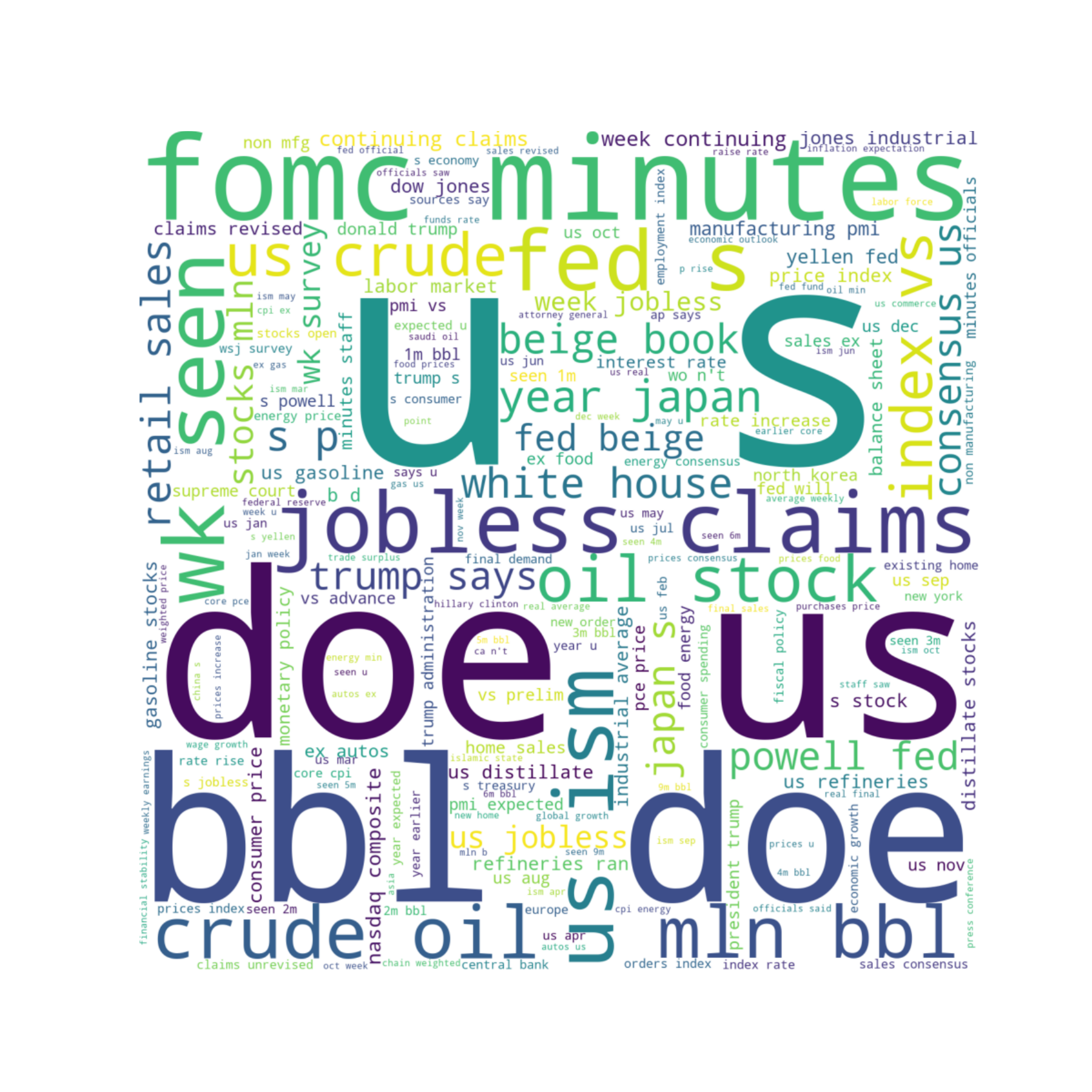}
	\caption{General Hot Headlines}
	\label{word_cloud_general_hot}
	\end{subfigure}
	\caption{Out-of-Sample Word Cloud}
	\begin{minipage}{1\linewidth} 
		\rule{\linewidth}{0.03em} \vspace{.15mm} \footnotesize
		{\emph{Notes:} (a) Word cloud of stock-related headlines for all 23 stocks together over the out-of-sample period. (b) Word cloud of general hot headlines over the out-of-sample period.}
	\end{minipage}  
	\label{word_cloud_all}  
\end{figure}

\Cref{token_distribution} shows the distribution of daily tokens. The number of daily tokens is calculated, and their histograms are plotted for stock-related news (left plot) and general hot news (right plot), using training data spanning 2{,}046 days. The vertical line represents the chosen maximum padding length, and ticker names are excluded from the left plot for clarity. The volume of general hot news is substantially larger than that of stock-related news; consequently, the maximum input length is increased from 500 to 2{,}000 tokens for general hot news. In addition, \Cref{word_cloud_all} presents out-of-sample word clouds for stock-related and general hot news headlines, where \Cref{word_cloud_stock_related} and \Cref{word_cloud_general_hot} show word clouds constructed from stock-related news headlines and general hot news headlines, respectively, over the out-of-sample period for all 23 stocks. As expected, stock-related news headlines are dominated by terms associated with specific firms and their operations, reflecting company-level events and developments. In contrast, the word cloud for general hot news highlights a much broader range of topics, including economic, financial, political, and geopolitical themes, consistent with its role in capturing market-wide information arrivals.

\section{Empirical Results} \label{results_section}
As a first step, we evaluate the complete set of HAR-family models described in \Cref{review_rv_subsection}. \Cref{appendix_HAR_results} reports their parameter estimates together with their in-sample and out-of-sample forecasting performance. Among these specifications, CHAR is used as the common reference model for reporting relative forecast performance using MSE and QLIKE.\footnote{The observation-level squared-error and quasi-likelihood (QLIKE) losses are defined as $\ell^{MSE}_{t} \equiv \left(RV_{t} - \widehat{RV}_{t}\right)^{2}$ and $\ell^{QLIKE}_{t} \equiv \frac{RV_{t}}{\widehat{RV}_{t}} - \log\!\left(\frac{RV_{t}}{\widehat{RV}_{t}}\right) - 1$, respectively, where $RV_{t}$ and $\widehat{RV}_{t}$ denote the RV and its corresponding forecast at time $t$. For each ticker, MSE and QLIKE are obtained by averaging the corresponding observation-level losses over the relevant evaluation period. As shown by \citet{patton2011volatility}, MSE, QLIKE, and their variants belong to the class of robust loss functions suitable for ranking RV forecasting models among competing HAR-type models. Throughout the out-of-sample evaluation, the observed volatility entering both MSE and QLIKE is the 5-minute realised variance defined in \Cref{main_RV_def_eq2}. This measure is used as the observed volatility proxy because the 5-minute sampling frequency provides a standard compromise between discretisation error and market-microstructure noise \citep{liu2015does}.} The out-of-sample performance of the complete HAR-family is nevertheless reported so that the relative ranking and robustness of the benchmark models can be evaluated without relying solely on CHAR. In addition, we report White's Reality Check (RC) \citep{white2000reality}, implemented using the stationary bootstrap procedure of \citet{politis1994stationary}. The test is conducted from the perspective of the model under evaluation, which is treated as the benchmark and compared with the competing models:
\begin{align}
H_{0} &: \mathop{\min}_{k=1,\ldots,n} \ \mathbb{E} \left[ L^{k}(RV,X) - L^{0}(RV,X) \right] \geq 0, \\
H_{1} &: \mathop{\min}_{k=1,\ldots,n} \ \mathbb{E} \left[ L^{k}(RV,X) - L^{0}(RV,X) \right] < 0.
\end{align}
Let \( L^{k} \) (for \( k = 1,\ldots,n \)) denote the loss of each competing model and let \( L^{0} \) denote the loss of the model under evaluation. Rejection of the null indicates that at least one competing model has statistically lower expected loss than the model under evaluation. The RC test is implemented using 999 stationary-bootstrap resamples and an average block length of five.\footnote{Our results show that the reported findings in this study are not sensitive to these choices; therefore, we follow the parameter values used in \citet{bollerslev2016exploiting}.} We implement the RC on a stock-by-stock basis. For ease of presentation, the reported RC statistic is the percentage of stocks for which the null hypothesis is not rejected at the specified significance level. Higher RC values therefore indicate that, for a larger fraction of stocks, there is insufficient statistical evidence that any competing model has lower expected loss. Importantly, a high RC value should not be interpreted as statistically significant outperformance of all competing models.

We further distinguish between normal and high volatility days when presenting the out-of-sample results. A day is defined as a high volatility day when RV for that day exceeds $Q3 + 1.5\,IQR$, where $IQR=Q3-Q1$, and $Q1$ and $Q3$ are, respectively, the empirical 25th and 75th percentiles of RV computed separately for each stock over the 1,604-day out-of-sample period.\footnote{The threshold multiplier of $1.5$ follows Tukey’s (boxplot) \emph{inner-fence} convention for flagging unusually large observations using a robust, nonparametric cutoff \citep{tukey1977eda,mcgill1978boxplots}. This convention is widely used in exploratory data analysis and standard software implementations, and it is particularly suitable for RV measures that are typically skewed and heavy-tailed.} Normal volatility days are defined as all trading days not classified as high volatility days.\footnote{This classification is used exclusively as an ex-post diagnostic of state-dependent forecasting performance. The subsequently realised volatility state is not known at the forecast origin and does not enter model estimation, model selection, or forecast construction. Accordingly, the conditional results should not be interpreted as the performance of an implementable regime-switching forecasting strategy. Applying this criterion separately to the 23 tickers classifies, on average, approximately 160 observations per ticker (about 10\% of the 1,604-day out-of-sample period) as high volatility days.  Additional descriptive evidence is provided in \Cref{appendix_high_vol_days_appendix}.} Turning to the out-of-sample results, CHAR also delivers the lowest average and median losses among the HAR-family specifications over the full evaluation period. This out-of-sample ranking provides evidence on the suitability of CHAR as a reference benchmark, but it is not used as an out-of-sample model-selection rule. For details on the in-sample and out-of-sample results, the reader is referred to the \Cref{appendix_HAR_results}

Moving to the NLP models, we use CHAR as the common HAR-family reference model for reporting relative forecast losses, consistent with its superior in-sample performance documented above. The performance difference between NLP model \(j\) and CHAR is computed across stocks as
\begin{equation}
\rho_{MSE,j}
= \operatorname{Mean}_{i=1,\ldots,23}
\left[
\frac{MSE_i\!\left(NLP_j\right)}{MSE_i\!\left(CHAR\right)}
\right],
\label{eq:MSE_ML_CHAR_comp}
\end{equation}
where the index \( i \) denotes an individual stock in the cross section, with a total of 23 stocks. Alternatively, the mean operator in \Cref{eq:MSE_ML_CHAR_comp} can be replaced by the median to obtain a measure that is more robust to outliers. The MSE can also be replaced by the QLIKE loss. For both MSE and QLIKE, a value below one indicates that the NLP model outperforms the CHAR benchmark, whereas a value above one implies superior performance of the CHAR model.\footnote{The use of CHAR in the denominator provides a common reference point across NLP specifications and should not be interpreted as selecting the benchmark retrospectively on the basis of the same out-of-sample losses used for forecast evaluation.} We also report results based on the RC as defined earlier. For the remainder of the analysis, we compare each NLP model against the full set of HAR-family of models, including AR, HAR, SHAR, HAR-J, CHAR, ARQ, HARQ, and HARQ-F.\footnote{We summarise out-of-sample forecasting performance from two complementary perspectives. Relative loss measures compare each NLP model directly with CHAR and quantify the magnitude of the forecast-error difference. By contrast, the reported RC statistic measures, on a stock-by-stock basis, the percentage of cases in which the Reality Check does not reject the model under evaluation in favour of a better-performing member of the HAR-family. Consequently, a model may exhibit a relative-loss ratio above one while retaining a high RC value if the observed loss difference is not statistically significant. A high RC value therefore indicates limited statistical evidence of inferiority to the competing benchmark set, rather than statistically significant outperformance of that set.}

\subsection{Forecasting Performance of NLP Models} \label{results_NLP_subsection}
The NLP models analysed in this study differ along two dimensions: the word embedding specification and the underlying news information set. As described in \Cref{fwe_section}, we consider domain-specific FinText word embeddings constructed using Word2Vec (CBOW), Word2Vec (skip-gram), FastText (CBOW), and FastText (skip-gram). In addition, we include two general-purpose word embeddings: Word2Vec Google \citep{mikolov2013efficient}, hereafter referred to as Google for simplicity, and WikiNews. As discussed in \Cref{nlp_ml_structure_subsection}, we also distinguish between stock-related news and general hot news as separate inputs.\footnote{\Cref{appendix_oos_volume_stability} documents the stability of news volume during the out-of-sample period. The monthly word count of both stock-related and general hot news remains broadly constant over time, with only a small number of highly visible stocks exhibiting persistently higher coverage.} \Cref{NLP_ML_primary_experiment_table_ticker_relatedX} reports out-of-sample forecasting results using stock-related news, while \Cref{NLP_ML_primary_experiment_table_general_hot} presents corresponding results based on general hot news. For each loss function, average and median values are reported as discussed in \Cref{results_section}. The tables also report out-of-sample RC results for six word embeddings, each evaluated with 25, 50, 75, and 100 filters, where a larger number of filters implies greater model complexity; RC denotes the percentage of tickers for which the Reality Check does not reject the NLP model in favour of a better-performing HAR-family model at the 5\% and 10\% significance levels based on MSE and QLIKE. Finally, the top, middle, and bottom rows correspond to the full out-of-sample period, normal volatility days, and high volatility days, respectively.\footnote{The alternative word embeddings and filter counts are treated as a specification and sensitivity analysis rather than as an out-of-sample model-selection exercise. We therefore report the full grid of configurations rather than selecting a single specification on the basis of its realised out-of-sample loss. When the discussion below refers to the `best-performing' or `lowest-loss' configuration, this terminology is descriptive and identifies the ex-post minimum among the specifications displayed in the table; it does not represent an implementable model-selection rule. Accordingly, the substantive conclusions are based primarily on patterns that persist across multiple embeddings and model-complexity settings rather than on the performance of a single ex-post selected configuration. This distinction is particularly important given the number of model specifications considered and reduces the risk of attributing undue importance to an isolated out-of-sample result.}

\Cref{NLP_ML_primary_experiment_table_ticker_relatedX} and \Cref{NLP_ML_primary_experiment_table_general_hot} jointly show that the predictive value of NLP models depends critically on both the information set and the volatility regime. For stock-related news, the lowest observed loss ratios are generally obtained from specifications based on FinText embeddings, and the broader pattern across configurations also tends to favour the specialised representations. Over the full out-of-sample period, FinText Word2Vec with the skip-gram architecture delivers the lowest MSE ratios, corresponding to an average deterioration of approximately 10.6\% (average MSE ratio of 1.106 using 25 filters) and a median deterioration of about 9.2\% (median MSE ratio of 1.092 using 50 filters) relative to the CHAR benchmark. In terms of QLIKE, WikiNews performs best, with average and median QLIKE ratios of 1.476 and 1.446 using 25 filters, implying loss increases of roughly 47.6\% and 44.6\%, respectively. Despite these higher loss ratios, the RC results indicate that the relative underperformance is not statistically decisive for a large fraction of stocks. In particular, under MSE, the RC rate exceeds 90\% for some FinText Word2Vec specifications, indicating that the Reality Check does not identify a significantly better HAR-family competitor for most stocks. When performance is decomposed by volatility regimes, the results are state dependent and differ somewhat across loss functions. Under MSE, stock-related news becomes substantially more competitive on high volatility days, with the lowest average ratio falling to 1.081. Under QLIKE, however, loss ratios are lower on normal volatility days than on high volatility days. At the same time, RC non-rejection rates rise substantially during high volatility days under both loss functions, indicating less statistical evidence that a HAR-family competitor has lower expected loss.

In contrast, NLP models based on general hot news deliver weaker and less robust performance in \Cref{NLP_ML_primary_experiment_table_general_hot}. Over the full out-of-sample period, FinText FastText with the CBOW architecture and 100 filters, attains average MSE and QLIKE ratios of 1.175 and 1.544, implying deteriorations of approximately 17.5\% and 54.4\%, respectively, with low RC non-rejection rates under QLIKE. Also, general hot news contributes primarily on normal volatility days, where Google achieves average MSE ratios below one, with a minimum of 0.956 corresponding to an improvement of approximately 4.4\%, and a QLIKE ratio of 1.285, implying a loss increase of 28.5\%. However, general hot news remains uncompetitive relative to CHAR on high volatility days, with best average MSE and QLIKE ratios of 1.168 and 1.622, corresponding to loss increases of approximately 16.8\% and 62.2\%, respectively. Moreover, when general hot news is employed, general-purpose embeddings exhibit competitive performance and, in some cases, such as WikiNews, outperform alternative models in terms of QLIKE ratios. This result likely reflects the broader and more diverse semantic coverage of general-purpose word embeddings, which better capture the heterogeneous language and rapidly evolving content characteristic of general hot news. These results suggest that the usefulness of embedding specialisation depends on the nature of the textual information set. Financially specialised representations tend to perform more favourably for firm-specific news, whereas general-purpose embeddings remain competitive when the input consists of broader and more heterogeneous market-wide news.\footnote{\Cref{appendix_NLP_Models_hyper} reports a set of robustness checks that examine the sensitivity of the NLP forecasts to key architectural and aggregation choices, including the length of the news input window and the filter sizes. We show that extending the input window beyond one day or increasing textual granularity does not lead to systematic improvements in forecasting performance. Overall, the results do not provide consistent evidence that extending the news window or increasing textual granularity improves forecasting performance across evaluation criteria.}

\afterpage{%
\clearpage
\thispagestyle{empty}

\atxy{\dimexpr\paperwidth-0.45in}{.5\paperheight}{%
  \rotatebox[origin=center]{90}{\thepage}
}

\begin{landscape}
\begin{table}
\centering
\begin{adjustbox}{max width={1.35\textwidth}}
\begin{threeparttable}
\centering
\setlength{\tabcolsep}{6pt}

\caption{Out-of-Sample RV Forecasting Performance: Stock-Related News}

\begin{tabular}{cccc|cccc|cccc|cccc|cccc|cccc|cccc}
\toprule
& & &
& \multicolumn{16}{c|}{\textbf{FinText}}
& \multicolumn{4}{c|}{\multirow{2}{*}{\raisebox{-0.6ex}{\textbf{Google}}}}
& \multicolumn{4}{c}{\multirow{2}{*}{\raisebox{-0.6ex}{\textbf{WikiNews}}}} \\
\cmidrule(lr){5-20}
& & &
& \multicolumn{4}{c|}{Word2Vec (CBOW)}
& \multicolumn{4}{c|}{Word2Vec (skip-gram)}
& \multicolumn{4}{c|}{FastText (CBOW)}
& \multicolumn{4}{c|}{FastText (skip-gram)}
&  &  &  &  &  &  &  &  \\
					{{\textbf{\small Full Out-of-Sample Period}}}  & {} & {$\alpha$} & {} & 25 & 50 & 75 & 100 & 25 & 50 & 75 & 100 & 25 & 50 & 75 & 100 & 25 & 50 & 75 & 100 & 25 & 50 & 75 & 100 & 25 & 50 & 75 & 100  \\  \midrule 
                    \addlinespace[0.25cm]
					MSE\tnote{a} & &  & Avg & 1.146 & 1.146 & 1.153 & 1.144 & 1.106 & 1.108 & 1.108 & 1.107 & 1.169 & 1.187 & 1.178 & 1.182 & 1.108 & 1.110 & 1.109 & 1.109 & 1.116 & 1.117 & 1.118 & 1.117 & 1.118 & 1.118 & 1.118 & 1.118 \\
					& &  & Med & 1.157 & 1.156 & 1.160 & 1.164 & 1.101 & 1.092 & 1.096 & 1.093 & 1.169 & 1.144 & 1.142 & 1.140 & 1.101 & 1.100 & 1.102 & 1.096 & 1.094 & 1.097 & 1.104 & 1.104 & 1.106 & 1.105 & 1.107 & 1.108  \\
					
					RC\tnote{c} & & 0.05 &  & 100 & 100 & 100 & 100 & 100 & 100 & 100 & 100 & 95.65 & 91.30 & 95.65 & 91.30 & 100 & 100 & 100 & 100 & 95.65 & 95.65 & 100 & 95.65 & 95.65 & 91.30 & 91.30 & 91.30 \\
					& & 0.10 &  & 91.30 & 86.96 & 91.30 & 86.96 & 82.61 & 82.61 & 82.61 & 82.61 & 73.91 & 78.26 & 65.22 & 65.22 & 86.96 & 86.96 & 86.96 & 86.96 & 65.22 & 65.22 & 65.22 & 65.22 & 56.52 & 60.87 & 60.87 & 60.87  \\ \addlinespace[0.25cm]

					QLIKE\tnote{b} & &  & Avg & 1.838 & 1.744 & 1.782 & 1.740 & 1.566 & 1.640 & 1.571 & 1.571 & 1.829 & 1.701 & 1.609 & 1.636 & 1.567 & 1.564 & 1.571 & 1.569 & 1.500 & 1.520 & 1.513 & 1.519 & 1.476 & 1.489 & 1.492 & 1.486 \\
					& &  & Med & 1.766 & 1.695 & 1.749 & 1.736 & 1.570 & 1.579 & 1.563 & 1.567 & 1.684 & 1.541 & 1.545 & 1.587 & 1.597 & 1.570 & 1.577 & 1.561 & 1.472 & 1.467 & 1.480 & 1.488 & 1.446 & 1.451 & 1.455 & 1.459 \\

					RC & & 0.05 &  & 30.44 & 43.48 & 26.09 & 30.44 & 39.13 & 34.78 & 39.13 & 30.44 & 26.09 & 17.39 & 21.74 & 21.74 & 34.78 & 34.78 & 34.78 & 30.44 & 30.44 & 30.44 & 30.44 & 30.44 & 21.74 & 21.74 & 21.74 & 21.74 \\
					& & 0.10 &  & 13.04 & 17.39 & 17.39 & 17.39 & 21.74 & 21.74 & 21.74 & 21.74 & 17.39 & 13.04 & 13.04 & 4.35 & 21.74 & 21.74 & 21.74 & 21.74 & 13.04 & 13.04 & 13.04 & 13.04 & 4.35 & 4.35 & 4.35 & 4.35 \\ \addlinespace[0.25cm]

					\midrule
					{\textbf{\small Normal Volatility Days}}  & {} & {} &   {} & {} & {} \\ 	
					\midrule
                    \addlinespace[0.25cm]
					MSE & &  & Avg & 2.383 & 2.426 & 2.194 & 2.102 & 1.683 & 1.703 & 1.702 & 1.690 & 1.763 & 1.450 & 1.337 & 1.300 & 1.685 & 1.715 & 1.708 & 1.705 & 1.510 & 1.503 & 1.497 & 1.507 & 1.384 & 1.381 & 1.383 & 1.384 \\
					& &  & Med & 2.051 & 2.044 & 1.902 & 1.872 & 1.592 & 1.562 & 1.595 & 1.609 & 1.620 & 1.359 & 1.295 & 1.342 & 1.605 & 1.587 & 1.576 & 1.574 & 1.379 & 1.391 & 1.404 & 1.408 & 1.297 & 1.312 & 1.311 & 1.314 \\
					
					RC & & 0.05 &  & 21.74 & 17.39 & 17.39 & 17.39 & 69.57 & 69.57 & 73.91 & 69.57 & 65.22 & 34.78 & 47.83 & 52.17 & 73.91 & 69.57 & 69.57 & 69.57 & 69.57 & 65.22 & 65.22 & 65.22 & 65.22 & 60.87 & 60.87 & 60.87 \\
					& & 0.10 &  & 17.39 & 17.39 & 13.04 & 17.39 & 69.57 & 65.22 & 65.22 & 65.22 & 43.48 & 34.78 & 30.44 & 39.13 & 69.57 & 65.22 & 65.22 & 60.87 & 69.57 & 60.87 & 60.87 & 60.87 & 60.87 & 60.87 & 60.87 & 56.52 \\ \addlinespace[0.25cm]
     
					QLIKE & &  & Avg & 1.584 & 1.562 & 1.586 & 1.534 & 1.345 & 1.370 & 1.363 & 1.354 & 1.515 & 1.415 & 1.421 & 1.434 & 1.342 & 1.340 & 1.351 & 1.357 & 1.329 & 1.327 & 1.340 & 1.330 & 1.337 & 1.341 & 1.339 & 1.334 \\
					& &  & Med & 1.534 & 1.502 & 1.511 & 1.481 & 1.343 & 1.342 & 1.346 & 1.327 & 1.451 & 1.389 & 1.413 & 1.423 & 1.341 & 1.330 & 1.347 & 1.349 & 1.362 & 1.358 & 1.364 & 1.357 & 1.371 & 1.370 & 1.364 & 1.368 \\				

					RC & & 0.05 &  & 0.00 & 0.00 & 0.00 & 0.00 & 4.35 & 4.35 & 4.35 & 4.35 & 4.35 & 0.00 & 4.35 & 4.35 & 4.35 & 4.35 & 4.35 & 4.35 & 4.35 & 4.35 & 4.35 & 4.35 & 0.00 & 4.35 & 4.35 & 0.00 \\
					& & 0.10 &  & 0.00 & 0.00 & 0.00 & 0.00 & 4.35 & 4.35 & 4.35 & 4.35 & 4.35 & 0.00 & 4.35 & 4.35 & 4.35 & 4.35 & 4.35 & 4.35 & 4.35 & 4.35 & 4.35 & 4.35 & 0.00 & 0.00 & 0.00 & 0.00 \\ \addlinespace[0.25cm]

					\midrule
					{\textbf{\small High Volatility Days}}  & {} & {} &   {} & {} & {} \\ 	
					\midrule	
                    \addlinespace[0.25cm]
					MSE & &  & Avg & 1.097 & 1.095 & 1.111 & 1.107 & 1.081 & 1.082 & 1.082 & 1.082 & 1.145 & 1.175 & 1.171 & 1.176 & 1.082 & 1.083 & 1.082 & 1.082 & 1.097 & 1.099 & 1.100 & 1.099 & 1.105 & 1.106 & 1.105 & 1.105 \\
					& &  & Med & 1.122 & 1.097 & 1.102 & 1.137 & 1.095 & 1.094 & 1.096 & 1.093 & 1.133 & 1.142 & 1.138 & 1.140 & 1.094 & 1.093 & 1.094 & 1.095 & 1.111 & 1.113 & 1.113 & 1.113 & 1.113 & 1.114 & 1.115 & 1.116 \\
					
					RC & & 0.05 &  & 100 & 100 & 100 & 100 & 95.65 & 95.65 & 95.65 & 95.65 & 95.65 & 95.65 & 100 & 100 & 95.65 & 95.65 & 95.65 & 95.65 & 91.30 & 91.30 & 91.30 & 91.30 & 82.61 & 82.61 & 82.61 & 82.61 \\
					& & 0.10 &  & 100 & 100 & 100 & 100 & 65.22 & 69.57 & 69.57 & 69.57 & 69.57 & 82.61 & 82.61 & 69.57 & 65.22 & 69.57 & 65.22 & 69.57 & 65.22 & 65.22 & 65.22 & 65.22 & 56.52 & 60.87 & 60.87 & 60.87 \\ [0.25cm]	
     
					QLIKE & &  & Avg & 2.123 & 1.947 & 1.996 & 1.965 & 1.804 & 1.960 & 1.794 & 1.812 & 2.135 & 1.997 & 1.796 & 1.823 & 1.804 & 1.813 & 1.788 & 1.804 & 1.665 & 1.704 & 1.675 & 1.699 & 1.616 & 1.637 & 1.649 & 1.641 \\
					& &  & Med & 1.918 & 1.855 & 1.917 & 1.843 & 1.749 & 1.771 & 1.788 & 1.784 & 1.967 & 1.848 & 1.694 & 1.770 & 1.819 & 1.758 & 1.725 & 1.859 & 1.609 & 1.628 & 1.634 & 1.657 & 1.614 & 1.637 & 1.618 & 1.614 \\
					
					RC & & 0.05 &  & 69.57 & 60.87 & 65.22 & 73.91 & 60.87 & 69.57 & 65.22 & 65.22 & 43.48 & 56.52 & 47.83 & 39.13 & 65.22 & 69.57 & 69.57 & 69.57 & 52.17 & 52.17 & 56.52 & 60.87 & 43.48 & 52.17 & 52.17 & 56.52 \\
					& & 0.10 &  & 39.13 & 43.48 & 52.17 & 47.83 & 43.48 & 43.48 & 43.48 & 43.48 & 21.74 & 34.78 & 21.74 & 8.70 & 43.48 & 43.48 & 43.48 & 43.48 & 34.78 & 39.13 & 39.13 & 43.48 & 13.04 & 30.44 & 30.44 & 30.44 \\ [0.25cm]		

					\bottomrule                                
				\end{tabular}
				\begin{tablenotes} [para,flushleft]
					\footnotesize
                        \textit{Notes:} \item[a] For each ticker, MSE is calculated over the relevant evaluation sample and expressed relative to the corresponding MSE of CHAR. The reported average (median) is the cross-sectional mean (median) of the 23 ticker-level MSE ratios. \item[b] For each ticker, QLIKE is calculated over the relevant evaluation sample and expressed relative to the corresponding QLIKE of CHAR. The reported average (median) is the cross-sectional mean (median) of the 23 ticker-level QLIKE ratios. \item[c] RC non-rejection denotes the percentage of tickers for which White's Reality Check does not reject, at the stated significance level, the null hypothesis that no competing HAR-family model has lower expected loss than the model under evaluation. A high value therefore indicates limited statistical evidence of inferiority.

				\end{tablenotes}
				\label{NLP_ML_primary_experiment_table_ticker_relatedX}
			\end{threeparttable}
		\end{adjustbox}
	\end{table}
\end{landscape}
}

\afterpage{%
\clearpage
\thispagestyle{empty}

\atxy{\dimexpr\paperwidth-0.45in}{.5\paperheight}{%
  \rotatebox[origin=center]{90}{\thepage}
}

\begin{landscape}
\begin{table}
\centering
\begin{adjustbox}{max width={1.35\textwidth}}
\begin{threeparttable}
\centering
\setlength{\tabcolsep}{6pt}

\caption{Out-of-Sample RV Forecasting Performance: General Hot News}

\begin{tabular}{cccc|cccc|cccc|cccc|cccc|cccc|cccc}
\toprule
& & &
& \multicolumn{16}{c|}{\textbf{FinText}}
& \multicolumn{4}{c|}{\multirow{2}{*}{\raisebox{-0.6ex}{\textbf{Google}}}}
& \multicolumn{4}{c}{\multirow{2}{*}{\raisebox{-0.6ex}{\textbf{WikiNews}}}} \\
\cmidrule(lr){5-20}
& & &
& \multicolumn{4}{c|}{Word2Vec (CBOW)}
& \multicolumn{4}{c|}{Word2Vec (skip-gram)}
& \multicolumn{4}{c|}{FastText (CBOW)}
& \multicolumn{4}{c|}{FastText (skip-gram)}
&  &  &  &  &  &  &  &  \\
					{{\textbf{\small Full Out-of-Sample Period}}}  & {} & {$\alpha$} & {} & 25 & 50 & 75 & 100 & 25 & 50 & 75 & 100 & 25 & 50 & 75 & 100 & 25 & 50 & 75 & 100 & 25 & 50 & 75 & 100 & 25 & 50 & 75 & 100  \\ \midrule
                    \addlinespace[0.25cm]
					MSE\tnote{a} & &  & Avg & 1.195 & 1.181 & 1.180 & 1.177 & 1.185 & 1.185 & 1.183 & 1.181 & 1.179 & 1.181 & 1.178 & 1.175 & 1.183 & 1.183 & 1.182 & 1.180 & 1.175 & 1.176 & 1.176 & 1.178 & 1.180 & 1.181 & 1.181 & 1.181 \\
					& &  & Med & 1.175 & 1.146 & 1.139 & 1.142 & 1.156 & 1.153 & 1.151 & 1.144 & 1.141 & 1.154 & 1.141 & 1.137 & 1.159 & 1.154 & 1.154 & 1.150 & 1.144 & 1.145 & 1.143 & 1.144 & 1.153 & 1.157 & 1.159 & 1.159 \\
					
					RC\tnote{c} & & 0.05 &  & 56.52 & 47.83 & 43.48 & 39.13 & 91.30 & 91.30 & 91.30 & 91.30 & 95.65 & 95.65 & 95.65 & 91.30 & 91.30 & 91.30 & 86.96 & 86.96 & 91.30 & 91.30 & 91.30 & 91.30 & 78.26 & 86.96 & 82.61 & 82.61 \\
					& & 0.10 &  & 21.74 & 13.04 & 13.04 & 17.39 & 43.48 & 34.78 & 39.13 & 39.13 & 39.13 & 39.13 & 39.13 & 39.13 & 34.78 & 34.78 & 34.78 & 39.13 & 43.48 & 39.13 & 34.78 & 39.13 & 26.09 & 30.44 & 30.44 & 34.78 \\ \addlinespace[0.25cm]

					QLIKE\tnote{b} & &  & Avg & 1.814 & 1.583 & 1.606 & 1.588 & 1.888 & 1.856 & 1.793 & 1.718 & 1.579 & 1.651 & 1.563 & 1.544 & 1.862 & 1.838 & 1.805 & 1.721 & 1.680 & 1.683 & 1.677 & 1.683 & 1.670 & 1.685 & 1.687 & 1.684 \\
					& &  & Med & 1.725 & 1.543 & 1.513 & 1.515 & 1.876 & 1.863 & 1.789 & 1.707 & 1.537 & 1.565 & 1.508 & 1.499 & 1.794 & 1.824 & 1.756 & 1.679 & 1.636 & 1.667 & 1.653 & 1.619 & 1.660 & 1.688 & 1.651 & 1.728 \\
					
					RC & & 0.05 &  & 0.00 & 0.00 & 0.00 & 0.00 & 4.35 & 4.35 & 4.35 & 4.35 & 4.35 & 4.35 & 4.35 & 4.35 & 4.35 & 4.35 & 4.35 & 4.35 & 4.35 & 4.35 & 4.35 & 4.35 & 0.00 & 0.00 & 0.00 & 4.35 \\
					& & 0.10 &  & 0.00 & 0.00 & 0.00 & 0.00 & 0.00 & 0.00 & 0.00 & 0.00 & 0.00 & 0.00 & 0.00 & 0.00 & 0.00 & 0.00 & 0.00 & 0.00 & 0.00 & 0.00 & 0.00 & 0.00 & 0.00 & 0.00 & 0.00 & 0.00 \\ \addlinespace[0.25cm] 
                     
					\midrule
					{\textbf{\small Normal Volatility Days}}  & {} & {} &   {} & {} & {} \\ 	
					\midrule	
                    \addlinespace[0.25cm]
					MSE & &  & Avg & 1.875 & 1.486 & 1.457 & 1.336 & 1.288 & 1.363 & 1.324 & 1.414 & 1.430 & 1.460 & 1.407 & 1.312 & 1.226 & 1.282 & 1.312 & 1.345 & 0.956 & 0.968 & 0.995 & 1.082 & 1.070 & 1.078 & 1.088 & 1.101 \\
					& &  & Med & 1.843 & 1.351 & 1.459 & 1.316 & 1.142 & 1.325 & 1.243 & 1.381 & 1.366 & 1.421 & 1.319 & 1.315 & 1.144 & 1.210 & 1.359 & 1.357 & 0.918 & 0.914 & 0.944 & 1.097 & 1.105 & 1.094 & 1.094 & 1.085 \\

					RC & & 0.05 &  & 4.35 & 8.70 & 8.70 & 8.70 & 73.91 & 47.83 & 52.17 & 47.83 & 78.26 & 73.91 & 73.91 & 73.91 & 86.96 & 86.96 & 86.96 & 82.61 & 73.91 & 52.17 & 52.17 & 43.48 & 56.52 & 39.13 & 47.83 & 43.48 \\
					& & 0.10 &  & 4.35 & 4.35 & 4.35 & 0.00 & 69.57 & 47.83 & 47.83 & 47.83 & 78.26 & 69.57 & 69.57 & 65.22 & 86.96 & 82.61 & 78.26 & 73.91 & 69.57 & 43.48 & 47.83 & 43.48 & 56.52 & 39.13 & 34.78 & 43.48 \\ \addlinespace[0.25cm]
     
					QLIKE & &  & Avg & 1.728 & 1.497 & 1.508 & 1.459 & 1.425 & 1.487 & 1.468 & 1.514 & 1.509 & 1.513 & 1.479 & 1.458 & 1.419 & 1.434 & 1.464 & 1.479 & 1.285 & 1.293 & 1.299 & 1.348 & 1.284 & 1.296 & 1.310 & 1.316 \\
					& &  & Med & 1.640 & 1.450 & 1.456 & 1.427 & 1.371 & 1.455 & 1.437 & 1.506 & 1.498 & 1.488 & 1.470 & 1.432 & 1.364 & 1.403 & 1.438 & 1.452 & 1.262 & 1.286 & 1.275 & 1.327 & 1.297 & 1.295 & 1.323 & 1.312 \\

					RC & & 0.05 &  & 0.00 & 0.00 & 0.00 & 0.00 & 8.70 & 4.35 & 4.35 & 4.35 & 4.35 & 4.35 & 4.35 & 4.35 & 4.35 & 4.35 & 4.35 & 4.35 & 8.70 & 4.35 & 4.35 & 4.35 & 0.00 & 0.00 & 0.00 & 0.00 \\
					& & 0.10 &  & 0.00 & 0.00 & 0.00 & 0.00 & 8.70 & 4.35 & 4.35 & 4.35 & 4.35 & 4.35 & 4.35 & 4.35 & 4.35 & 4.35 & 4.35 & 4.35 & 4.35 & 4.35 & 4.35 & 4.35 & 0.00 & 0.00 & 0.00 & 0.00 \\ \addlinespace[0.25cm]

					\midrule
					{\textbf{\small High Volatility Days}}  & {} & {} &   {} & {} & {} \\ 	
					\midrule	
                    \addlinespace[0.25cm]
					MSE & &  & Avg & 1.168 & 1.168 & 1.169 & 1.169 & 1.180 & 1.178 & 1.178 & 1.172 & 1.168 & 1.169 & 1.169 & 1.168 & 1.181 & 1.179 & 1.177 & 1.174 & 1.182 & 1.182 & 1.181 & 1.181 & 1.183 & 1.183 & 1.183 & 1.182 \\
					& &  & Med & 1.138 & 1.139 & 1.140 & 1.137 & 1.139 & 1.135 & 1.140 & 1.137 & 1.138 & 1.137 & 1.137 & 1.137 & 1.154 & 1.152 & 1.159 & 1.151 & 1.153 & 1.153 & 1.150 & 1.150 & 1.152 & 1.152 & 1.152 & 1.153 \\
					
					RC & & 0.05 &  & 65.22 & 91.30 & 91.30 & 86.96 & 69.57 & 69.57 & 69.57 & 69.57 & 69.57 & 73.91 & 73.91 & 73.91 & 60.87 & 60.87 & 60.87 & 60.87 & 69.57 & 69.57 & 69.57 & 69.57 & 56.52 & 82.61 & 73.91 & 69.57 \\
					& & 0.10 &  & 43.48 & 47.83 & 47.83 & 52.17 & 34.78 & 39.13 & 39.13 & 39.13 & 34.78 & 34.78 & 34.78 & 39.13 & 30.44 & 34.78 & 34.78 & 34.78 & 34.78 & 39.13 & 39.13 & 39.13 & 30.44 & 39.13 & 39.13 & 39.13 \\ \addlinespace[0.25cm]
     
					QLIKE & &  & Avg & 1.907 & 1.660 & 1.699 & 1.705 & 2.370 & 2.245 & 2.149 & 1.936 & 1.648 & 1.782 & 1.642 & 1.622 & 2.326 & 2.253 & 2.168 & 1.968 & 2.068 & 2.075 & 2.058 & 2.017 & 2.049 & 2.063 & 2.058 & 2.050 \\
					& &  & Med & 1.709 & 1.653 & 1.654 & 1.612 & 2.246 & 2.194 & 2.052 & 1.896 & 1.692 & 1.620 & 1.555 & 1.547 & 2.144 & 2.325 & 2.151 & 1.912 & 1.934 & 2.048 & 1.961 & 1.899 & 1.987 & 1.945 & 1.920 & 2.029 \\
					
					RC & & 0.05 &  & 8.70 & 4.35 & 8.70 & 17.39 & 17.39 & 17.39 & 21.74 & 21.74 & 17.39 & 17.39 & 21.74 & 26.09 & 13.04 & 13.04 & 17.39 & 17.39 & 21.74 & 17.39 & 21.74 & 21.74 & 8.70 & 17.39 & 4.35 & 4.35 \\
					& & 0.10 &  & 0.00 & 0.00 & 4.35 & 4.35 & 8.70 & 8.70 & 8.70 & 8.70 & 8.70 & 8.70 & 8.70 & 8.70 & 0.00 & 0.00 & 0.00 & 0.00 & 4.35 & 4.35 & 4.35 & 8.70 & 0.00 & 0.00 & 0.00 & 0.00 \\ \addlinespace[0.25cm]

					\bottomrule                                
				\end{tabular}
				\begin{tablenotes} [para,flushleft]
					\footnotesize
                        \textit{Notes:} \item[a] For each ticker, MSE is calculated over the relevant evaluation sample and expressed relative to the corresponding MSE of CHAR. The reported average (median) is the cross-sectional mean (median) of the 23 ticker-level MSE ratios. \item[b] For each ticker, QLIKE is calculated over the relevant evaluation sample and expressed relative to the corresponding QLIKE of CHAR. The reported average (median) is the cross-sectional mean (median) of the 23 ticker-level QLIKE ratios. \item[c] RC non-rejection denotes the percentage of tickers for which White's Reality Check does not reject, at the stated significance level, the null hypothesis that no competing HAR-family model has lower expected loss than the model under evaluation. A high value therefore indicates limited statistical evidence of inferiority.

				\end{tablenotes}
				\label{NLP_ML_primary_experiment_table_general_hot}
			\end{threeparttable}
		\end{adjustbox}
	\end{table}
\end{landscape}
}
\newpage

\subsection{Forecasting Performance of Ensemble Models} \label{Results_Ensemble_Subsection}
Although \Cref{results_NLP_subsection} shows that NLP models deliver promising results and that news content contains meaningful information for predicting RV across different market regimes, these models underperform the HAR-family of models in terms of standard forecasting metrics, particularly with respect to the magnitude of performance improvements. Combining forecasts from multiple models provides diversification gains and can improve predictive performance when individual forecasts rely on heterogeneous information sets or are subject to model uncertainty \citep{bates1969combination,timmermann2006forecast,becker2008combination}. This naturally raises the question of whether, rather than serving as a replacement for the HAR-family of models, news-based NLP forecasts can provide complementary information. To this end, we define a straightforward ensemble model that combines forecasts from CHAR, which serves as the common HAR-family reference model based on the pre-evaluation benchmark comparison described in \Cref{results_section}, with forecasts generated by the NLP models. For each day, we compute the average of the two forecasts.\footnote{The equal-weight combination is deliberately simple and is not intended to represent an optimal weighting scheme. The equal weights are fixed ex ante and are not selected or optimised using out-of-sample forecasting performance, thereby avoiding an additional layer of specification search. The purpose of the exercise is to provide a transparent assessment of whether forecasts constructed from heterogeneous information sets contain complementary predictive information rather than to identify an optimal forecast-combination rule. A related approach is used by \citet{hong2025forecasting}, who assess the incremental information in textual narratives by combining narrative-based and macro-data-based forecasts, including through a simple average of forecasts constructed from the two information sets. Because forecast averaging can also generate diversification or shrinkage gains, improvements in the ensemble are interpreted as evidence consistent with forecast complementarity rather than as formal identification of incremental textual information.} This approach integrates the persistent dynamics captured by historical RV with short term information conveyed by recent news, potentially enhancing the robustness of daily RV forecasts.

The ensemble results for stock-related news in \Cref{NLP_ML_primary_experiment_table_stock_related_ensemble} show that combining the NLP and CHAR forecasts produces lower full-sample loss ratios than CHAR for most FinText-based specifications. Over the full out-of-sample period, most FinText-based ensemble specifications achieve MSE ratios below one, with the strongest performance delivered by FinText Word2Vec with the skip-gram architecture, for which the average MSE ratio declines to 0.961 and the median ratio to 0.980, corresponding to improvements of approximately 3.9\% and 2.0\% relative to CHAR. QLIKE ratios also fall substantially, reaching values as low as 0.937 on average and around 0.970 at the median, which implies loss reductions of roughly 6.3\% to 3.0\%. These lower loss ratios are accompanied by uniformly high RC non-rejection rates under MSE, which reach 100\% at both the 5\% and 10\% significance levels across nearly all FinText specifications, and remain high under QLIKE, frequently exceeding 90\%, indicating that the Reality Check rarely identifies a significantly better HAR-family competitor. The regime-specific results differ across loss functions. Under MSE, the largest improvements occur on normal volatility days, where the average loss ratio falls to as low as 0.914, corresponding to an improvement of approximately 8.6\% relative to CHAR. Under QLIKE, the largest improvements occur during high volatility days, where the average loss ratio falls to as low as 0.763, representing an approximate 23.7\% reduction and accompanied by near-universal RC non-rejection rates. As in the standalone NLP analysis, variation in the number of filters has only a limited effect on ensemble performance.

For general hot news in \Cref{NLP_ML_primary_experiment_table_general_ensemble}, the ensemble model delivers more mixed and regime-dependent results. Over the full out-of-sample period, average MSE ratios remain close to one, typically ranging between 1.018 and 1.023, corresponding to a deterioration of approximately 1.8\% to 2.3\% relative to the CHAR benchmark, while average QLIKE ratios are moderately above one, implying losses that are roughly 7\% to 12\% higher. This indicates that, when evaluated across all trading days, there is limited evidence that general hot news provides incremental predictive information beyond historical RV dynamics. However, under MSE, the ensemble exhibits substantial improvements on normal volatility days. Several specifications achieve average MSE ratios well below one, with the best performance observed for Google, where the minimum average MSE ratio of 0.765 corresponds to an improvement of approximately 23.5\% relative to CHAR. These gains are accompanied by high RC rates under MSE, often exceeding 80\%, indicating that the Reality Check rarely identifies a HAR-family competitor with significantly lower expected loss. In contrast, during high volatility days the ensemble based on general hot news performs close to, but generally not better than, the CHAR benchmark, with average MSE ratios around 1.02, implying performance losses of about 2\%, and only modest and specification-dependent improvements under QLIKE. As with stock-related news, increasing the number of filters does not generate systematic performance gains, indicating that higher model complexity plays a secondary role.\footnote{\Cref{appendix_LM_Dicionary} evaluates simpler textual signals by augmenting CHAR with Loughran and McDonald (LM) sentiment measures from \citet{loughran2011liability} and a `News Count' variable. The results show that news intensity itself contains meaningful predictive information, whereas most dictionary-based sentiment measures produce only limited improvements over CHAR. The NLP-based ensemble achieves stronger performance in several dimensions. However, the comparison should be interpreted cautiously because the textual signals enter the forecasting framework differently: LM variables and `News Count' augment CHAR directly, whereas the NLP signal is incorporated through a combination of forecasts.} \footnote{\Cref{appendix_nonlinear} evaluates whether nonlinear reformulations of the HAR-family of models, holding the information set fixed, can account for the forecasting improvements and higher realised utility achieved by the proposed NLP and ensemble models. While nonlinear HAR specifications show improvements in MSE performance during normal volatility days, they fail to deliver robust performance across volatility regimes and generate inferior realised utility.}

Taken together, the ensemble results clarify and strengthen the conclusions drawn from the standalone NLP analysis in \Cref{results_NLP_subsection}. While the NLP models typically underperformed the best HAR-family specifications in terms of loss ratios, the ensemble approach converts the informational content embedded in news text into economically and quantitatively meaningful reductions in forecast loss. This complementarity is strongest for stock-related news over the full out-of-sample period, while its regime-specific pattern differs across loss functions: MSE gains are larger on normal volatility days, whereas QLIKE gains are substantially larger on high volatility days. For general hot news, MSE gains are concentrated on normal volatility days, while modest and specification-dependent QLIKE improvements appear during high volatility days. Across both news types, the ensemble evidence is consistent with the view that forecasts constructed from different information sets can contain complementary predictive information, while the limited sensitivity to the number of filters indicates that increasing NLP model complexity alone is not the main determinant of the results. This interpretation is consistent with the forecast combination literature, where \citet{becker2008combination} show in the context of volatility forecasting that combining forecasts can improve predictive accuracy relative to individual models. More importantly, these results are consistent with news-based signals containing incremental predictive information for RV rather than fully substituting for volatility-history dynamics. Finally, specialised word embeddings tend to perform more favourably for stock-related news, while general-purpose word embeddings are especially competitive for general hot news, particularly during normal volatility days.

\afterpage{%
\clearpage
\thispagestyle{empty}

\atxy{\dimexpr\paperwidth-0.45in}{.5\paperheight}{%
  \rotatebox[origin=center]{90}{\thepage}
}

\begin{landscape}
\begin{table}
\centering
\begin{adjustbox}{max width={1.35\textwidth}}
\begin{threeparttable}
\centering
\setlength{\tabcolsep}{6pt}

\caption{Out-of-Sample RV Forecasting Performance: Stock-Related News (Ensemble)}

\begin{tabular}{cccc|cccc|cccc|cccc|cccc|cccc|cccc}
\toprule
& & &
& \multicolumn{16}{c|}{\textbf{FinText}}
& \multicolumn{4}{c|}{\multirow{2}{*}{\raisebox{-0.6ex}{\textbf{Google}}}}
& \multicolumn{4}{c}{\multirow{2}{*}{\raisebox{-0.6ex}{\textbf{WikiNews}}}} \\
\cmidrule(lr){5-20}
& & &
& \multicolumn{4}{c|}{Word2Vec (CBOW)}
& \multicolumn{4}{c|}{Word2Vec (skip-gram)}
& \multicolumn{4}{c|}{FastText (CBOW)}
& \multicolumn{4}{c|}{FastText (skip-gram)}
&  &  &  &  &  &  &  &  \\
					{{\textbf{\small Full Out-of-Sample Period}}}  & {} & {$\alpha$} & {} & 25 & 50 & 75 & 100 & 25 & 50 & 75 & 100 & 25 & 50 & 75 & 100 & 25 & 50 & 75 & 100 & 25 & 50 & 75 & 100 & 25 & 50 & 75 & 100  \\  \midrule 
                    \addlinespace[0.25cm]
					MSE\tnote{a} & &  & Avg & 0.971 & 0.973 & 0.978 & 0.977 & 0.961 & 0.961 & 0.961 & 0.962 & 1.003 & 1.019 & 1.017 & 1.020 & 0.962 & 0.962 & 0.962 & 0.962 & 0.974 & 0.975 & 0.977 & 0.975 & 0.980 & 0.980 & 0.980 & 0.980 \\
					& &  & Med & 0.993 & 0.992 & 0.988 & 0.995 & 0.981 & 0.980 & 0.981 & 0.980 & 1.027 & 1.024 & 1.020 & 1.025 & 0.984 & 0.982 & 0.984 & 0.984 & 0.993 & 0.993 & 0.994 & 0.993 & 1.000 & 1.001 & 1.001 & 1.001  \\

					RC\tnote{c} & & 0.05 &  & 100 & 100 & 100 & 100 & 100 & 100 & 100 & 100 & 100 & 100 & 100 & 100 & 100 & 100 & 100 & 100 & 100 & 100 & 100 & 100 & 95.65 & 95.65 & 95.65 & 95.65 \\
					& & 0.10 &  & 100 & 100 & 100 & 100 & 100 & 100 & 100 & 100 & 100 & 100 & 100 & 100 & 100 & 100 & 100 & 100 & 100 & 100 & 100 & 100 & 91.30 & 95.65 & 95.65 & 95.65  \\ \addlinespace[0.25cm]
     
					QLIKE\tnote{b} & &  & Avg & 0.972 & 0.965 & 0.991 & 1.002 & 0.937 & 0.938 & 0.937 & 0.939 & 1.038 & 1.086 & 1.088 & 1.096 & 0.938 & 0.938 & 0.939 & 0.940 & 0.959 & 0.959 & 0.966 & 0.959 & 0.967 & 0.970 & 0.968 & 0.968 \\
					& &  & Med & 1.015 & 0.994 & 1.011 & 1.036 & 0.969 & 0.968 & 0.970 & 0.972 & 1.058 & 1.079 & 1.076 & 1.081 & 0.969 & 0.972 & 0.970 & 0.972 & 0.983 & 0.983 & 0.989 & 0.988 & 0.996 & 1.006 & 1.010 & 1.000 \\
					
					RC & & 0.05 &  & 95.65 & 95.65 & 86.96 & 82.61 & 95.65 & 95.65 & 95.65 & 95.65 & 91.30 & 91.30 & 91.30 & 86.96 & 91.30 & 91.30 & 91.30 & 91.30 & 95.65 & 95.65 & 95.65 & 95.65 & 56.52 & 34.78 & 34.78 & 39.13 \\
					& & 0.10 &  & 82.61 & 91.30 & 78.26 & 73.91 & 95.65 & 95.65 & 95.65 & 95.65 & 86.96 & 86.96 & 91.30 & 86.96 & 91.30 & 91.30 & 91.30 & 91.30 & 91.30 & 91.30 & 95.65 & 91.30 & 47.83 & 30.44 & 34.78 & 21.74 \\ \addlinespace[0.25cm]

					\midrule
					{\textbf{\small Normal Volatility Days}}  & {} & {} &   {} & {} & {} \\ 	
					\midrule
                    \addlinespace[0.25cm]
					MSE & &  & Avg & 1.164 & 1.181 & 1.127 & 1.107 & 0.976 & 0.981 & 0.982 & 0.979 & 1.009 & 0.942 & 0.923 & 0.919 & 0.977 & 0.985 & 0.984 & 0.982 & 0.946 & 0.942 & 0.942 & 0.943 & 0.916 & 0.914 & 0.915 & 0.915 \\
					& &  & Med & 1.134 & 1.145 & 1.018 & 1.058 & 0.917 & 0.923 & 0.917 & 0.911 & 1.036 & 0.926 & 0.946 & 0.936 & 0.906 & 0.926 & 0.901 & 0.903 & 0.861 & 0.860 & 0.864 & 0.859 & 0.863 & 0.860 & 0.860 & 0.861 \\
					
					RC & & 0.05 &  & 78.26 & 82.61 & 82.61 & 86.96 & 78.26 & 78.26 & 78.26 & 78.26 & 86.96 & 86.96 & 86.96 & 86.96 & 82.61 & 82.61 & 82.61 & 82.61 & 78.26 & 78.26 & 78.26 & 82.61 & 82.61 & 91.30 & 91.30 & 91.30 \\
					& & 0.10 &  & 69.57 & 65.22 & 78.26 & 82.61 & 78.26 & 78.26 & 78.26 & 78.26 & 78.26 & 86.96 & 82.61 & 86.96 & 78.26 & 78.26 & 82.61 & 82.61 & 78.26 & 78.26 & 78.26 & 78.26 & 78.26 & 86.96 & 91.30 & 91.30 \\ \addlinespace[0.25cm]

					QLIKE & &  & Avg & 1.125 & 1.133 & 1.142 & 1.149 & 1.112 & 1.113 & 1.115 & 1.114 & 1.123 & 1.164 & 1.176 & 1.182 & 1.111 & 1.113 & 1.115 & 1.114 & 1.130 & 1.126 & 1.131 & 1.127 & 1.136 & 1.135 & 1.134 & 1.134 \\
					& &  & Med & 1.127 & 1.134 & 1.143 & 1.151 & 1.106 & 1.100 & 1.102 & 1.098 & 1.110 & 1.157 & 1.167 & 1.179 & 1.109 & 1.105 & 1.102 & 1.104 & 1.126 & 1.119 & 1.122 & 1.123 & 1.137 & 1.129 & 1.126 & 1.128 \\
					
					RC & & 0.05 &  & 30.44 & 26.09 & 21.74 & 21.74 & 21.74 & 21.74 & 26.09 & 21.74 & 21.74 & 26.09 & 21.74 & 21.74 & 21.74 & 21.74 & 21.74 & 21.74 & 21.74 & 21.74 & 21.74 & 21.74 & 30.44 & 17.39 & 13.04 & 13.04 \\
					& & 0.10 &  & 21.74 & 21.74 & 21.74 & 13.04 & 21.74 & 21.74 & 21.74 & 21.74 & 21.74 & 21.74 & 17.39 & 21.74 & 21.74 & 21.74 & 21.74 & 21.74 & 21.74 & 21.74 & 21.74 & 21.74 & 21.74 & 13.04 & 13.04 & 8.70 \\ \addlinespace[0.25cm]

					\midrule
					{\textbf{\small High Volatility Days}}  & {} & {} &   {} & {} & {} \\ 	
					\midrule	
                    \addlinespace[0.25cm]
					MSE & &  & Avg & 0.962 & 0.964 & 0.971 & 0.972 & 0.959 & 0.959 & 0.959 & 0.960 & 1.002 & 1.021 & 1.020 & 1.023 & 0.960 & 0.960 & 0.960 & 0.960 & 0.974 & 0.975 & 0.977 & 0.975 & 0.981 & 0.982 & 0.981 & 0.981 \\
					& &  & Med & 0.976 & 0.982 & 0.985 & 0.987 & 0.973 & 0.975 & 0.973 & 0.972 & 1.023 & 1.025 & 1.026 & 1.028 & 0.975 & 0.975 & 0.971 & 0.976 & 0.992 & 0.993 & 0.993 & 0.993 & 1.001 & 1.003 & 1.003 & 1.001 \\
					
					RC & & 0.05 &  & 100 & 100 & 100 & 100 & 100 & 100 & 100 & 100 & 100 & 100 & 100 & 100 & 100 & 100 & 100 & 100 & 100 & 100 & 100 & 100 & 95.65 & 95.65 & 95.65 & 95.65 \\
					& & 0.10 &  & 100 & 100 & 100 & 100 & 100 & 100 & 100 & 100 & 100 & 100 & 100 & 100 & 100 & 100 & 100 & 100 & 100 & 100 & 100 & 100 & 95.65 & 91.30 & 95.65 & 91.30 \\ [0.25cm]	
     
					QLIKE & &  & Avg & 0.821 & 0.798 & 0.840 & 0.859 & 0.765 & 0.766 & 0.763 & 0.767 & 0.952 & 1.006 & 0.996 & 1.006 & 0.769 & 0.767 & 0.765 & 0.768 & 0.788 & 0.792 & 0.802 & 0.792 & 0.800 & 0.806 & 0.804 & 0.803 \\
					& &  & Med & 0.858 & 0.832 & 0.875 & 0.895 & 0.787 & 0.776 & 0.775 & 0.788 & 0.963 & 0.992 & 0.983 & 0.991 & 0.799 & 0.783 & 0.806 & 0.807 & 0.810 & 0.830 & 0.835 & 0.831 & 0.822 & 0.825 & 0.826 & 0.823 \\
					
					RC & & 0.05 &  & 100 & 100 & 100 & 100 & 100 & 100 & 100 & 100 & 100 & 100 & 100 & 100 & 100 & 100 & 100 & 100 & 100 & 100 & 100 & 100 & 95.65 & 91.30 & 91.30 & 95.65 \\
					& & 0.10 &  & 100 & 100 & 100 & 100 & 100 & 100 & 100 & 100 & 100 & 100 & 100 & 100 & 100 & 100 & 100 & 100 & 100 & 100 & 100 & 100 & 95.65 & 91.30 & 91.30 & 91.30 \\ [0.25cm]		
                    
					\bottomrule                                
				\end{tabular}
				\begin{tablenotes} [para,flushleft]
					\footnotesize
                        \textit{Notes:} \item[a] For each ticker, MSE is calculated over the relevant evaluation sample and expressed relative to the corresponding MSE of CHAR. The reported average (median) is the cross-sectional mean (median) of the 23 ticker-level MSE ratios. \item[b] For each ticker, QLIKE is calculated over the relevant evaluation sample and expressed relative to the corresponding QLIKE of CHAR. The reported average (median) is the cross-sectional mean (median) of the 23 ticker-level QLIKE ratios. \item[c] RC non-rejection denotes the percentage of tickers for which White's Reality Check does not reject, at the stated significance level, the null hypothesis that no competing HAR-family model has lower expected loss than the model under evaluation. A high value therefore indicates limited statistical evidence of inferiority.

				\end{tablenotes}
				\label{NLP_ML_primary_experiment_table_stock_related_ensemble}
			\end{threeparttable}
		\end{adjustbox}
	\end{table}
\end{landscape}
}

\afterpage{%
\clearpage
\thispagestyle{empty}

\atxy{\dimexpr\paperwidth-0.45in}{.5\paperheight}{%
  \rotatebox[origin=center]{90}{\thepage}
}

\begin{landscape}
\begin{table}
\centering
\begin{adjustbox}{max width={1.35\textwidth}}
\begin{threeparttable}
\centering
\setlength{\tabcolsep}{6pt}

\caption{Out-of-Sample RV Forecasting Performance: General Hot News (Ensemble)}

\begin{tabular}{cccc|cccc|cccc|cccc|cccc|cccc|cccc}
\toprule
& & &
& \multicolumn{16}{c|}{\textbf{FinText}}
& \multicolumn{4}{c|}{\multirow{2}{*}{\raisebox{-0.6ex}{\textbf{Google}}}}
& \multicolumn{4}{c}{\multirow{2}{*}{\raisebox{-0.6ex}{\textbf{WikiNews}}}} \\
\cmidrule(lr){5-20}
& & &
& \multicolumn{4}{c|}{Word2Vec (CBOW)}
& \multicolumn{4}{c|}{Word2Vec (skip-gram)}
& \multicolumn{4}{c|}{FastText (CBOW)}
& \multicolumn{4}{c|}{FastText (skip-gram)}
&  &  &  &  &  &  &  &  \\
					{{\textbf{\small Full Out-of-Sample Period}}}  & {} & {$\alpha$} & {} & 25 & 50 & 75 & 100 & 25 & 50 & 75 & 100 & 25 & 50 & 75 & 100 & 25 & 50 & 75 & 100 & 25 & 50 & 75 & 100 & 25 & 50 & 75 & 100  \\ \midrule
                    \addlinespace[0.25cm]
					MSE\tnote{a} & &  & Avg & 1.023 & 1.020 & 1.020 & 1.019 & 1.020 & 1.020 & 1.020 & 1.019 & 1.020 & 1.020 & 1.020 & 1.019 & 1.020 & 1.020 & 1.020 & 1.019 & 1.018 & 1.018 & 1.018 & 1.019 & 1.019 & 1.020 & 1.019 & 1.020 \\
					& &  & Med & 1.029 & 1.021 & 1.022 & 1.023 & 1.022 & 1.023 & 1.022 & 1.024 & 1.023 & 1.027 & 1.021 & 1.021 & 1.024 & 1.022 & 1.023 & 1.022 & 1.023 & 1.022 & 1.021 & 1.023 & 1.023 & 1.024 & 1.024 & 1.024 \\
					
					RC\tnote{c} & & 0.05 &  & 91.30 & 95.65 & 95.65 & 95.65 & 95.65 & 95.65 & 95.65 & 95.65 & 95.65 & 95.65 & 95.65 & 95.65 & 95.65 & 95.65 & 95.65 & 95.65 & 95.65 & 95.65 & 95.65 & 95.65 & 95.65 & 95.65 & 95.65 & 95.65 \\
					& & 0.10 &  & 91.30 & 95.65 & 95.65 & 95.65 & 95.65 & 95.65 & 95.65 & 95.65 & 95.65 & 95.65 & 95.65 & 95.65 & 95.65 & 95.65 & 95.65 & 95.65 & 95.65 & 95.65 & 95.65 & 95.65 & 95.65 & 91.30 & 95.65 & 95.65 \\ \addlinespace[0.25cm]

					QLIKE\tnote{b} & &  & Avg & 1.117 & 1.106 & 1.110 & 1.105 & 1.097 & 1.102 & 1.097 & 1.107 & 1.105 & 1.113 & 1.107 & 1.106 & 1.091 & 1.099 & 1.100 & 1.100 & 1.077 & 1.078 & 1.080 & 1.088 & 1.089 & 1.090 & 1.089 & 1.090 \\
					& &  & Med & 1.107 & 1.098 & 1.098 & 1.094 & 1.099 & 1.099 & 1.096 & 1.091 & 1.091 & 1.111 & 1.097 & 1.091 & 1.087 & 1.091 & 1.089 & 1.083 & 1.082 & 1.079 & 1.078 & 1.082 & 1.081 & 1.078 & 1.078 & 1.080 \\
					
					RC & & 0.05 &  & 26.09 & 21.74 & 26.09 & 26.09 & 39.13 & 26.09 & 30.44 & 34.78 & 52.17 & 52.17 & 34.78 & 39.13 & 30.44 & 34.78 & 34.78 & 30.44 & 43.48 & 26.09 & 39.13 & 43.48 & 26.09 & 21.74 & 26.09 & 26.09 \\
					& & 0.10 &  & 21.74 & 21.74 & 26.09 & 26.09 & 21.74 & 21.74 & 26.09 & 26.09 & 34.78 & 30.44 & 34.78 & 26.09 & 26.09 & 17.39 & 21.74 & 21.74 & 30.44 & 17.39 & 26.09 & 30.44 & 26.09 & 21.74 & 21.74 & 26.09 \\ \addlinespace[0.25cm] 
                     
					\midrule
					{\textbf{\small Normal Volatility Days}}  & {} & {} &   {} & {} & {} \\ 	
					\midrule	
                    \addlinespace[0.25cm]
					MSE & &  & Avg & 1.097 & 0.991 & 0.984 & 0.947 & 0.872 & 0.910 & 0.903 & 0.956 & 0.973 & 0.983 & 0.970 & 0.946 & 0.853 & 0.883 & 0.906 & 0.927 & 0.765 & 0.770 & 0.781 & 0.818 & 0.806 & 0.806 & 0.811 & 0.815 \\
					& &  & Med & 1.167 & 0.928 & 0.971 & 0.942 & 0.837 & 0.899 & 0.901 & 0.936 & 0.968 & 0.962 & 0.937 & 0.947 & 0.850 & 0.869 & 0.931 & 0.923 & 0.747 & 0.735 & 0.747 & 0.818 & 0.826 & 0.823 & 0.832 & 0.828 \\
					
					RC & & 0.05 &  & 69.57 & 82.61 & 78.26 & 82.61 & 95.65 & 91.30 & 91.30 & 86.96 & 100 & 100 & 100 & 95.65 & 100 & 100 & 100 & 100 & 95.65 & 91.30 & 95.65 & 95.65 & 82.61 & 82.61 & 82.61 & 82.61 \\
					& & 0.10 &  & 65.22 & 78.26 & 78.26 & 82.61 & 95.65 & 91.30 & 91.30 & 86.96 & 100 & 100 & 100 & 95.65 & 95.65 & 100 & 95.65 & 95.65 & 95.65 & 91.30 & 91.30 & 82.61 & 82.61 & 78.26 & 82.61 & 78.26 \\ \addlinespace[0.25cm]

					QLIKE & &  & Avg & 1.235 & 1.223 & 1.226 & 1.212 & 1.106 & 1.149 & 1.150 & 1.206 & 1.220 & 1.224 & 1.224 & 1.225 & 1.104 & 1.137 & 1.158 & 1.176 & 1.071 & 1.077 & 1.086 & 1.117 & 1.091 & 1.091 & 1.094 & 1.097 \\
					& &  & Med & 1.227 & 1.208 & 1.220 & 1.215 & 1.114 & 1.121 & 1.140 & 1.200 & 1.211 & 1.212 & 1.216 & 1.212 & 1.093 & 1.125 & 1.155 & 1.176 & 1.068 & 1.074 & 1.101 & 1.107 & 1.105 & 1.105 & 1.105 & 1.113 \\
					
					RC & & 0.05 &  & 0.00 & 0.00 & 0.00 & 4.35 & 17.39 & 4.35 & 4.35 & 0.00 & 39.13 & 39.13 & 39.13 & 21.74 & 30.44 & 26.09 & 30.44 & 30.44 & 17.39 & 8.70 & 8.70 & 8.70 & 0.00 & 0.00 & 0.00 & 0.00 \\
					& & 0.10 &  & 0.00 & 0.00 & 0.00 & 4.35 & 8.70 & 4.35 & 4.35 & 0.00 & 39.13 & 39.13 & 39.13 & 21.74 & 30.44 & 26.09 & 26.09 & 30.44 & 13.04 & 8.70 & 8.70 & 4.35 & 0.00 & 0.00 & 0.00 & 0.00 \\ \addlinespace[0.25cm]	

					\midrule
					{\textbf{\small High Volatility Days}}  & {} & {} &   {} & {} & {} \\ 	
					\midrule	
                    \addlinespace[0.25cm]
					MSE & &  & Avg & 1.020 & 1.020 & 1.021 & 1.021 & 1.025 & 1.024 & 1.024 & 1.022 & 1.021 & 1.021 & 1.021 & 1.021 & 1.025 & 1.025 & 1.024 & 1.023 & 1.026 & 1.026 & 1.026 & 1.026 & 1.027 & 1.027 & 1.027 & 1.026 \\
					& &  & Med & 1.025 & 1.025 & 1.027 & 1.027 & 1.030 & 1.029 & 1.029 & 1.027 & 1.026 & 1.027 & 1.024 & 1.025 & 1.032 & 1.033 & 1.032 & 1.028 & 1.033 & 1.031 & 1.032 & 1.031 & 1.032 & 1.034 & 1.034 & 1.033 \\
					
					RC & & 0.05 &  & 95.65 & 95.65 & 95.65 & 95.65 & 95.65 & 95.65 & 95.65 & 95.65 & 91.30 & 86.96 & 91.30 & 91.30 & 86.96 & 86.96 & 86.96 & 82.61 & 95.65 & 95.65 & 95.65 & 95.65 & 95.65 & 95.65 & 95.65 & 95.65 \\
					& & 0.10 &  & 95.65 & 91.30 & 91.30 & 91.30 & 86.96 & 86.96 & 82.61 & 86.96 & 73.91 & 69.57 & 78.26 & 73.91 & 69.57 & 73.91 & 73.91 & 73.91 & 86.96 & 82.61 & 82.61 & 86.96 & 91.30 & 91.30 & 91.30 & 91.30 \\ \addlinespace[0.25cm]

					QLIKE & &  & Avg & 0.995 & 0.985 & 0.988 & 0.992 & 1.086 & 1.055 & 1.043 & 1.006 & 0.987 & 0.998 & 0.987 & 0.983 & 1.076 & 1.060 & 1.041 & 1.022 & 1.077 & 1.073 & 1.069 & 1.055 & 1.080 & 1.083 & 1.078 & 1.078 \\
					& &  & Med & 0.977 & 0.974 & 0.982 & 0.962 & 1.067 & 1.071 & 1.041 & 1.008 & 0.972 & 0.991 & 0.980 & 0.968 & 1.050 & 1.071 & 1.048 & 0.999 & 1.064 & 1.056 & 1.047 & 1.034 & 1.056 & 1.071 & 1.061 & 1.069 \\
					
					RC & & 0.05 &  & 100 & 100 & 100 & 100 & 82.61 & 91.30 & 95.65 & 95.65 & 78.26 & 78.26 & 78.26 & 86.96 & 73.91 & 73.91 & 73.91 & 73.91 & 82.61 & 91.30 & 86.96 & 91.30 & 95.65 & 95.65 & 100 & 100 \\
					& & 0.10 &  & 95.65 & 91.30 & 95.65 & 91.30 & 73.91 & 78.26 & 82.61 & 91.30 & 65.22 & 73.91 & 73.91 & 82.61 & 69.57 & 73.91 & 73.91 & 69.57 & 73.91 & 78.26 & 82.61 & 91.30 & 95.65 & 91.30 & 95.65 & 95.65 \\ \addlinespace[0.25cm]

					\bottomrule                                
				\end{tabular}
				\begin{tablenotes} [para,flushleft]
					\footnotesize
                        \textit{Notes:} \item[a] For each ticker, MSE is calculated over the relevant evaluation sample and expressed relative to the corresponding MSE of CHAR. The reported average (median) is the cross-sectional mean (median) of the 23 ticker-level MSE ratios. \item[b] For each ticker, QLIKE is calculated over the relevant evaluation sample and expressed relative to the corresponding QLIKE of CHAR. The reported average (median) is the cross-sectional mean (median) of the 23 ticker-level QLIKE ratios. \item[c] RC non-rejection denotes the percentage of tickers for which White's Reality Check does not reject, at the stated significance level, the null hypothesis that no competing HAR-family model has lower expected loss than the model under evaluation. A high value therefore indicates limited statistical evidence of inferiority.
				\end{tablenotes}
				\label{NLP_ML_primary_experiment_table_general_ensemble}
			\end{threeparttable}
		\end{adjustbox}
	\end{table}
\end{landscape}
}
\newpage

\subsection{Economic Evaluation} \label{economic_gain_section}
We complement the statistical forecast evaluation with an economic evaluation using the utility-based approach developed in \citet{bollerslev2018risk}. This setting considers a mean--variance investor who follows a risk-targeting strategy when trading assets with a constant Sharpe ratio (SR). Positions in each asset are dynamically scaled to maintain a fixed target level of volatility. In this environment, the quality of volatility forecasts plays a critical role in determining investor utility. Accurate forecasts enable the investor to closely adhere to the desired risk profile, whereas uncertainty induced by volatility-of-volatility generates fluctuations around the target risk level, leading to suboptimal portfolio scaling and lower expected utility.

The empirical assessment of expected utility is implemented by computing realised utility using out-of-sample volatility forecasts. Following this approach, we employ the utility-of-wealth (UoW) measure, which captures the realised utility associated with forecasts of RV. The measure is defined as
\begin{equation} \label{utility_function}
UoW^{M} = \frac{1}{T} \sum_{t=1}^{T} \left[ 8\% \frac{\sqrt{RV_{t+1}}}{\sqrt{\mathbb{E}^{M}_{t}\!\left( RV_{t+1} \right)}} - 4\% \frac{RV_{t+1}}{\mathbb{E}^{M}_{t}\!\left( RV_{t+1} \right)} \right],
\end{equation}
where $RV_{t+1}$ denotes the RV at time $t+1$, $\mathbb{E}^{M}_{t}\!\left( RV_{t+1} \right)$ represents the expectation from model $M$, that is, the RV forecast generated by model $M$ at time $t$, and $T$ denotes the number of days in the out-of-sample period. The coefficients 8\% and 4\% capture investor preferences for reward and penalty, respectively, and are calibrated using economically plausible assumptions regarding portfolio performance and risk aversion, as outlined in \citet{bollerslev2018risk}.\footnote{The calibration assumes an annualised SR of 0.4, a coefficient of risk aversion equal to 2, and an investor targeting an annualised volatility of 20\%.} Under this calibration, a perfectly specified risk model that accurately forecasts RV achieves a realised utility of 4\%. Realised utility is computed separately for each ticker over the out-of-sample period, and the reported values are the cross-sectional averages of the resulting ticker-level utilities across the 23 tickers.

\Cref{realised_utility_models} reports the realised utility of the NLP models presented in \Cref{results_NLP_subsection} in the left panel and the ensemble models discussed in \Cref{Results_Ensemble_Subsection} in the right panel. As a benchmark, the CHAR model attains a realised utility of 2.7540\%.\footnote{The realised utilities for the HAR-family of models are AR (2.4642\%), HAR (2.6277\%), HAR-J (2.7502\%), CHAR (2.7540\%), SHAR (2.6262\%), ARQ (2.5611\%), HARQ (2.6591\%), and HARQ-F (2.5837\%). Among these specifications, the CHAR model delivers the highest realised utility, consistent with the results reported in \Cref{results_section}.} For ease of presentation, the reported realised utilities are averaged across model complexity settings. The realised utility results provide a complementary economic perspective on the statistical forecasting evidence in \Cref{results_NLP_subsection} and \Cref{Results_Ensemble_Subsection}. When news is used in isolation (left panel), all NLP models deliver realised utilities substantially below CHAR, ranging from 1.6280\% to 2.1047\% for stock-related news and from 1.5550\% to 2.0165\% for general hot news. In this sense, the realised utility approach corroborates the earlier conclusion that news-only models, while informative, do not match the realised utility delivered by HAR-type benchmarks.

\begin{table}
    \centering
    \begin{adjustbox}{max width=0.84\textwidth}
        \begin{threeparttable}
            \caption{Realised Utility of NLP and Ensemble Models}
            \begin{tabular}{l@{\hspace{62pt}}cl@{\hspace{62pt}}c}
                \toprule
                \multicolumn{2}{c}{NLP Models} &
                \multicolumn{2}{c}{Ensemble Models} \\
                \cmidrule(r){1-2} \cmidrule(l){3-4}
                {Model} & {Realised Utility} & {Model} & {Realised Utility} \\
                \midrule
                \multicolumn{4}{l}{\textit{Stock-related news}} \\
                Word2Vec (CBOW)        & 1.6280 & Word2Vec (CBOW) & 2.8677 \\
                Word2Vec (skip-gram)   & 1.8949 & Word2Vec (skip-gram) & 2.9321 \\
                FastText (CBOW)        & 1.7772 & FastText (CBOW) & 2.7148 \\
                FastText (skip-gram)   & 1.9347 & FastText (skip-gram) & 2.9299 \\
                Google                 & 2.0437 & Google & 2.9013 \\
                WikiNews               & 2.1047 & WikiNews & 2.8928 \\
                \midrule
                \multicolumn{4}{l}{\textit{General hot news}} \\
                Word2Vec (CBOW)        & 1.9180 & Word2Vec (CBOW) & 2.6866 \\
                Word2Vec (skip-gram)   & 1.5550 & Word2Vec (skip-gram) & 2.6675 \\
                FastText (CBOW)        & 2.0165 & FastText (CBOW) & 2.6878 \\
                FastText (skip-gram)   & 1.7363 & FastText (skip-gram) & 2.6684 \\
                Google                 & 1.7334 & Google & 2.6722 \\
                WikiNews               & 1.7363 & WikiNews & 2.6578 \\
                \bottomrule
            \end{tabular}
            \label{realised_utility_models}
            \begin{tablenotes}[para,flushleft]
                \footnotesize
                \item \textit{Notes:} This table reports realised utility values from the utility-based approach for the NLP models in the left panel and the ensemble models in the right panel. In \Cref{utility_function}, the maximum attainable realised utility is 4\%. For ease of presentation, the reported realised utilities are averaged across model complexity settings with 25, 50, 75, and 100 filters. All values are expressed in percentage terms.
            \end{tablenotes}
        \end{threeparttable}
    \end{adjustbox}
\end{table}

By contrast, higher realised utility is obtained when stock-related news is combined with the RV dynamics captured by CHAR: five of the six ensemble variants exceed the CHAR utility, with the best-performing Word2Vec (skip-gram) ensemble attaining 2.9321\%, compared with 2.7540\% for CHAR, a gross difference of 0.1781 percentage points.\footnote{This magnitude should be interpreted cautiously because the utility calculation abstracts from transaction costs and other implementation frictions and does not provide statistical inference for the utility difference. We therefore interpret the result as evidence of potential economic value rather than as an estimate of net implementable investor gain.} Importantly, the utility ranking across word embedding algorithms is broadly consistent with the forecasting performance results. In contrast, general hot news ensembles remain below CHAR in realised utility (between 2.6578\% and 2.6878\%), which matches the full out-of-sample evidence in \Cref{NLP_ML_primary_experiment_table_general_ensemble} showing loss ratios modestly above one despite improvements on normal volatility days. Overall, realised utility confirms that news is best viewed as a complementary signal, with the highest realised utility obtained when stock-related news is combined with the HAR-family benchmark.\footnote{The results presented in \Cref{appendix_LM_Dicionary} show that the LM sentiment models yield realised utilities that are largely indistinguishable from the CHAR benchmark, while the News Count specification delivers a modest improvement. By contrast, most stock-related ensemble models deliver higher realised utilities than CHAR, with the best-performing ensemble also exceeding the dictionary-based alternatives. This is consistent with potential additional economic value from the NLP-based forecasting framework, although the comparison does not isolate semantic content because News Count enters CHAR directly whereas the NLP signal is incorporated through forecast combination.}

\subsection{Explainable AI (XAI)} \label{explainable_ai_section}
Over recent years, considerable efforts have been made to better understand ML models, which are often described as black-box. In this study, we explore one of the prominent XAI methods, Shapley additive explanations (SHAP), to analyse the explanatory power of specific phrases in RV forecasting. \citet{lundberg2017unified} propose the SHAP method based on coalition game theory. Shapley values, denoted by $\phi_i$, measure the contribution of individual model features to the model output. In our implementation, SHAP is applied to the fully connected output block, so the relevant inputs are the pooled convolutional features in $\mathbf{z}_t$ rather than the raw tokens themselves. Specifically:
\begin{equation}  \label{shap_main}
\phi_{i} = \frac{1}{\abs{N}!} \; \sum_{S\subseteq N\setminus\{i\}}^{}
\abs{S}! \;\; (\abs{N}-\abs{S}-1)! \;\; [f(S\cup\{i\})-f(S)],
\end{equation}
\noindent where $f(S \cup \{i\}) - f(S)$ captures the marginal contribution of adding pooled feature $i$ to the coalition $S$, and $N$ denotes the set of pooled output-block input features. The Shapley weighting accounts for all possible coalitions of the remaining model inputs.\footnote{The Shapley values $\phi_{i}$ indicate the magnitude and sign of the average contribution of feature $i$; these satisfy three properties, namely local accuracy (additivity), which means that the model output is fully accounted for by the baseline value and the individual feature attributions; missingness (nonexistence or null effect), which implies that a missing feature has no attributed impact ($\phi_{i} = 0$); and consistency, which means that if a change in a specific feature has a greater impact on the first model compared to the second model, the importance of this feature should be higher for the first model than for the second model.} As features are added to the coalition, changes in the model output reflect their relevance. The advantages of the SHAP method include a solid theoretical foundation in game theory and no requirement for differentiable models. However, it is computationally intensive and, like other permutation-based approaches, does not consider feature dependencies, potentially leading to misleading results.\footnote{Following \citet{zhao2020shap}, SHAP is applied to the fully connected output block of the model. In this instance, the inputs are outputs from the global max pooling layer, and the final output is the forecasted RV in \Cref{NN_detailed_rep}. Consequently, the number of inputs is equal to $3q$, where $q$ is the number of filters, and 3 corresponds to the three filter sizes (unigram, bigram, and trigram). Applying SHAP to the full model can lead to misleading results because non-existent words may mistakenly receive high scores. However, by applying SHAP specifically to the fully connected output block, the attributions are computed over the pooled features extracted by the convolutional filters. For economic interpretation, each pooled feature is mapped back to the n-gram that generates the corresponding maximum convolutional activation. Moreover, the duplication of filters is common in CNNs, particularly for large $q$ values \citep{roychowdhury2017reducing}, which could result in similar n-grams being generated. To prevent this, a de-duplication step is implemented by removing repeated n-grams and recalculating the Shapley value by summing the corresponding Shapley values. \citet{zhao2020shap} propose two steps for de-duplication, namely `exact de-duplication' (employed in this study) and `merge de-duplication'. In `merge de-duplication', overlapping n-grams are merged, and their Shapley value is determined by summing the Shapley values of the constituent n-grams. Given the generally shorter input text length, our experiments indicate that incorporating `exact de-duplication' alone is sufficient to achieve more granular results.} Here, we use a high-speed approximation algorithm, Deep SHAP based on DeepLIFT \citep{shrikumar2017learning}, to approximate Shapley values, with the DeepLIFT target defined as the pre-activation of the final output layer.\footnote{An overview of the DeepLIFT is provided in \Cref{appendix_deeplift}.}

For each stock, we obtained the Shapley values for the constituent n-grams of the textual information used to forecast RV during the out-of-sample period. For each out-of-sample forecast, we retain the five n-grams with the highest absolute Shapley values. We then aggregate these top-five lists within each stock and retain the unique n-grams that appear at least once. For each n-gram, we count the number of stocks \(t\), where \(t=1,\ldots,23\), in which it appears in at least one top-five list. An n-gram with \(t=23\) therefore appears among the top-five forecast attributions for all 23 stocks.\footnote{For the explainability analysis, we use FinText FastText (skip-gram) for stock-related news and WikiNews for general hot news, with 50 filters in both cases. These specifications are used to provide a representative and computationally tractable XAI analysis rather than to define a separate out-of-sample model-selection exercise. The forecasting results in \Cref{results_NLP_subsection} show that the main qualitative conclusions are not highly sensitive to the number of filters, and related patterns are observed across alternative embedding specifications. We therefore hold the number of filters fixed at 50 in the explainability exercise to facilitate comparability across the two news information sets.}

We examine the primary forecast-attributed n-grams for stock-related news in \Cref{RV_movers_stock_related} and general hot news in \Cref{RV_movers_general_hot} across the entire sample of 23 stocks during the out-of-sample period. As explained, we have a merged list of n-grams appearing in the forecast-level top-five lists for each stock, together with the number of stocks in which each n-gram appears. An n-gram with $t=23$ indicates that the specific n-gram appears in at least one forecast-level top-five list for each of the 23 stocks. For stock-related news, a threshold of $t=13$ repetitions is set to select the n-grams that appear among the most important forecast attributions for more than half of the 23 stocks. For general hot news, a threshold of $t=20$ repetitions is applied.\footnote{Further analysis indicates that varying $t$ does not substantially alter the defined groups of n-grams for both stock-related and general hot news.} There is no difference in importance among the n-grams within each group and across groups in these tables.\footnote{Numerical n-grams hold a prominent position in both tables, underscoring the importance of retaining numerical values during data cleaning, as discussed in \Cref{Appendix_data_cleaning}. While numerical information is fully retained in the modelling stage, numerical n-grams are anonymised in \Cref{RV_movers_stock_related} and \Cref{RV_movers_general_hot} for readability and are represented by \textit{NUMBERS}.}

In \Cref{RV_movers_stock_related}, we grouped the stock-related forecast-attributed n-grams into `Analyst opinion', `Event', `Verb', `Market', `Abbreviation', `Country/Company', `Announcement', `Numeric', `Calendar', `Insider', and `Mixed'.\footnote{It is important to note that this classification is based on our judgement. However, we believe that reassigning some of the n-grams between groups would not alter the fundamental informational content of these groups.} The key findings can be summarised as follows: (i) `Analyst opinion' and `Event' contain the majority of high-SHAP n-grams. This outcome is unsurprising, as it clearly demonstrates the importance of analyst opinions concerning a company's earnings calls and financial reports. Among others, popular n-grams in this group include \textit{registers}, \textit{announces}, \textit{files}, \textit{raises}, and \textit{surrenders}. (ii) `Market' includes market-related n-grams such as \textit{stocks to buy}, \textit{premarket}, and \textit{stock market opens}. (iii) \textit{China} is the only country in the high-SHAP n-gram list for stock-related news, underscoring the relevance of news related to this country. The remaining classes contain fewer n-grams and exhibit less commonality; however, it is evident that these n-grams in stock-related news convey varying levels of information about stocks, the market, and the economy.

\afterpage{%
	\clearpage
	\thispagestyle{empty}
	\atxy{\dimexpr\paperwidth-0.45in}{.5\paperheight}{\rotatebox[origin=center]{90}{\thepage}}
\begin{landscape}
	\begin{table}
		\centering
		\begin{adjustbox}{max width=685pt}
			\begin{threeparttable}
				\centering
				\setlength{\tabcolsep}{4pt}
				\caption{High-SHAP N-grams (Stock-Related News)}
    \begin{tabular}{|l|l|l|l|l|l|l|r|}
    \toprule
    \rowcolor[rgb]{ .949,  .949,  .949} \multicolumn{1}{|c|}{\textbf{Analyst opinion}} & \cellcolor[rgb]{ 1,  1,  1}outperform by & \cellcolor[rgb]{ 1,  1,  1}price target & \cellcolor[rgb]{ 1,  1,  1}of earnings & \cellcolor[rgb]{ 1,  1,  1}sees \#q adj & \cellcolor[rgb]{ 1,  1,  1}jumps & \cellcolor[rgb]{ 1,  1,  1}rtgs & \multicolumn{1}{l|}{\cellcolor[rgb]{ 1,  1,  1}\textit{NUMBERS}\tnote{c}} \\
    \midrule
    equal-weight & neutral by & price target raised & earnings call & other events > & rises & chmn  & \multicolumn{1}{c|}{\cellcolor[rgb]{ .949,  .949,  .949}\textbf{Calendar}} \\
    \midrule
    overweight & equal-weight by & price target announced & earning call transcripts & filing > & sues  & dir   & \multicolumn{1}{l|}{week ended} \\
    \midrule
    neutral & equal-weight from & price target cut & transcript & report & unveils & corp  & \multicolumn{1}{l|}{week ended MONTH\tnote{d}} \\
    \midrule
    cut to neutral & overweight by & target announced & transcript, > & compensation filing & \multicolumn{1}{c|}{\cellcolor[rgb]{ .949,  .949,  .949}\textbf{Market}} & shrs  & \multicolumn{1}{l|}{ended MONTH} \\
    \midrule
    maintained at equal-weight & outperform from & target raised & events > & profit & stock market opens & yr    & \multicolumn{1}{l|}{review for week} \\
    \midrule
    maintained at overweight & overweight from equal-weight & target announced at & earning season & revenue & premarket & \multicolumn{1}{c|}{\cellcolor[rgb]{ .949,  .949,  .949}\textbf{Country/Company}} & \multicolumn{1}{l|}{for week ended} \\
    \midrule
    maintained at outperform & overweight from neutral & announces at & earnings preview & \multicolumn{1}{c|}{\cellcolor[rgb]{ .949,  .949,  .949}\textbf{Verb}} & stock falls & china & \multicolumn{1}{c|}{\cellcolor[rgb]{ .949,  .949,  .949}\textbf{Insider}} \\
    \midrule
    maintained at neutral & equal-weight from overweight & cut to hold & earnings tomorrow & sees  & stock surges & goldman sachs & \multicolumn{1}{l|}{insider review} \\
    \midrule
    from neutral & perform from outperform & raised to neutral & earnings DAY & announces & stock soars & moody's & \multicolumn{1}{l|}{insider review for} \\
    \midrule
    from hold & to neutral & raised to outperform & u.s. earnings DAY & registers & stocks to watch & credit suisse & \multicolumn{1}{l|}{insider sales} \\
    \midrule
    buy from neutral & to neutral from & raised to buy & u.s. earnings & appoints & stocks to buy & nasdaq & \multicolumn{1}{l|}{substantial insider sales} \\
    \midrule
    buy from hold & to outperform & outlook & earnings beat & assigns & to watch & cfa   & \multicolumn{1}{c|}{\cellcolor[rgb]{ .949,  .949,  .949}\textbf{Mixed}} \\
    \midrule
    hold from buy & to outperform from & outlook stable & reports earnings tomorrow & soars & tech stocks & wsj   & \multicolumn{1}{l|}{long-term} \\
    \midrule
    from equal-weight & neutral from & outlk & reports earnings & surrenders & chip stocks & \multicolumn{1}{c|}{\cellcolor[rgb]{ .949,  .949,  .949}\textbf{Announcement}} & \multicolumn{1}{l|}{technology} \\
    \midrule
    from outperform & neutral from buy & analyst says & for u.s. earnings & release & morning movers & announces completion of & \multicolumn{1}{l|}{growth} \\
    \midrule
    from neutral by & neutral from overweight & from hold & conference (transcript) & update & morning report & moody's announces & \multicolumn{1}{l|}{sales} \\
    \midrule
    from overweight & outperform from & \multicolumn{1}{c|}{\cellcolor[rgb]{ .949,  .949,  .949}\textbf{Event}} & holders, \#q & acquires & on the street & agreement & \multicolumn{1}{l|}{sales: morning} \\
    \midrule
    at outperform & outperform from neutral & \#\tnote{a} \smallskip q   & files 8k & backs & \multicolumn{1}{c|}{\cellcolor[rgb]{ .949,  .949,  .949}\textbf{Abbreviation}} & definitive agreement & \multicolumn{1}{l|}{shares} \\
    \midrule
    at overweight & initiated at outperform & fourth quarter & 13f   & declares & inst  & entry into definitive & \multicolumn{1}{l|}{correction} \\
    \midrule
    at overweight by & initiated at neutral & second quarter & dividend & files & inst holders & deal  & \multicolumn{1}{l|}{price} \\
    \midrule
    at equal-weight & initiated at equal-weight & quarter & cash dividend & launches & inc   & \multicolumn{1}{c|}{\cellcolor[rgb]{ .949,  .949,  .949}\textbf{Numeric}} & \multicolumn{1}{l|}{stake} \\
    \midrule
    at neutral & overweight by keybanc & >\tnote{b}     & rev   & completes & mgmt  & billion & \multicolumn{1}{l|}{results} \\
    \midrule
    at neutral by & overweight by morgan & \#q, YEAR & eps   & raises & exec  & bln   & \multicolumn{1}{l|}{new} \\
    \midrule
    at equal-weight by & equal-weight by morgan & holders \#q & adj eps & boosts & exec mgmt & million & \multicolumn{1}{l|}{vs} \\
    \midrule
    at outperform by & outperform & earnings & \#q adj eps & gains & changes exec mgmt & mln   &  \\
    \bottomrule

    \end{tabular}%
				\begin{tablenotes} [para,flushleft]
					\footnotesize
					\item \textit{Notes:} \item[a] \textit{\#} indicates number. \item[b] \textit{>} commonly indicates quantities in news stories, especially the news headlines about financial reports and earning calls. \item[c] For clarity, the numerical n-grams are removed and replaced by \textit{NUMBERS} in this table.  \item[d] \textit{MONTH} represents different months.
				\end{tablenotes}
				\label{RV_movers_stock_related}
			\end{threeparttable}
		\end{adjustbox}
	\end{table}
\end{landscape}
}

\Cref{RV_movers_general_hot} presents the high-SHAP n-grams in general hot news, grouped into `Person', `Place', `Legal entity', `Level', `Verb', `Index', `Data', `Numeric', and `Mixed', and is revealing in several ways. The `Person' category is dominated by prominent political and policy figures, including U.S. presidents Donald Trump, Barack Obama, and Joe Biden, New York City mayor Bill de Blasio, senior political actors such as Mark Meadows, Mitch McConnell, Mike Pompeo, and Nancy Pelosi, as well as key Federal Reserve officials, including Chair Jerome Powell, Janet Yellen, and Federal Reserve Bank presidents James Bullard of St. Louis, Loretta Mester of Cleveland, John C. Williams of New York, Neel Kashkari of Minneapolis, and Raphael Bostic of Atlanta, underscoring the importance assigned by the model to political and monetary policy communication in forecasting RV. International political figures such as Dominic Cummings in the United Kingdom and Kim Jong-un of North Korea also appear among the important high-SHAP n-grams, while the presence of Jim Cramer, a prominent financial media commentator, is notable but unsurprising given the influence of news coverage on market sentiment. In contrast, the appearance of William G. Kaelin, a Nobel Prize-winning physician-scientist, and Reinhard Genzel, an astrophysicist, does not align with the economic and financial focus of the remaining n-grams in this category, illustrating that XAI methods are not error-free and that their outputs must be interpreted with caution.

\afterpage{%
	\clearpage
	\thispagestyle{empty}
	\atxy{\dimexpr\paperwidth-0.45in}{.5\paperheight}{\rotatebox[origin=center]{90}{\thepage}}
\begin{landscape}
	\begin{table}
		\centering
		\begin{adjustbox}{max width=685pt}
			\begin{threeparttable}
				\centering
				\setlength{\tabcolsep}{4pt}
				\caption{High-SHAP N-grams (General Hot News)}
    \begin{tabular}{|l|l|l|l|l|l|r|}
     \toprule
    \rowcolor[rgb]{ .949,  .949,  .949} \multicolumn{1}{|c|}{\textbf{Person}} & \cellcolor[rgb]{ 1,  1,  1}saudis & \cellcolor[rgb]{ 1,  1,  1}fall \#\% & \cellcolor[rgb]{ 1,  1,  1}\% rate & \cellcolor[rgb]{ 1,  1,  1}ranks & \cellcolor[rgb]{ 1,  1,  1}payroll-tax cut & \cellcolor[rgb]{ 1,  1,  1}vacancies \\
    \midrule
    trump (ref. Donald Trump)\tnote{a} & japan & fell \#\% & \% food & correct & shutdown & problems \\
    \midrule
    donald (ref. Donald Trump) & eu    & raising \#\% & high  & win   & offering & major \\
    \midrule
    obama (ref. Barack Obama) & u.k   & above \$\tnote{c} & still high & update & speech & minor \\
    \midrule
    joe biden (ref. Joe Biden) & spain & at \$ & higher & search & trade speech & source \\
    \midrule
    biden (ref. Joe Biden) & asia  & in \$ & min   & influence & hearing & week (wk) \\
    \midrule
    powell (ref. Jerome Powell) & north korea & of \$ & dip   & becoming & us crude & global \\
    \midrule
    yellen (ref. Janet Yellen) & \multicolumn{1}{c|}{\cellcolor[rgb]{ .949,  .949,  .949}\textbf{Legal entity}} & on \$ & up    & \multicolumn{1}{c|}{\cellcolor[rgb]{ .949,  .949,  .949}\textbf{Index}} & gold  & approval \\
    \midrule
    de blasio (ref. Bill de Blasio) & gop (ref. Republican Party) & than \$ & down  & s\&p500 down & crisis & leverage \\
    \midrule
    cummings (ref. Dominic Cummings) & ism (ref. Institute of Supply Management) & from \$ & low   & s\&p500 falls & coalitions & standard \\
    \midrule
    bullard (ref. James B. Bullard) & doe (ref. Department of Energy) & to \$ & least & s\&p500 drops & airstrike & tremendous \\
    \midrule
    cramer (ref. Jim Cramer) & opec  & up to \$ & \multicolumn{1}{c|}{\cellcolor[rgb]{ .949,  .949,  .949}\textbf{Verb}} & s\&p500 gains & shifting coalitions & obstruction \\
    \midrule
    mester (ref. Loretta J. Mester) & fed (ref. Federal Reserve) & around \$ & seen  & s\&p500 rises & campaign staffer & analysis \\
    \midrule
    meadows (ref. Mark Meadows) & omb (ref. Office of Management and Budget) & by \# & sees  & s\&p500 up & law   & statement \\
    \midrule
    williams (ref. John C. Williams) & health organization & \% over & forces & s\&p500 adds & lawmakers & sources \\
    \midrule
    mcconnell (ref. Mitch McConnell) & u.s. treasury & \% target & achieve & s\&p500 climbs & attorney general & adversity sources \\
    \midrule
    pompeo (ref. Mike Pompeo) & treasury & \% from & cut   & s\&p500 & trump administration & advisor \\
    \midrule
    pelosi (ref. Nancy Pelosi) & cdc (ref. Centers for Disease Control and Prevention) & \% rate & grows & \multicolumn{1}{c|}{\cellcolor[rgb]{ .949,  .949,  .949}\textbf{Data}} & compensation & emails \\
    \midrule
    kim (ref. Kim Jong-un) & occ (ref. Options Clearing Corporation) & \% vs & resigns & ex-autos (ref. Retail Sales ex Autos) & banks & state tv \\
    \midrule
    kashkari (ref. Neel Kashkari) & sec (ref. Securities and Exchange Commission) & \% through & says  & inflation & economy & companies \\
    \midrule
    bostic (ref. Raphael Bostic) & wsj (ref. The Wall Street Journal) & about \$ & charges & jobs  & police & opioid companies \\
    \midrule
    g kaelin (ref. William G. Kaelin) & st. louis (ref. Federal Reserve Bank of St. Louis) & under \$ & dies  & gdp   & dodd\_frank & relationships \\
    \midrule
    genzel (ref. Reinhard Genzel) & bloomberg & for \$ & passes & deficit & rico's & sru\_related problems \\
    \midrule
    \rowcolor[rgb]{ .949,  .949,  .949} \multicolumn{1}{|c|}{\textbf{Place}} & \cellcolor[rgb]{ 1,  1,  1}wolfe (ref. Wolfe Research) & \cellcolor[rgb]{ 1,  1,  1}yr to \$ & \cellcolor[rgb]{ 1,  1,  1}propose & \multicolumn{1}{c|}{\textbf{Numeric}} & \cellcolor[rgb]{ 1,  1,  1}corridors & \cellcolor[rgb]{ 1,  1,  1}first \\
    \midrule
    us    & \multicolumn{1}{c|}{\cellcolor[rgb]{ .949,  .949,  .949}\textbf{Level}} & \% on \$ & releases & trillion & officials & chance \\
    \midrule
    korea & below \#\tnote{b} \ \% & \% to \$ & triggers & billion & organization & groups \\
    \midrule
    syria & above \#\% & \% at \# & proceed & million & pm    & house \\
    \midrule
    israel & to \#\% & \% in & totaled & mln bbl & program & direction \\
    \midrule
    china & up \#\% & \% on & transferred & \textit{NUMBERS}\tnote{d} & schools & dreamers' \\
    \midrule
    iran  & achieve \#\% & \% to & pass  & \multicolumn{1}{c|}{\cellcolor[rgb]{ .949,  .949,  .949}\textbf{Mixed}} & prison & others \\
    \midrule
    russia & slide \#\% & \% on year & talk  & spac  & groups & speaker \\
    \bottomrule

    \end{tabular}%
				\begin{tablenotes} [para,flushleft]
					\footnotesize
					\item \textit{Notes:} \item[a] The most likely full phrase to which the specified n-gram refers. \item[b] \textit{\#} indicates number. \item[c] The symbol \textit{\$} denotes monetary values in U.S. dollars. \item[d] For clarity, the numerical n-grams are removed and replaced by \textit{NUMBERS} in this table.
				\end{tablenotes}
				\label{RV_movers_general_hot}
			\end{threeparttable}
		\end{adjustbox}
	\end{table}
\end{landscape}
}

Further analysis of \Cref{RV_movers_general_hot} highlights the significance of places in our analysis and underscores the role of a group of countries, including \textit{China}, among the high-SHAP n-grams. The term ‘Legal entity’ encompasses a variety of major offices, departments, commissions, and companies. The presence of the \textit{health organization} and \textit{CDC} (centres for disease control and prevention) is likely attributable to the COVID-19 pandemic. The next significant class, with a high number of n-grams, is labelled ‘Level’, encompassing a range of levels and changes, percentages, currency values, and specific terms like \textit{below}, \textit{above}, \textit{fall}, and \textit{under}, all referencing certain quantitative expectations. Moving to the subsequent groups, similar to stock-related news, the ‘Verb’ and ‘Numeric’ groups in general hot news underscore the relevance of these particular verbs and numbers as high-SHAP n-grams. The ‘Mixed’ group includes a variety of n-grams, such as \textit{SPAC}\footnote{Special-purpose acquisition company.}, \textit{Payroll-tax cut}, \textit{Airstrike}, \textit{Shutdown}, \textit{Coalitions}, \textit{Crisis}, \textit{Trade Speech}, \textit{Attorney General}, and \textit{Hearing}. Finally, the ‘Index’ and ‘Data’ groups underscore the importance of changes in the S\&P500 index and various financial and economic indicators such as \textit{inflation}, \textit{GDP}, and \textit{deficit} as key n-grams. A detailed analysis of all n-grams is beyond the scope of this study; however, we believe that most of these n-grams align with expectations.

These results are broadly consistent with the view that volatility reflects variation in the rate at which economically relevant information is incorporated into prices. The recurrent importance of stock-related n-grams classified as ‘Analyst opinion’, ‘Event’, and ‘Market’, which include recommendation changes, earnings-related events and transcripts, corporate filings and announcements, and market-opening or premarket references, accords with classic empirical evidence that trading volume is closely linked to the magnitude of price changes and return variability \citep{karpoff1987relation}, that volume accounts for a non-trivial share of conditional heteroskedasticity in daily stock returns by proxying for the intensity of information arrivals \citep{lamoureux1990heteroskedasticity}, and with no-arbitrage theory implying that price volatility scales with the rate of information flow \citep{ross1989information}. The prominence of ‘Analyst opinion’ terminology is similarly consistent with event-study evidence showing that changes in brokerage recommendations convey information that is rapidly incorporated into prices and generate sharp contemporaneous market reactions \citep{womack1996brokerage}. Finally, the salience of general hot-news n-grams classified as ‘Person’, ‘Place’, ‘Legal entity’, ‘Index’, and ‘Data’, tied to macroeconomic, monetary policy, and geopolitical developments, aligns with a large literature demonstrating that scheduled macroeconomic announcements and unanticipated policy actions generate sharp contemporaneous market reactions in volatility \citep{ederington1993markets,andersen2007real,bernanke2005explains}, and with text-based evidence that news-implied measures of economic and political risk co-move closely with implied volatility and broader measures of market uncertainty \citep{manela2017news,caldara2022measuring}.

Overall, the SHAP-based XAI analysis indicates that the NLP models assign importance to economically meaningful and interpretable information from news text. By attributing RV forecasts to specific n-grams, the results show that stock-related news is primarily associated with analyst opinions, earnings-related events, and firm-specific announcements, while general hot news is dominated by macroeconomic indicators, political developments, and policy communication. This clear separation is consistent with the distinct regime-dependent forecasting patterns of the two news information sets, although their relative performance across volatility states varies with the loss function. Importantly, the SHAP attributions provide transparency comparable to dictionary-based models and show that the forecasts can be related to identifiable and economically interpretable textual signals, thereby strengthening the interpretability of the proposed framework.\footnote{The results in \Cref{appendix_LM_Dicionary} show that the NLP models rely on a heterogeneous set of economically interpretable n-grams whose contributions to RV forecasts vary in sign and magnitude across stocks and news types. In contrast to dictionary-based models, which impose a fixed and uniform sentiment interpretation, the SHAP analysis reveals substantial cross-sectional and contextual heterogeneity in how identical words affect RV forecasts.}

\section{Conclusions} \label{conclusions_section}
This study develops and evaluates a news-driven RV forecasting framework that uses an embedding-based NLP framework to transform news into predictive signals. The core empirical design tests whether a text-only NLP forecaster can produce competitive one-day-ahead RV forecasts relative to established volatility-history benchmarks, and whether combining news-based signals with standard RV models yields incremental predictive value. We further compare general-purpose and specialised language representations and apply explainability analysis to attribute forecast variation to specific phrases and themes in the news.

Overall, the evidence indicates that news contains useful information for RV forecasting. A news-only NLP model contains predictive information, but generally does not outperform the strongest HAR-family benchmark over the full out-of-sample period according to MSE and QLIKE. Predictive content varies with coverage, with stock-related news typically more informative than general news. When combined with a strong RV benchmark, particularly for stock-related news, the news-based signal improves statistical performance and realised utility, supporting the interpretation that textual information is most useful as a complement to, rather than a substitute for, established econometric models. The explainability analysis makes it possible to trace RV forecasts back to the underlying news content and to distinguish clearly between stock-related news and general hot news. Stock-related news is mainly associated with analyst opinions and firm-level events and announcements, while general hot news is linked to broader macroeconomic and policy-related categories, including political actors, institutions, economic indicators, and market-wide developments.

Future research could integrate news text and numerical inputs, such as lagged RV, within a unified model to exploit complementarities across information sources and to improve robustness across regimes. It would also be valuable to assess whether news-based signals enhance cross-asset and bespoke-volatility frameworks \citep{bollerslev2018risk, patton2022bespoke}. Another direction is to account for potential changes in the semantic meaning of words over time, which are not explicitly modelled in this study due to computational limitations. Moreover, this line of research can be extended to related finance applications, including return forecasting and credit risk assessment. The development of broader and larger specialised language models for finance represents another promising avenue for future research.
\clearpage
\doublespacing
\appendix

\section{Appendix}
\label{Appendix}

\setcounter{table}{0}
\renewcommand{\thetable}{A\arabic{table}}

\setcounter{figure}{0}
\renewcommand{\thefigure}{A\arabic{figure}}
\setcounter{equation}{0}
\renewcommand{\theequation}{A\arabic{equation}}

\subsection{Data Cleaning Details} \label{Appendix_data_cleaning}
All duplicate news stories, as well as those without headlines and bodies, are removed as an initial filtering step. Given the heterogeneous nature of raw news data, extensive pre-processing of the textual content is required to eliminate redundant characters, sentences, and structural artefacts that do not convey semantic information. This cleaning process aims to standardise the text, reduce noise, and ensure that only meaningful content is retained for subsequent analysis. \Cref{table_textual_cleaning_rules} provides a structured overview of the textual data cleaning rules applied in this study. Each rule is implemented using a regular expression and may include multiple variations in order to capture differences in formatting and presentation across news sources; however, for clarity and brevity, only one representative variation of each rule is displayed. \emph{XX} denotes an arbitrary sequence of characters of variable length, used as a placeholder to illustrate the general form of the corresponding regular expression pattern.

The text cleaning procedures are grouped into five main categories: 1) Primary, 2) Begins with, 3) Ends with, 4) General, and 5) Final checks. The \emph{Primary} rules focus on fundamental transformations of the raw data, including extracting the body of news from extensible markup language (XML), removing XML-encoding characters (XMLENCOD), converting XML content into plain text through parsing, converting upper-case letters into lower-case, and removing embedded tables. These steps ensure that the core textual content is isolated and normalised before applying more specific pattern-based cleaning rules. This category forms the foundation of the cleaning pipeline and prepares the text for further refinement.

The remaining categories target recurring non-informative patterns commonly found in news articles. The \emph{Begins with} and \emph{Ends with} rules remove boilerplate segments that appear at the start or end of articles, such as copyright notices, contact information, source attributions, subscription prompts, and references to external content. The \emph{General} category addresses patterns that may occur anywhere within the text, including social media references, repeated disclaimers, attachments, and promotional content. Finally, the \emph{Final checks} perform a last-stage clean-up by removing links, email addresses, phone numbers, short news stories containing fewer than 25 characters, and leading or trailing spaces. All five categories of cleaning rules are applied separately to both news headlines and news bodies to ensure consistent pre-processing across different textual components.

Due to the importance of numerical information in accounting and finance, all numbers are preserved. This design choice is essential for maintaining sentence integrity, particularly in settings where token order conveys economically meaningful information rather than merely contributing to word counts. For example, removing numbers from the sentence `Over 540{,}000 apps wiped from Apple App Store in Q3 reaching lowest number in 7 years' yields `Over apps wiped from Apple App Store in Q reaching lowest number in years,' which materially alters the meaning and weakens the informational content.

\vspace{2cm}

    \begin{table}[htbp]
        \centering
        \caption{Textual Data Cleaning Rules}
        \label{table_textual_cleaning_rules}
        \resizebox{\textwidth}{!}{%
            \begin{tabular}{ll}
                \toprule
                \multicolumn{2}{c}{\textbf{Primary}} \\
                \midrule
                \multicolumn{1}{l|}{Extracting body of news from XML} & Removing XML-Encoding Characters (XMLENCOD) \\
                \multicolumn{1}{l|}{Converting XML to text (parsing)} & Converting uppercase letters to lowercase letters \\
                \multicolumn{1}{l|}{Removing tables} & \\
                \midrule
                \multicolumn{2}{c}{\textbf{Begins with}} \\
                \midrule
                \multicolumn{1}{l|}{(END) XX} & (email|e-mail): XX \\
                \multicolumn{1}{l|}{for (more|further) (information|from marketwatch), please visit: XX} & (phone|fax|contact|dgap-ad-hoc|dgap-news): XX \\
                \multicolumn{1}{l|}{(EMAIL; @XX)} & image available: XX \\
                \multicolumn{1}{l|}{copyright XXXX, XX URL} & source: XX \\
                \multicolumn{1}{l|}{(more to follow) XX} & to read more, visit: XX \\
                \multicolumn{1}{l|}{end of (message|corporate news) XX} & (view source|view original content) (with|on) XX \\
                \multicolumn{1}{l|}{source: XX URL} & (investor relations|investor contact) XX \\
                \multicolumn{1}{l|}{XX contributed to this article XX} & like us on XX \\
                \multicolumn{1}{l|}{view source version on XX} & (copyright|(c)|©) XX \\
                \multicolumn{1}{l|}{(=————————————————————)} & XX can be found at URL XX \\
                \multicolumn{1}{l|}{view original content with multimedia XX} & by dow jones newswires XX \\
                \multicolumn{1}{l|}{readers can alert XX} & (write to|follow) XX at EMAIL \\
                \multicolumn{1}{l|}{view original content: XX} & (phone|tel|telephone|mobile|contact|inquiries|comment): XX \\
                \multicolumn{1}{l|}{media inquiries: XX} & (contact information|media contact|contact client services|internet) XX \\
                \multicolumn{1}{l|}{readers: send feedback to XX} & click here to subscribe to XX \\
                \multicolumn{1}{l|}{follow us on XX} & to learn more about XX \\
                \multicolumn{1}{l|}{contact(s): XX} & (website|web site): URL XX \\
                \multicolumn{1}{l|}{please refer to URL XX} & contact us in XX \\
                \multicolumn{1}{l|}{find out more at URL XX} & to receive news releases by (e-mail|email) XX \\
                \multicolumn{1}{l|}{XX enquiries: XX} & full story at XX \\
                \multicolumn{1}{l|}{-by XX, dow jones newswires XX} & \\
                \midrule
                \multicolumn{2}{c}{\textbf{Ends with}} \\
                \midrule
                \multicolumn{1}{l|}{(more to follow)} & view original content XX: \\
                \multicolumn{1}{l|}{(fax|tel|contact|dgap-ad-hoc|dgap-news):} & (contacts|web site): \\
                \multicolumn{1}{l|}{ratings actions from baystreet:} & (=|- -|\_|·|-) \\
                \multicolumn{1}{l|}{cannot parse story} & for notes, kindly refer \\
                \multicolumn{1}{l|}{lipper indexes: to subscribe to} & following is the related link: \\
                \multicolumn{1}{l|}{for full details, please click on} & \\
                \midrule
                \multicolumn{2}{c}{\textbf{General}} \\
                \midrule
                \multicolumn{1}{l|}{(linkedin|facebook|fb): XX} & (URL (and|\&) XX) \\
                \multicolumn{1}{l|}{(twitter|ig): XX} & (EMAIL (and|\&) EMAIL) \\
                \multicolumn{1}{l|}{(attachment|attachments): XX} & this information was brought to you by XX \\
                \multicolumn{1}{l|}{please visit XX} & write to EMAIL \\
                \multicolumn{1}{l|}{follow us on XX} & to receive our XX URL \\
                \multicolumn{1}{l|}{All rights reserved} & more at, XX URL \\
                \midrule
                \multicolumn{2}{c}{\textbf{Final checks}} \\
                \midrule
                \multicolumn{1}{l|}{Removing links and emails} & Removing news shorter than 25 characters \\
                \multicolumn{1}{l|}{Removing both the leading and the trailing space(s)} & Removing phone numbers \\
                \bottomrule
            \end{tabular}%
        }
    \end{table}
\clearpage

\subsection{Word Embedding Development Details} \label{Appendix_word_embedding_details}
After cleaning the raw news data following the steps described in \Cref{Appendix_data_cleaning}, each headline and news body is tokenised into an ordered sequence $(w_1,\ldots,w_T)$, as defined in \Cref{eq:token_sequence}. Each token takes a value in the vocabulary $\mathcal{V}$ defined in \Cref{eq:vocabulary}, and is ultimately mapped to a vector representation $\mathbf{e}_w \in \mathbb{R}^d$ as introduced in \Cref{eq:word_vector}. Many economic concepts are expressed as multi-word phrases rather than individual words, and treating their components separately may dilute their economic meaning. To address this issue, frequently occurring two-word expressions are identified using the phrase-detection procedure of \citet{mikolov2013distributed}. This method evaluates adjacent word pairs based on their empirical co-occurrence frequency and retains sufficiently strong phrase associations. In the empirical implementation, a threshold value of ten is applied to the phrase score so that only strongly associated word pairs are retained. These bigrams are then merged into a single token by joining the two words with an underscore (e.g., \texttt{financial\_statement}), allowing the word embedding models described in \Cref{word2vec_subsection} and \Cref{fasttext_subsection} to learn a unified vector representation for economically meaningful concepts.

To reduce estimation noise and improve the stability of the embedding matrix $E$ defined in \Cref{eq:embedding_matrix}, tokens that appear fewer than five times in the entire corpus are removed. Very infrequent tokens provide insufficient contextual information for reliable vector estimation and may introduce unnecessary variance. Following pre-processing, Word2Vec and FastText models are estimated using a context window of size five, as defined in \Cref{word2vec_subsection}. This choice implies that the representation of each token is learned from the five tokens preceding and the five tokens following it in the text, so that local co-occurrence patterns drive the estimation of the embedding matrix $E$.

Model estimation relies on negative sampling, which is introduced conceptually in \Cref{word2vec_subsection} as an approximation to the full softmax likelihood in \Cref{eq:word2vec_prob}. Rather than comparing each observed token–context pair with the entire vocabulary, the model contrasts it with a small number of artificially generated noise pairs. In this study, five negative samples are used for each observed pair. Negative samples are drawn from a smoothed unigram distribution,
\begin{equation}
P_n(w)\propto C(w)^{0.75},
\end{equation}
following \citet{mikolov2013distributed}, where $P_n(w)$ denotes the probability of drawing token $w$ as a negative sample and $C(w)$ is the total number of occurrences of token $w$ in the training corpus. The exponent of $0.75$ smooths the empirical frequency distribution by down weighting extremely common tokens, such as function words, while still assigning relatively higher probability to frequent and economically relevant terms, thereby improving estimation stability and representation quality.

All models are trained for five epochs, meaning that the entire corpus is processed five times during estimation. The learning rate is initialised at $0.025$ and decays linearly to a minimum value of $0.0001$ to ensure numerical stability and convergence. Each token is represented by a 300-dimensional vector, corresponding to the word embedding dimension $d$ in \Cref{eq:word_vector}, which is a commonly adopted standard in general-purpose word embeddings, including the off-the-shelf Word2Vec and FastText word embeddings used in this study.

\clearpage

\subsection{Gold-Standard Financial Benchmark Details} \label{Appendix_gold_benchmak}
Each analogy follows the structure A is to B as C is to X, where the objective is to recover X by preserving the same financial relationship. For example, the analogy \textit{AAPL} is to \textit{Cupertino} as \textit{MSFT} is to \textit{Redmond} tests whether a word embedding links a firm’s ticker to its headquarters city; \textit{Apple} is to \textit{AAPL} as \textit{Amazon} is to \textit{AMZN} captures the mapping between company names and tickers; \textit{Amazon} is to \textit{1994} as \textit{Google} is to \textit{1998} relates firms to their incorporation years; \textit{Microsoft} is to \textit{NASDAQ} as \textit{IBM} is to \textit{NYSE} links firms to their primary stock exchanges; \textit{Tesla} is to \textit{California} as \textit{Boeing} is to \textit{Illinois} reflects US headquarters states; \textit{HSBC} is to \textit{UK} as \textit{JPMorgan} is to \textit{US} captures countries of headquarters; and \textit{Toyota} is to \textit{Japan} as \textit{Alibaba} is to \textit{China} extends this relationship to a global setting.

The benchmark is constructed in several systematic steps. First, firm-level data are collected separately for the United States, the United Kingdom, China, and Japan. Firms are filtered to retain only publicly listed companies with complete and consistent information on key attributes such as name, ticker, headquarters location, exchange, country, and incorporation year. Company names are standardised by removing legal suffixes and formatting inconsistencies to ensure that each entity is represented by a single, unambiguous token. To avoid confounding effects, only unigram representations are retained, and firms with duplicate or ambiguous identifiers are excluded.

Second, firms are classified by size using the Orbis size classification, and the benchmark focuses exclusively on very large companies to ensure reliable coverage in the underlying text corpora. The benchmark includes firms from four countries: the United States, the United Kingdom, China, and Japan, representing major equity markets. For US-only benchmarks, the top 20 very large US companies are selected and used across all US benchmarks, covering relationships involving company names, tickers, headquarters cities, incorporation years, stock exchanges, and US states. To extend the evaluation beyond a single country while maintaining sufficient firm coverage, a cross-country benchmark combines the top 10 very large firms from the United States and the United Kingdom. Finally, to assess whether word embeddings capture country-level relationships in a broader international setting, a global benchmark is constructed using the top 5 very large firms from each of the four countries.

Third, each benchmark group is defined by a specific financial relationship: ticker to headquarters city, company name to ticker, company name to incorporation year, company name to stock exchange, company name to US headquarters state, company name to country of headquarters in a US--UK setting, and company name to country of headquarters in a global setting. Within each group, all ordered permutations of the selected firms are generated to form analogy questions. This means that every firm is systematically paired with every other firm in the same group, ensuring that each financial relationship is tested exhaustively rather than through a small number of hand-picked examples. For each permutation, the known attribute of one firm is used to infer the corresponding attribute of another firm under the same relationship structure. This procedure yields 380 unique analogies per group and 2,660 analogies in total. For evaluation, a prediction is considered correct if the true answer appears among the top five candidates returned by the word embedding model. Analogy questions that cannot be evaluated by a given embedding, including because a required token is unavailable, are retained and counted as incorrect, so each embedding is evaluated using the same 380 questions per group. We found five to be a fair and balanced choice, as a smaller cutoff would make the benchmark more restrictive, while a larger cutoff would make it more permissive.

\clearpage
\subsection{RV Cleaning Procedures} \label{appendix_RV_cleaning}
We implemented the relevant data-cleaning procedures proposed by \citet{barndorff2009realized}, adapting them to a LOB setting. The procedures applied below use the following notation: $P$ denotes the full dataset, $Q$ refers exclusively to quote data, and $T$ denotes trade data only.
\begin{itemize}
\item \textbf{P2:} Delete entries with a bid, ask or transaction price equal to zero.
\item \textbf{T4:} Delete entries with prices that are above the `ask' plus the bid-ask spread, or below the `bid' minus the bid-ask spread.
\item \textbf{Q1:} When multiple quotes share the same timestamp, they are replaced by a single entry using the median bid price, median ask price, the sum of all volumes, and the last snapshot of the LOB is selected as the LOB associated with the merged message data.
For messages with different directions (buy or sell), the message data and the LOB with the same direction are grouped according to the buy side or sell side. The procedure mentioned above is then applied to the message data and the last snapshot of the LOB of the group. 
\item \textbf{Q2}: Delete entries for which the spread is negative.
\item \textbf{Q3}: Delete entries for which the spread is more than 50 times the median spread on that day.
\item \textbf{Q4}: Delete entries for which the mid-quote deviated by more than 10 mean absolute deviations from a rolling centred median (excluding the observation under consideration) of 50 observations (25 observations before and 25 after).
\end{itemize}
\Cref{data_cleaning_table} presents the summary statistics of the data cleaning process. This table indicates that approximately 40\% of the samples (ticks) were discarded in the cleaning phase. Notably, T4 was responsible for the removal of a significant portion of the data, amounting to nearly 89.33\% (i.e., 35.78/40.05) of the total data excluded. The filtering rules are applied uniformly across all stocks and all time periods. While filtering reduces the number of observations, it is designed to remove stale or erroneous records rather than information-bearing trades. To assess the impact of these cleaning procedures on the calculated RVs, we further analysed the LOB data without implementing the cleaning procedures. The comparative statistics and the correlation between RVs calculated with and without these data cleaning procedures are detailed in \Cref{descriptive_stat_table_wo_cleaning}. It is evident that the descriptive statistics without the application of cleaning procedures closely align with those presented in \Cref{descriptive_stat_table}. In all instances, the correlation remains significantly high, thereby affirming the robustness of the calculated RVs.

\clearpage

\begin{table}
	\centering
	\begin{adjustbox}{max width=0.9\textwidth}
		\begin{threeparttable}
			
			\caption{Data Cleaning Summary Statistics}
			
			\begin{tabular}{cccccccccc} \toprule
				
				{Name} & {Ticker} & {Sample Size} & {Removed (\%)}  & {P2 (\%)} & {T4 (\%)} & {Q1 (\%)} & {Q2 (\%)} & {Q3 (\%)} & {Q4 (\%)} \\ \midrule
                {Apple} & AAPL & 4174971328 & 34.22 & 0.00 & 29.88 & 4.34 & 0.00 & 0.00 & 0.00 \\
                {Microsoft} & MSFT & 3827824574 & 35.88 & 0.01 & 30.18 & 5.69 & 0.00 & 0.00 & 0.00 \\
                {Intel} &  INTC & 2807965330 & 38.59 & 0.01 & 31.79 & 6.78 & 0.01 & 0.00 & 0.01 \\
                {Comcast} &  CMCSA & 2390133817 & 45.18 & 0.01 & 39.59 & 5.58 & 0.00 & 0.00 & 0.01 \\
                {Qualcomm} & QCOM & 2086295132 & 41.46 & 0.00 & 36.46 & 4.98 & 0.00 & 0.00 & 0.01 \\
                {Cisco Systems} &  CSCO & 2296179428 & 40.46 & 0.01 & 33.50 & 6.94 & 0.00 & 0.00 & 0.01 \\
                {eBay} & EBAY & 1683001942 & 40.73 & 0.01 & 35.68 & 5.03 & 0.00 & 0.00 & 0.01 \\
                {Gilead Sciences} & GILD & 1404574567 & 41.68 & 0.00 & 38.41 & 3.25 & 0.00 & 0.00 & 0.01 \\
                {Texas Instruments} & TXN & 1485049597 & 39.45 & 0.00 & 35.14 & 4.29 & 0.00 & 0.00 & 0.01 \\
                {Amazon.com} & AMZN & 1201210867 & 23.06 & 0.00 & 19.89 & 3.15 & 0.00 & 0.00 & 0.02 \\
                {Starbucks} & SBUX & 1564221129 & 44.04 & 0.01 & 39.95 & 4.07 & 0.00 & 0.00 & 0.01 \\
                {Nvidia} &  NVDA & 1548447223 & 35.47 & 0.01 & 30.40 & 5.05 & 0.00 & 0.00 & 0.01 \\
                {Micron Technology} & MU & 2110482619 & 35.99 & 0.00 & 31.11 & 4.86 & 0.00 & 0.00 & 0.01 \\
                {Applied Materials} & AMAT & 1616466522 & 39.70 & 0.01 & 34.41 & 5.27 & 0.00 & 0.00 & 0.01 \\
                {NetApp} & NTAP & 1015914054 & 44.99 & 0.01 & 41.14 & 3.82 & 0.00 & 0.00 & 0.02 \\
                {Adobe} & ADBE & 1083392595 & 37.76 & 0.01 & 34.35 & 3.39 & 0.00 & 0.00 & 0.02 \\
                {Xilinx} & XLNX & 1172584895 & 40.24 & 0.01 & 36.97 & 3.26 & 0.00 & 0.00 & 0.02 \\
                {Amgen} & AMGN & 863464001 & 38.62 & 0.01 & 34.73 & 3.86 & 0.00 & 0.00 & 0.02 \\
                {Vodafone Group} & VOD & 1012861232 & 47.20 & 0.01 & 44.23 & 2.95 & 0.00 & 0.00 & 0.02 \\
                {Cognizant} & CTSH & 928987253 & 46.22 & 0.01 & 43.28 & 2.91 & 0.00 & 0.00 & 0.02 \\
                {KLA Corporation} & KLAC & 783931409 & 42.63 & 0.01 & 39.83 & 2.77 & 0.00 & 0.00 & 0.02 \\
                {Paccar} & PCAR & 775954122 & 45.74 & 0.01 & 42.98 & 2.73 & 0.00 & 0.00 & 0.03 \\
                {Autodesk} & ADSK & 803552017 & 41.73 & 0.01 & 38.96 & 2.74 & 0.00 & 0.00 & 0.02 \\

				\midrule
				{Average}  &  {}  &  & 40.05 & 0.01 & 35.78 & 4.25 & 0.00 & 0.00 & 0.01 \\
				\bottomrule
			\end{tabular}
			\begin{tablenotes}[para,flushleft]
				\footnotesize \item \textit{Notes:} 
				\textbf{P2:} Delete entries with a bid, ask or transaction price equal to zero, \textbf{T4:} Delete entries with prices that are above the `ask' plus the bid-ask spread, or below the `bid' minus the bid-ask spread, \textbf{Q1:} When multiple quotes share the same timestamp, they are replaced by a single entry using the median bid price, median ask price, the sum of all volumes, and the last snapshot of the LOB is selected as the LOB associated with the merged message data. For messages with different directions (buy or sell), the message data and the LOB with the same direction are grouped according to the buy side or sell side. The procedure mentioned above is then applied to the message data, and the last snapshot of the LOB of the group, \textbf{Q2}: Delete entries for which the spread is negative, \textbf{Q3}: Delete entries for which the spread is more than 50 times the median spread on that day, \textbf{Q4}: Delete entries for which the mid-quote deviated by more than 10 mean absolute deviations from a rolling centred median (excluding the observation under consideration) of 50 observations (25 observations before and 25 after). 
			\end{tablenotes}
			\label{data_cleaning_table}
		\end{threeparttable}
	\end{adjustbox}
\end{table}

\begin{table}[h]
	\centering
	\begin{adjustbox}{max width=0.9\textwidth}
		\begin{threeparttable}
			\caption{RV Descriptive Statistics (Without Cleaning)}
			\begin{tabular}{ccccccccccc} \toprule
				{Ticker} & {Min} & {Max} & {\nth{1} Quartile} & {Median} & {\nth{3} Quartile} & {Mean} & {STD} & {Kurtosis} & {Skewness} &{Correlation\tnote{a}}\\ \midrule

                {AAPL}  & 0.101 & 229.529 & 0.898 & 1.737 & 3.702 & 4.623 & 12.579 & 110.333 & 9.093 & 0.9999 \\
                {MSFT}  & 0.096 & 216.486 & 0.828 & 1.458 & 2.810 & 3.240 & 8.119 & 194.967 & 11.294 & 0.9997 \\
                {INTC}  & 0.030 & 318.118 & 1.099 & 1.876 & 3.592 & 4.300 & 11.615 & 295.828 & 14.000 & 0.9999 \\
                {CMCSA}  & 0.006 & 237.387 & 0.913 & 1.631 & 3.344 & 3.833 & 9.773 & 190.221 & 11.459 & 0.9994 \\
				{QCOM}  & 0.123 & 368.449 & 1.025 & 1.980 & 4.140 & 5.076 & 15.363 & 197.106 & 12.012 & 0.9999 \\
				{CSCO}  & 0.038 & 343.946 & 0.884 & 1.564 & 3.031 & 4.117 & 13.170 & 213.266 & 12.287 & 0.9999 \\
				{EBAY}  & 0.215 & 259.723 & 1.328 & 2.263 & 4.361 & 5.111 & 12.745 & 142.560 & 10.028 & 0.9974 \\
				{GILD}  & 0.063 & 261.664 & 1.170 & 1.895 & 3.375 & 4.312 & 12.933 & 184.066 & 12.094 & 0.9999 \\
				{TXN}   & 0.183 & 289.765 & 1.046 & 1.895 & 3.713 & 4.006 & 9.928 & 310.664 & 14.275 & 0.9986 \\
				{AMZN}  & 0.066 & 551.566 & 1.307 & 2.342 & 4.833 & 6.203 & 19.355 & 246.159 & 12.802 & 0.9998 \\
				{SBUX}  & 0.048 & 265.554 & 0.864 & 1.594 & 3.441 & 4.209 & 11.227 & 161.691 & 10.615 & 0.9997 \\        
                {NVDA}  & 0.159 & 1104.483 & 2.280 & 4.342 & 9.098 & 9.760 & 30.112 & 586.751 & 20.055 & 1.0000 \\
                {MU}    & 0.288 & 484.388 & 3.575 & 6.281 & 11.964 & 12.821 & 25.726 & 89.478 & 7.966 & 0.9990 \\
                {AMAT}  & 0.312 & 529.508 & 1.773 & 3.031 & 5.730 & 6.014 & 14.615 & 526.901 & 18.237 & 0.9999 \\
                {NTAP}  & 0.114 & 463.545 & 1.508 & 2.606 & 5.180 & 6.301 & 18.020 & 201.869 & 11.942 & 0.9998 \\
                {ADBE}  & 0.120 & 575.498 & 1.096 & 2.001 & 3.874 & 4.810 & 14.784 & 661.002 & 20.309 & 0.9998 \\
                {XLNX}  & 0.231 & 265.372 & 1.300 & 2.374 & 4.791 & 5.022 & 11.950 & 193.951 & 11.739 & 0.9990 \\
                {AMGN}  & 0.039 & 212.485 & 0.962 & 1.580 & 2.860 & 3.311 & 9.225 & 202.622 & 12.391 & 0.9995 \\
                {VOD}   & 0.043 & 217.091 & 0.684 & 1.334 & 3.081 & 3.693 & 9.500 & 149.567 & 10.122 & 0.9998 \\
                {CTSH}  & 0.186 & 493.255 & 0.979 & 1.743 & 4.085 & 5.218 & 15.843 & 339.283 & 14.667 & 0.9997 \\
                {KLAC}  & 0.149 & 499.806 & 1.451 & 2.701 & 5.412 & 5.820 & 16.308 & 384.071 & 16.516 & 0.9933 \\
                {PCAR}  & 0.029 & 389.021 & 1.157 & 2.172 & 4.657 & 5.137 & 12.123 & 309.547 & 12.928 & 0.9998 \\
                {ADSK}  & 0.268 & 696.615 & 1.642 & 2.770 & 5.176 & 6.660 & 22.503 & 389.706 & 16.604 & 0.9999 \\
				\bottomrule
			\end{tabular}
			\begin{tablenotes}[para,flushleft]
				\footnotesize
				\textit{Notes:}\item[a] Correlation between two sets of calculated RVs with and without the implementation of the cleaning procedures. The RV descriptive statistics computed on the cleaned dataset are presented in \Cref{descriptive_stat_table}.
			\end{tablenotes}
			\label{descriptive_stat_table_wo_cleaning}
		\end{threeparttable}
	\end{adjustbox}
\end{table}

\clearpage
\subsection{HAR-Family of Models: In-Sample and Out-of-Sample Results} \label{appendix_HAR_results}
All HAR-family of models are estimated following the rolling-window forecasting design described in \Cref{review_rv_subsection}. The in-sample period spans from 27 July 2007 to 11 September 2015 and contains 2{,}046 daily observations for each stock, while the out-of-sample period extends from 14 September 2015 to 27 January 2022 and contains 1{,}604 forecast observations. For each of the 23 stocks, model parameters are estimated by ordinary least squares (OLS) using a fixed-length rolling window of 2{,}046 observations. At each out-of-sample date, the estimation window is advanced by one day, the model is re-estimated using the most recent observations, and a one-step-ahead forecast of RV is generated. This procedure yields 1{,}604 rolling estimations per stock and a total of 36{,}892 estimations per model across the full cross-section. Also, following \citet{bollerslev2016exploiting}, when the forecasted RV exceeds (falls below) the maximum (minimum) value observed in the estimation sample, it is replaced with the sample mean of RV from the estimation period.

The rolling-window estimation diagnostics in \Cref{CHARx_1lag_model_coefs} summarise coefficient estimates and within-estimation-window performance measures obtained from 36{,}892 rolling regressions across assets and time. The table shows that coefficient estimates largely conform to expected volatility dynamics. The AR specification delivers the weakest fit, whereas introducing heterogeneous horizons via HAR leads to a clear improvement in adjusted $R^{2}$ and loss measures, with further extensions providing additional but uneven gains. Among all specifications, the CHAR model stands out as the strongest overall in-sample specification, achieving the highest average and median adjusted $R^{2}$, the lowest average and median MSE, and the lowest average QLIKE; HAR-J records a slightly lower median QLIKE. The out-of-sample forecasting results reported in \Cref{har_outofsample_table_base} provide a comprehensive comparison of the HAR-family of models using both loss-based metrics and the RC across the full out-of-sample period and across different volatility regimes. The RC statistics report the percentage of tickers for which the Reality Check does not identify a competing specification with significantly lower expected loss, where each model is evaluated relative to all other HAR-family of models under the chosen loss function at the 5\% and 10\% significance levels. Over the full out-of-sample period, clear performance differences emerge. The AR specification is among the weakest performers under both MSE and QLIKE, while HAR leads to sizeable improvements. Further extensions yield additional gains, but the ranking is consistent across average and median losses: the CHAR model delivers the lowest average and median MSE and QLIKE among all competitors. This loss ranking is accompanied by high RC non-rejection rates, where CHAR achieves rates of $(100\%,100\%)$ across tickers at both the 5\% and 10\% significance levels under both loss functions, indicating that the RC does not identify a significantly better alternative specification. The dominance of CHAR is even more pronounced on normal volatility days, where it attains substantially lower losses than all other models and very high RC non-rejection rates. During high volatility days, forecast losses increase sharply for all models and relative performance becomes more heterogeneous, with HAR-J exhibiting slightly lower average losses, particularly under MSE. Nevertheless, CHAR remains among the top-performing models and achieves among the highest RC non-rejection rates under MSE and the highest rates under QLIKE.

One natural question that arises in the empirical analysis concerns the sensitivity of the forecasting results to the choice of the estimation window size. To examine how changes in the window length affect out-of-sample predictive performance, we conduct an additional robustness exercise in which the estimation sample is reduced to 1{,}023 observations, while the out-of-sample evaluation period is held fixed. The resulting out-of-sample forecasts are reported in \Cref{har_outofsample_table_base_short}. Compared with the baseline results based on the larger estimation window, shortening the estimation window produces mixed changes in forecast losses. Under MSE, most specifications deteriorate, whereas under QLIKE several specifications improve and others deteriorate. Importantly, however, the broad relative ranking of the models is largely preserved. CHAR remains the strongest overall specification, attaining the lowest average MSE and QLIKE and the lowest median MSE, while HAR-J records a slightly lower median QLIKE. Accordingly, CHAR remains an appropriate common HAR-family reference model under the shorter estimation window. Due to computational constraints, this robustness test is not extended to the remainder of the analyses in this study. Nevertheless, at least for the benchmark models, the results indicate that the main findings are robust to changes in the length of the in-sample estimation window.

\begin{table}
	\centering
	\begin{adjustbox}{max width=400pt}  
		\begin{threeparttable}	

			\scriptsize
			\centering
		\captionsetup{format=myformat}
		\caption{Parameter Estimates of HAR-family of Models}	
			\begin{tabular}{ccccccccc} \toprule
				\label{CHARx_1lag_model_coefs}	
				{}  & {AR} & {HAR} & {HAR-J} & {CHAR} & {SHAR} & {ARQ} & {HARQ}  & {HARQ-F} \\ \midrule
                \multirow{4}{*}{$\beta_{0}$}
                & 3.6436\tnote{a} & 1.8528 & 1.4401 & 1.2273 & 1.8444 & 2.8734 & 1.6127 & 0.6807 \\
                & 3.6436\tnote{b} & 1.8530 & 1.4408 & 1.2479 & 1.8447 & 2.8734 & 1.6135 & 0.8452 \\ 
                & (1.6305)\tnote{c} & (0.9216) & (0.6446) & (0.6357) & (0.9213) & (1.3585) & (0.8758) & (0.8382) \\ 
                & (100, 0)\tnote{d} & (100, 0) & (99.9, 0.1)  & (98.7, 1.3)  & (99.9, 0.1)  &  (100, 0)  & (99.9, 0.1)  & (77.2, 22.8) \\[0.2cm]

				\multirow{4}{*}{$\beta_{1}$}
				{} & 0.1428 & 0.0394 & 0.6544 &       &       & 0.3653 & 0.2136 & 0.1695  \\
                {} & 0.1428 & 0.0421 & 0.6544 &       &       & 0.3653 & 0.2136 & 0.1702 \\
				{} &  (0.1090) & (0.0382) & (0.2743) & {} & {} & (0.2028) & (0.1093) & (0.1193) \\ 
				{} &  (100, 0) & (93.8, 6.2) & (100, 0) & {} & {} & (100, 0) & (100, 0) & (99.3, 0.7) \\[0.2cm]

                \multirow{4}{*}{$\beta^+_{1}$}
				{} &  &       &       &       & 0.0625 &       &       &  \\
                {} &  &       &       &       & 0.0638 &       &       &  \\
				{} & {} & {} & {} & {} & (0.0942) & {} & {} & {} \\
				{} & {} & {} & {} & {} & (92.8, 7.2) & {} & {} & {} \\[0.2cm]

                \multirow{4}{*}{$\beta^-_{1}$}
				{} & &       &       &       & 0.0346 &       &       & \\
                {} & &       &       &       & 0.0397 &       &       & \\
				{} &  {} & {} & {} & {} & (0.0390) & {} & {} & {} \\ 
				{} & {} & {} & {} & {} & (88.4, 11.6) & {} & {} & {} \\[0.2cm]

				\multirow{4}{*}{$\beta^w$}
				{} & & 0.1471 & 0.0523 &       & 0.1449 &       & 0.1123 & 0.2076\\
                {} & & 0.1642 & 0.0925 &       & 0.1616 &       & 0.1341 & 0.2197\\
				{} &  {} & (0.1805) & (0.1305) & {} & (0.1774) & {} & (0.1568) & (0.1731) \\ 
				{} & {} & (77.2, 22.8) & (60.2, 39.8) & {} & (77.0, 23.0) & {} & (72.6, 27.4) & (89.6, 10.4) \\[0.2cm]

				\multirow{4}{*}{$\beta^m$}
				{} &  & 0.3566 & 0.2510 &       & 0.3527 &       & 0.3090 & 0.5367 \\
                {} &  & 0.3566 & 0.2529 &       & 0.3530 &       & 0.3097 & 0.5368 \\
				{} &  {} & (0.1754) & (0.1562) & {} & (0.1767) & {} & (0.1675) & (0.1956) \\ 
				{} & {} & (100, 0) & (97.1, 2.9) & {} & (98.7, 1.3) & {} & (98.1, 1.9) & (100, 0) \\[0.2cm]

				\multirow{4}{*}{$\beta_{jump}$}
				{} & &       & -0.7353 &       &       &       &       & \\
                {} & &       & 0.7353 &       &       &       &       & \\
				{} &  {} & {} & (0.2840) & {} & {} & {} & {} & {} \\ 
				{} & {} & {} & (0, 100) & {} & {} & {} & {} & {} \\[0.2cm]

				\multirow{4}{*}{$\beta_{BPV}^{d}$}
				{} & &       &       & 0.4136 &       &       &       & \\
                {} & &       &       & 0.4161 &       &       &       & \\
				{} &  {} & {} & {} & (0.2526) & {} & {} & {} & {} \\ 
				{} & {} & {} & {} & (98.3, 1.7) & {} & {} & {} & {} \\[0.2cm]

				\multirow{4}{*}{$\beta^w_{BPV}$}
				{} & &       &       & 0.3434 &       &       &       & \\
                {} & &       &       & 0.3697 &       &       &       & \\
				{} &  {} & {} & {} & (0.2926) & {} & {} & {} & {} \\ 
				{} & {} & {} & {} & (91.2, 8.8) & {} & {} & {} & {} \\[0.2cm]

				\multirow{4}{*}{$\beta^m_{BPV}$}
				{} & &       &       & 0.4336 &       &       &       & \\
                {} & &       &       & 0.4433 &       &       &       & \\
				{} &  {} & {} & {} & (0.3329) & {} & {} & {} & {} \\ 
				{} & {} & {} & {} & (92.5, 7.5) & {} & {} & {} & {} \\[0.2cm] 

				\multirow{4}{*}{$\beta_{1Q}$}
				{} & &       &       &       &       & -0.0004 & -0.0003 & -0.0002 \\
                {} & &       &       &       &       & 0.0004 & 0.0003 & 0.0002 \\
				{} &  {} & {} & {} & {} & {} & (0.0004) & (0.0003) & (0.0003) \\ 
				{} & {} & {} & {} & {} & {} & (0.4, 99.6) & (0.6, 99.4) & (2.0, 98.0) \\[0.2cm]	

				\multirow{4}{*}{$\beta^w_{Q}$}
				{} & &       &       &       &       &       &       & -0.0008 \\
                {} & &       &       &       &       &       &       & 0.0011 \\
				{} &  {} & {} & {} & {} & {} & {} & {} & (0.0013) \\ 	
				{} & {} & {} & {} & {} & {} & {} & {} & (22.0, 78.0) \\[0.2cm]

				\multirow{4}{*}{$\beta^m_{Q}$}
				{} & &       &       &       &       &       &       & -0.0064 \\
                {} & &       &       &       &       &       &       & 0.0068 \\
				{} &  {} & {} & {} & {} & {} & {} & {} & (0.0080) \\ 
				{} & {} & {} & {} & {} & {} & {} & {} & (7.4, 92.6) \\[0.2cm]

				\hline\hline \rule{0pt}{2.7ex}
				$Adj. \ R^{2} (avg)$ & 0.0318 & 0.0721 & 0.0984 & 0.1000 & 0.0731 & 0.0583 & 0.0818 & 0.0871 \\

                $Adj. \ R^{2} (med)$ & 0.0104 & 0.0404 & 0.0673 & 0.0692 & 0.0406 & 0.0406 & 0.0505 & 0.0552 \\
				
				$MSE \ (avg)$ & 173.3761 & 165.6939 & 161.0308 & 160.7043 & 165.4823 & 168.4723 & 163.9610 & 162.7886\\ 
				$MSE \ (med)$ & 127.6872 & 124.0916 & 120.5695 & 119.9260 & 124.0427 & 125.2517 & 121.4500 & 120.8497 \\
				
				$QLIKE \ (avg)$ & 0.6139 & 0.5619 & 0.5177 & 0.5167 & 0.5615 & 0.5773 & 0.5481 & 0.5599 \\ 
				$QLIKE \ (med)$ & 0.6223 & 0.5670 & 0.5187 & 0.5236 & 0.5656 & 0.5813 & 0.5452 & 0.5737 \\ \bottomrule

			\end{tabular}
			\begin{tablenotes}[para,flushleft]
				\footnotesize
				\item \textit{Notes:} For each parameter in each model, the $\text{mean of coefficients}^{[a]}$, the $\text{mean of absolute coefficients}^{[b]}$, the $\text{standard deviation}^{[c]}$, and the proportions of positive and negative $\text{coefficients}^{[d]}$ are computed using 36,892 daily estimations (1,604 days $\times$ 23 stocks). \item `HAR-J' is HAR with jump, `CHAR' stands for continuous HAR, and `SHAR' denotes semivariance HAR. `ARQ' and `HARQ' are quarticity-based specifications following \citet{bollerslev2016exploiting}, while `HARQ-F' denotes the further HARQ extension considered in this study. $\beta_{0}$ is the constant term, $\beta_{1}$ is the coefficient of the first lag of RV, $\beta^+_{1}$ and $\beta^-_{1}$ are the coefficients of the first lag of RV for the positive and negative returns, $\beta^w$ and $\beta^m$ are the coefficients of the daily average of RV over the last week and last month, $\beta_{jump}$ is the coefficient of the jump term of `HAR-J' model, $\beta_{BPV}^{d}$, $\beta^w_{BPV}$, and $\beta^m_{BPV}$ are the coefficients of continuous terms of the `CHAR' model, and $\beta_{1Q}$ denotes the coefficient on the interaction $\sqrt{RQ_t}RV_t$, while $\beta^w_{Q}$ and $\beta^m_{Q}$ denote the coefficients on the weekly and monthly averages of realised quarticity, respectively. For each stock, $MSE$ and $QLIKE$ are initially calculated as the average (or median) within-estimation-window errors. These values are subsequently aggregated across all 23 stocks, using either the overall average (or median).
     
			\end{tablenotes}
		\end{threeparttable}	
	\end{adjustbox}
\end{table}

\clearpage

\afterpage{%
	\clearpage
	\thispagestyle{empty}
	\atxy{\dimexpr\paperwidth-0.45in}{.5\paperheight}{\rotatebox[origin=center]{90}{\thepage}}
\begin{landscape}
\vspace*{0.01cm}
	\centering
	\begin{adjustbox}{max width=630pt}  
		\begin{threeparttable}
			\centering
			\setlength{\tabcolsep}{12.3pt}
			\caption{HAR-family Out-of-Sample Forecasting Performance}
			\begin{tabular}{cccccccccc} \toprule
				{{\textbf{\small Full Out-of-Sample Period}}}  & & {AR} & {HAR} & {HAR-J} & {CHAR} & {SHAR} & {ARQ} & {HARQ} & {HARQ-F} \\ \midrule
				
				MSE & Avg  & 218.957 & 208.345 & 199.070 & 198.924 & 209.478 & 210.459 & 204.707 & 207.962 \\
				& Med &  144.744 & 137.001 & 130.463 & 129.495 & 138.829 & 142.448 & 136.673 & 147.108  \\
				RC\tnote{a} &  & \textit{(86.96,52.17)} & \textit{(95.65,73.91)} & \textit{(100,95.65)} & \textit{(100,100)} & \textit{(95.65,73.91)} & \textit{(69.57,47.83)} & \textit{(86.96,69.57)} & \textit{(95.65,82.61)}  \\[0.05cm]
				
				QLIKE & Avg  & 0.693 & 0.608 & 0.552 & 0.543 & 0.608 & 0.641 & 0.590 & 0.607 \\ 
				& Med  & 0.687 & 0.612 & 0.560 & 0.537 & 0.606 & 0.650 & 0.591 & 0.601  \\
				RC &  & \textit{(0,0)} & \textit{(0,0)} & \textit{(86.96,86.96)} & \textit{(100,100)} & \textit{(0,0)} & \textit{(8.70,0)} & \textit{(8.70,8.70)} & \textit{(17.39,13.04)}  \\
				
				\midrule		

				{{\textbf{\small Normal Volatility Days}}}  & &  &  &  &  &  &  &  &  \\ \midrule

				MSE & Avg  & 11.745 & 11.228 & 9.030 & 7.730 & 12.230 & 10.115 & 10.364 & 10.808 \\
				& Med &  7.798 & 7.000 & 5.947 & 5.091 & 7.489 & 7.478 & 7.101 & 7.392 \\
				RC &  & \textit{(56.52,47.83)} & \textit{(60.87,52.17)} & \textit{(69.57,52.17)} & \textit{(95.65,95.65)} & \textit{(78.26,56.52)} & \textit{(30.44,26.09)} & \textit{(47.83,26.09)} & \textit{(30.44,26.09)}  \\[0.05cm]
				
				QLIKE & Avg  & 0.399 & 0.340 & 0.313 & 0.298 & 0.339 & 0.364 & 0.327 & 0.318 \\ 
				& Med  & 0.362 & 0.327 & 0.296 & 0.269 & 0.330 & 0.339 & 0.314 & 0.303 \\
				RC &  & \textit{(0,0)} & \textit{(4.35,4.35)} & \textit{(34.78,21.74)} & \textit{(95.65,95.65)} & \textit{(0,0)} & \textit{(0,0)} & \textit{(17.39,13.04)} & \textit{(56.52,43.48)}  \\
				
				\midrule
				{\textbf{\small High Volatility Days}}  & {} & {} & {} & {} & {}  & {} \\ \midrule

				MSE & Avg  & 2131.361 & 2028.059 & 1954.202 & 1964.630 & 2030.282 & 2060.872 & 1999.278 & 2027.084 \\
				& Med &  1293.253 & 1223.783 & 1171.940 & 1163.489 & 1236.025 & 1266.151 & 1218.390 & 1334.427  \\
				RC &  & \textit{(91.30,47.83)} & \textit{(91.30,69.57)} & \textit{(100,100)} & \textit{(100,91.30)} & \textit{(95.65,69.57)} & \textit{(91.30,69.57)} & \textit{(91.30,91.30)} & \textit{(100,100)}  \\[0.05cm]
				
				QLIKE & Avg &  3.329 & 3.008 & 2.690 & 2.739 & 3.015 & 3.126 & 2.946 & 3.184  \\	
				& Med &  3.275 & 2.905 & 2.618 & 2.691 & 2.914 & 3.109 & 2.902 & 3.108  \\
				RC &  & \textit{(73.91,43.48)} & \textit{(56.52,52.17)} & \textit{(95.65,95.65)} & \textit{(100,100)} & \textit{(56.52,47.83)} & \textit{(82.61,69.57)} & \textit{(52.17,43.48)} & \textit{(52.17,26.09)}  \\			 
				
				\bottomrule			
			\end{tabular}
			\begin{tablenotes}[para,flushleft]
				\footnotesize
				\item \textit{Notes:} Following \Cref{har_general_spec}, the modelling functions are $f_{AR}({RV}_{t})$, 
				$f_{HAR}({RV}_{t}, \overline{RV}^w_t, \overline{RV}^m_t)$, 
				$f_{HAR-J}(RV_{t}, \overline{RV}^w_t, \overline{RV}^m_t, J_{t})$, $f_{CHAR}(BPV_{t}, \overline{BPV}^w_t, \overline{BPV}^m_t)$, $f_{SHAR}(RV_{t}^{+}, RV_{t}^{-}, \overline{RV}^w_t, \overline{RV}^m_t)$, $f_{ARQ}\!\left(RV_t,\sqrt{RQ_t}RV_t\right)$, $f_{HARQ}\!\left(RV_t,\overline{RV}^w_t,\overline{RV}^m_t, \sqrt{RQ_t}RV_t \right)$, and $f_{HARQ-F}\!\left(RV_t,\overline{RV}^w_t,\overline{RV}^m_t, \sqrt{RQ_t}RV_t, \overline{RQ}^w_t, \overline{RQ}^m_t \right)$, where $RV_t$ is the realised variance, $\overline{RV}^w_t$ is the average of $RV_t$ for the last week, $\overline{RV}^m_t$ is the average $RV_t$ of the last month, $J_{t}$ is the jump component, $BPV_{t}$, $\overline{BPV}^w_t$, and $\overline{BPV}^m_t$ are the bipower variation, the average $BPV_{t}$ of the last week, and the average $BPV_{t}$ of the last month. $RV_{t}^{+}$ and $RV_{t}^{-}$  denote the RV aggregated from positive and negative intraday returns. $RQ_t$, $\overline{RQ}^w_t$, and $\overline{RQ}^m_t$ denote realised quarticity and its weekly and monthly averages, respectively. In the HARQ specification, demeaned square-root daily quarticity interacts with daily RV to account for time-varying measurement error, while HARQ-F additionally includes the weekly and monthly averages of realised quarticity.
				\item[a] The RC denotes the percentage of tickers for which the Reality Check does not reject, at the 5\% (or 10\%) significance level, the null hypothesis that no competing HAR-family model, excluding the model under assessment, has lower expected loss than the specified model.
			\end{tablenotes}
			\label{har_outofsample_table_base}
		\end{threeparttable}
	\end{adjustbox}

\end{landscape}
    \clearpage
}

\afterpage{%
	\clearpage
	\thispagestyle{empty}
	\atxy{\dimexpr\paperwidth-0.45in}{.5\paperheight}{\rotatebox[origin=center]{90}{\thepage}}
\begin{landscape}
\vspace*{0.1cm}
	\centering
	\begin{adjustbox}{max width=610pt}  
		\begin{threeparttable}
			\centering
			\setlength{\tabcolsep}{12.3pt}
			\caption{HAR-family Out-of-Sample Forecasting Performance (1,023 Estimation Sample Size)}
			\begin{tabular}{cccccccccc} \toprule
				{{\textbf{\small Full Out-of-Sample Period}}}  & & {AR} & {HAR} & {HAR-J} & {CHAR} & {SHAR} & {ARQ} & {HARQ} & {HARQ-F} \\ \midrule
				
				MSE & Avg  & 217.611 & 209.478 & 200.948 & 200.631 & 211.665 & 206.889 & 204.719 & 212.313 \\
				& Med &  144.957 & 139.762 & 136.629 & 135.86 & 141.447 & 138.695 & 137.645 & 143.170  \\
				RC\tnote{a} &  & (95.65,69.57) & (95.65,86.96) & \textit{(100,100)} & \textit{(100,100)} & \textit{(95.65,86.96)} & \textit{(95.65,86.96)} & \textit{(95.65,91.30)} & \textit{(91.30,91.30)}  \\[0.05cm]
				
				QLIKE & Avg  & 0.661 & 0.604 & 0.557 & 0.549 & 0.607 & 0.619 & 0.586 & 0.646 \\ 
				& Med  & 0.643 & 0.602 & 0.551 & 0.556 & 0.602 & 0.612 & 0.589 & 0.644  \\
				RC &  & \textit{(4.35,0)} & (17.39,4.35) & \textit{(95.65,95.65)} & \textit{(100,100)} & \textit{(8.70,0)} & (26.09,8.70) & \textit{(30.44,21.74)} & \textit{(39.13,21.74)}  \\
				
				\midrule		

				{{\textbf{\small Normal Volatility Days}}}  & &  &  &  &  &  &  &  &  \\ \midrule

				MSE & Avg  & 10.552 & 11.637 & 9.632 & 8.462 & 13.874 & 9.015 & 10.758 & 15.372 \\
				& Med &  6.269 & 7.457 & 5.592 & 5.116 & 8.326 & 6.821 & 6.917 & 8.336  \\
				RC &  & \textit{(100,91.30)} & \textit{(82.61,52.17)} & \textit{(69.57,56.52)} & \textit{(91.30,82.61)} & \textit{(86.96,60.87)} & \textit{(73.91,56.52)} & \textit{(60.87,39.13)} & \textit{(34.78,26.09)}  \\[0.05cm]
				
				QLIKE & Avg  & 0.342 & 0.324 & 0.304 & 0.292 & 0.323 & 0.319 & 0.312 & 0.331 \\ 
				& Med  & 0.316 & 0.306 & 0.294 & 0.279 & 0.308 & 0.299 & 0.300 & 0.327 \\
				RC &  & \textit{(4.35,4.35)} & \textit{(4.35,4.35)} & \textit{(47.83,39.13)} & \textit{(95.65,95.65)} & \textit{(4.35,0)} & \textit{(30.44,30.44)} & \textit{(34.78,26.09)} & \textit{(34.78,21.74)}  \\
				
				\midrule
				{\textbf{\small High Volatility Days}}  & {} & {} & {} & {} & {}  & {} \\ \midrule

				MSE & Avg  & 2127.649 & 2035.158 & 1966.077 & 1973.312 & 2037.025 & 2034.627 & 1995.393 & 2030.749 \\
				& Med &  1307.010 & 1252.836 & 1232.057 & 1225.938 & 1257.860 & 1236.097 & 1226.840 & 1313.707  \\
				RC &  & \textit{(86.96,60.87)} & \textit{(82.61,69.57)} & \textit{(100,100)} & \textit{(100,100)} & \textit{(91.30,73.91)} & \textit{(82.61,73.91)} & \textit{(95.65,86.96)} & \textit{(100,91.30)}  \\[0.05cm]
				
				QLIKE & Avg &  3.514 & 3.110 & 2.810 & 2.841 & 3.139 & 3.303 & 3.037 & 3.451  \\	
				& Med &  3.486 & 3.058 & 2.731 & 2.778 & 3.038 & 3.305 & 2.923 & 3.339  \\
				RC &  & \textit{(34.78,26.09)} & \textit{(52.17,39.13)} & \textit{(100,100)} & \textit{(100,100)} & \textit{(52.17,39.13)} & \textit{(56.52,39.13)} & \textit{(56.52,47.83)} & \textit{(82.61,43.48)}  \\			 
				
				\bottomrule			
			\end{tabular}
			\begin{tablenotes}[para,flushleft]
				\footnotesize
				\item \textit{Notes:} Following \Cref{har_general_spec}, the modelling functions are $f_{AR}({RV}_{t})$, 
				$f_{HAR}({RV}_{t}, \overline{RV}^w_t, \overline{RV}^m_t)$, 
				$f_{HAR-J}(RV_{t}, \overline{RV}^w_t, \overline{RV}^m_t, J_{t})$, $f_{CHAR}(BPV_{t}, \overline{BPV}^w_t, \overline{BPV}^m_t)$, $f_{SHAR}(RV_{t}^{+}, RV_{t}^{-}, \overline{RV}^w_t, \overline{RV}^m_t)$, $f_{ARQ}\!\left(RV_t,\sqrt{RQ_t}RV_t\right)$, $f_{HARQ}\!\left(RV_t,\overline{RV}^w_t,\overline{RV}^m_t, \sqrt{RQ_t}RV_t \right)$, and $f_{HARQ-F}\!\left(RV_t,\overline{RV}^w_t,\overline{RV}^m_t, \sqrt{RQ_t}RV_t, \overline{RQ}^w_t, \overline{RQ}^m_t \right)$, where $RV_t$ is the realised variance, $\overline{RV}^w_t$ is the average of $RV_t$ for the last week, $\overline{RV}^m_t$ is the average $RV_t$ of the last month, $J_{t}$ is the jump component, $BPV_{t}$, $\overline{BPV}^w_t$, and $\overline{BPV}^m_t$ are the bipower variation, the average $BPV_{t}$ of the last week, and the average $BPV_{t}$ of the last month. $RV_{t}^{+}$ and $RV_{t}^{-}$  denote the RV aggregated from positive and negative intraday returns. $RQ_t$, $\overline{RQ}^w_t$, and $\overline{RQ}^m_t$ denote realised quarticity and its weekly and monthly averages, respectively. In the HARQ specification, demeaned square-root daily quarticity interacts with daily RV to account for time-varying measurement error, while HARQ-F additionally includes the weekly and monthly averages of realised quarticity.
				\item[a] The RC denotes the percentage of tickers for which the Reality Check does not reject, at the 5\% (or 10\%) significance level, the null hypothesis that no competing HAR-family model, excluding the model under assessment, has lower expected loss than the specified model.
			\end{tablenotes}
			\label{har_outofsample_table_base_short}
		\end{threeparttable}
	\end{adjustbox}
\end{landscape}
    \clearpage
}

\clearpage
\subsection{Stability of News Volume During the Out-of-Sample Period} \label{appendix_oos_volume_stability}
\begin{figure}[H]
	\centering
	\includegraphics[scale=0.30, trim = {0cm 1.0cm 0cm 0cm}]{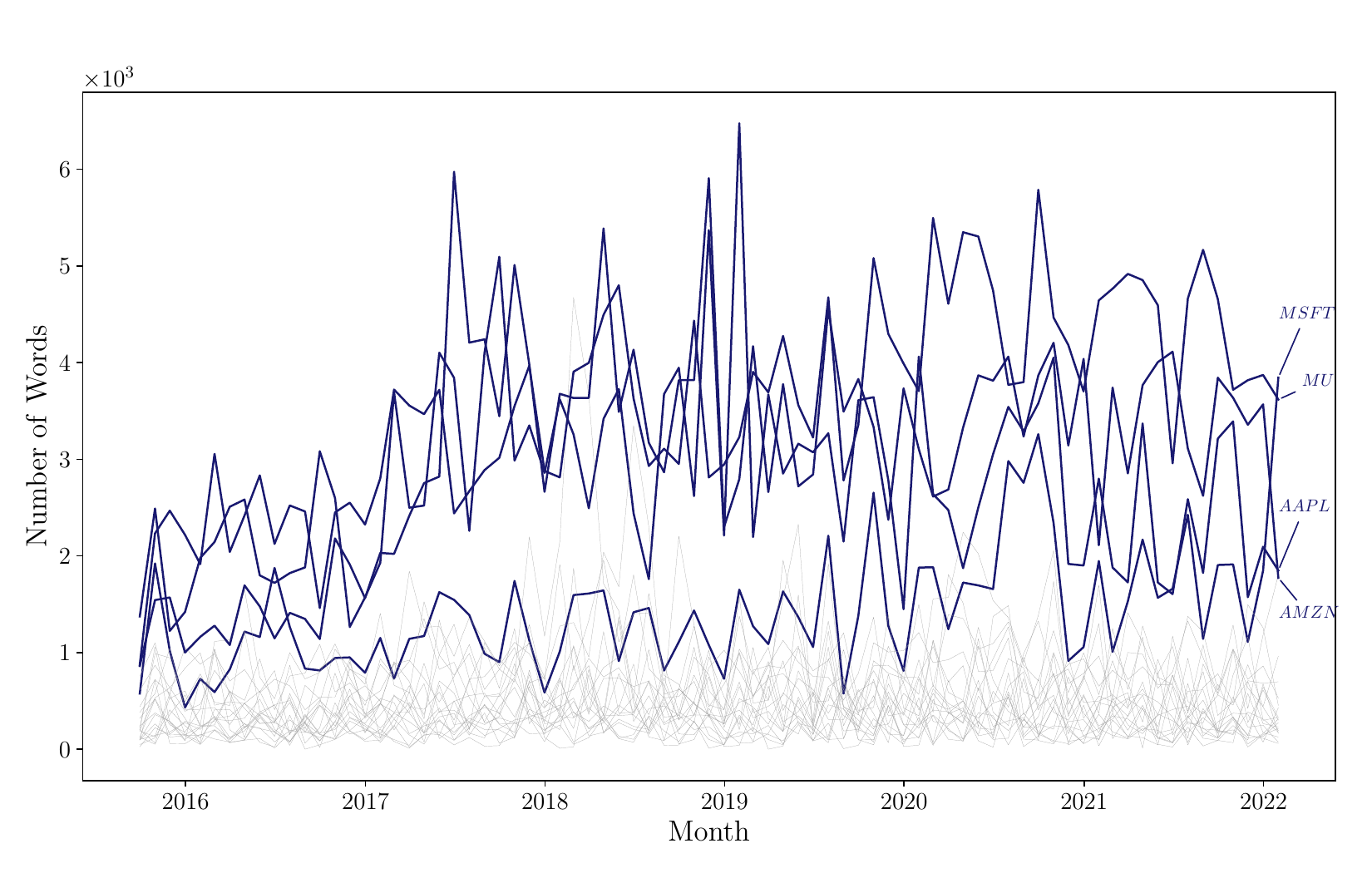}
	\caption{Stock-Related News Word Count During the Out-of-Sample Period}
	\begin{minipage}{0.8\linewidth} 
		\rule{\linewidth}{0.03em} \vspace{.15mm} \footnotesize
		{\emph{Notes:} The figure presents the monthly word count of stock-related news stories during the out-of-sample period. To enhance clarity, we have used bold lines in a different colour and labelled only those lines that deviate from the general trend, as most stocks follow similar patterns.}
		
	\end{minipage}  
	\label{RV_number_of_words_trent_stock_related}
\end{figure}

\begin{figure}[H]
	\centering
	\includegraphics[scale=0.30, trim = {0cm 1.0cm 0cm 0cm}]{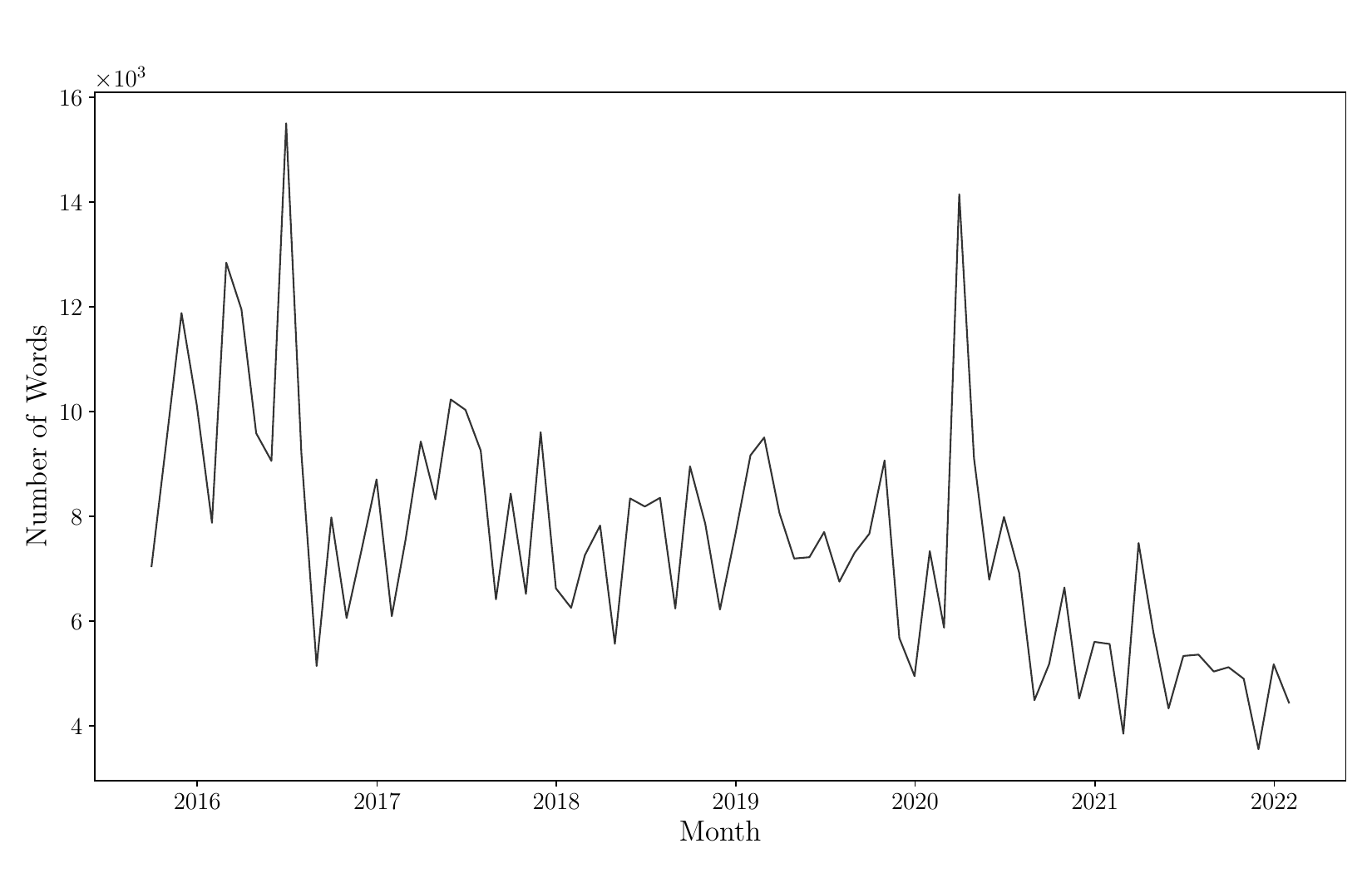}
	\caption{General Hot News Word Count During the Out-of-Sample Period}
	\begin{minipage}{0.8\linewidth} 
		\rule{\linewidth}{0.03em} \vspace{.15mm} \footnotesize
		{\emph{Notes:} The figure presents the monthly word count of general hot news stories during the out-of-sample period.}
		
	\end{minipage}  
	\label{RV_number_of_words_trent_hot_general}
\end{figure}

\Cref{RV_number_of_words_trent_stock_related} and \Cref{RV_number_of_words_trent_hot_general} provide complementary evidence on the evolution of news volume over the out-of-sample period. \Cref{RV_number_of_words_trent_stock_related} shows that the monthly word count of stock-related news is broadly stable across time for the majority of stocks, indicating a relatively constant level of stock-related media attention. To improve readability, only series that deviate from this common pattern are highlighted, revealing that a small number of highly visible stocks, such as MU, AMZN, AAPL, and MSFT, exhibit persistently higher word counts rather than short-lived spikes. \Cref{RV_number_of_words_trent_hot_general} similarly indicates that the aggregate volume of general hot news remains stable at the monthly frequency, suggesting no systematic expansion or contraction in overall news production during the evaluation window. Taken together, these figures imply that subsequent empirical results are unlikely to be mechanically driven by changes in the quantity of textual data, and instead reflect variation in the informational content or relevance of news.


\subsection{Evolution of High Volatility Day Counts Over Time} \label{appendix_high_vol_days_appendix}

\begin{figure}[H]
    \tiny
    \centering
    \includegraphics[scale=0.29, clip, trim={0cm 0cm 0cm 2cm}]{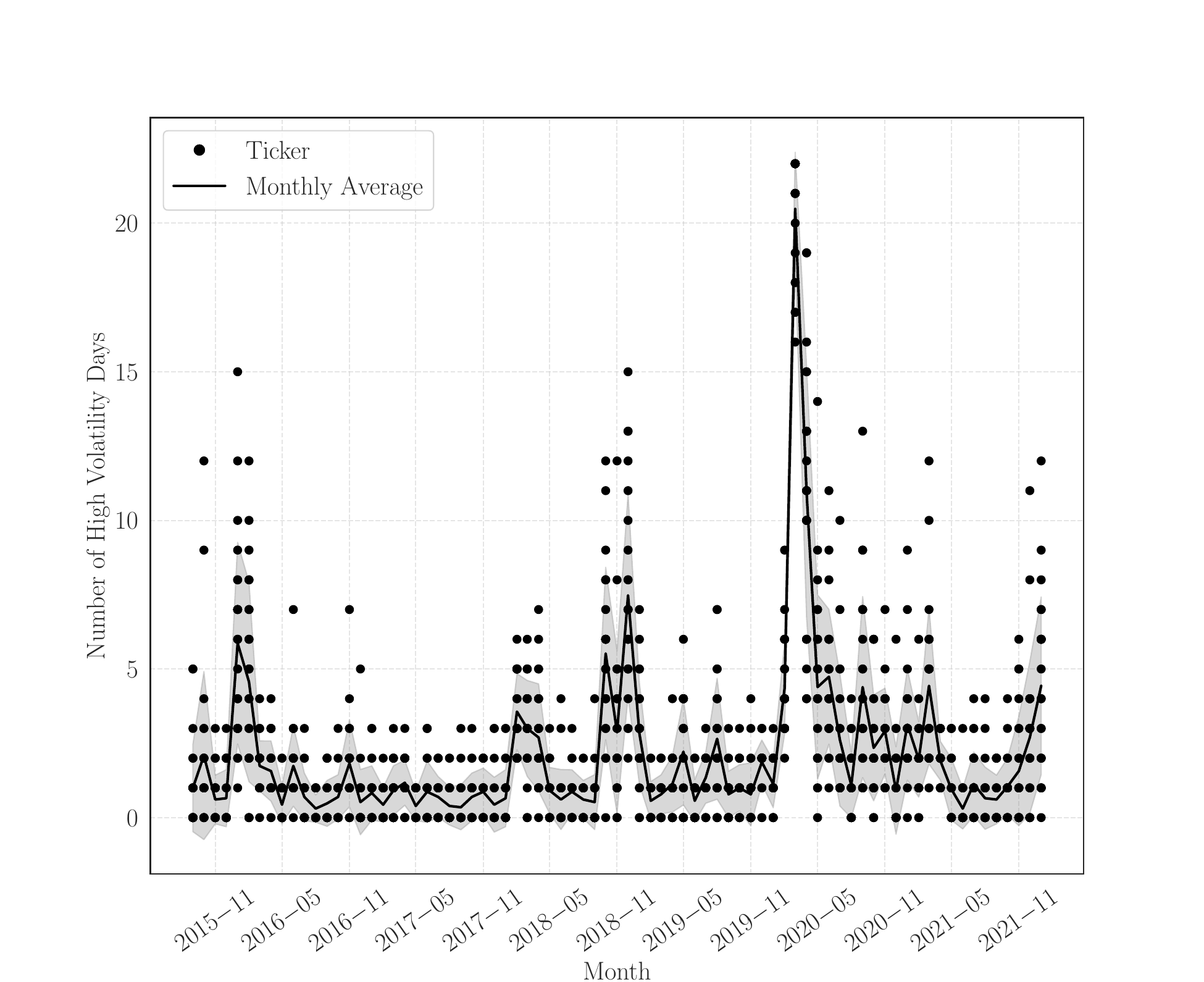}
    \caption{Number of High Volatility Days per Month over the Out-of-Sample Period}
    \label{normal_vs_high_RV}
    \begin{minipage}{\linewidth}
        \rule{\linewidth}{0.02em}\vspace{1.001mm}
        {\footnotesize
        \emph{Notes:} The x-axis represents the out-of-sample months, whereas the y-axis shows the count of days identified as high volatility days for each ticker in the respective month. Black-filled circles depict the tickers. The trend on a monthly average basis is represented by the line, and the areas shaded around this monthly average line signify the standard deviation.}
    \end{minipage}
\end{figure}

\Cref{normal_vs_high_RV} presents the evolution of the monthly frequency of high volatility days across individual tickers during the out-of-sample period. The horizontal axis denotes calendar months, while the vertical axis reports the number of days classified as high volatility within each month for a given ticker. Individual tickers are displayed as black markers, allowing for a granular view of cross-sectional dispersion, whereas the solid line summarises the average monthly count across all assets. The shaded band surrounding this line represents one standard deviation, providing a measure of cross-ticker variability in high volatility occurrences over time.

The figure reveals pronounced temporal variation in the incidence of high volatility days, with clear spikes during periods of elevated market stress, most notably around the COVID-19 pandemic. Importantly, however, high volatility days are not exclusively concentrated in these extreme episodes. Instead, they occur throughout the sample, indicating that elevated volatility is a recurrent feature rather than a phenomenon confined to crisis periods. Additionally, the substantial month-to-month and cross-sectional dispersion underscores heterogeneity in volatility dynamics across assets, suggesting that both systematic shocks and idiosyncratic factors contribute to the observed distribution of high volatility days.

\vspace*{\fill}

\vspace*{\fill}

\clearpage
\subsection{NLP-based RV forecasting: Model Structure Details} \label{appendix_RV_model}
Following \citet{DBLP:journals/corr/Kim14f}, let $X_i \in \mathbb{R}^{d}$ be the $d$-dimensional token vector corresponding to the $i$\textsuperscript{th} token in the news headline, with $d = 300$. Daily stock-related news input sequences (concatenated headlines) with fewer than 500 tokens are post-padded using index zero to ensure a fixed input length.\footnote{The padding value referenced here applies exclusively to stock-related news. For general hot news, a padding value of 2,000 is used. Additional details are provided in \Cref{nlp_ml_structure_subsection}.} Let $X_{i:i+j}$ refer to the concatenation of token vectors $X_i, X_{i+1}, \dots , X_{i+j}$ as follows:
\begin{equation}
\label{sentence_cnn_eq}
X_{i:i+j} = X_{i} \oplus X_{i+1} \oplus \dots \oplus X_{i+j},
\end{equation}
where $\oplus$ denotes the concatenation operator. A convolution operation involves a filter $W \in \mathbb{R}^{hd}$, which is applied over a window
of size $h$ tokens to produce a new feature:
\begin{equation}
\label{C_n_cnn_eq}
C_{i}=f(W \cdot X_{i:i+h-1}+b),
\end{equation}
where $b \in \mathbb{R}$ is a bias term and $f(\cdot)=\mathrm{ReLU}(\cdot)$ is the activation function. This filter is applied to each possible window of tokens
$\{X_{1:h}, X_{2:h+1}, \dots, X_{n-h+1:n}\}$ to produce a feature map
$C \in \mathbb{R}^{n-h+1}$:
\begin{equation}
\label{C_cnn_eq}
C = \{C_{1}, C_{2}, \dots, C_{n-h+1}\}.
\end{equation}

As the next step, global max-pooling is applied to the feature map:
\begin{equation}
\hat{C} = \max \{ C_1, C_2, \dots, C_{n-h+1} \},
\end{equation}
which retains the most informative feature detected by the filter \citep{collobert2011natural}. For each of the three window sizes, $q$ convolutional filters are employed. After global max pooling, the pooled outputs are concatenated to form the feature vector
\begin{equation}
\mathbf{z}_t =
\begin{bmatrix}
\hat{C}^{(1)} &
\hat{C}^{(2)} &
\dots &
\hat{C}^{(3q)}
\end{bmatrix}^{\top}
\in \mathbb{R}^{3q},
\end{equation}
where $q\in\{25,50,75,100\}$ denotes the number of filters employed for each window size. A dropout rate of 0.5 is applied to $\mathbf{z}_t$ before the fully connected output layer. The forecast of next-day RV is then
\begin{equation}
\widehat{RV}_{t+1}
=
\mathrm{ReLU}
\left(
\mathbf{w}^{\top}\mathbf{z}_t+b
\right),
\end{equation}
where $\mathbf{w}$ and $b$ denote the parameters of the output layer. The ReLU activation ensures non-negativity of the raw RV forecast.

Following the forecast-screening rule applied to the HAR-family models, any out-of-sample NLP forecast falling below the minimum or above the maximum RV observed in the corresponding rolling estimation window is replaced by the estimation-window mean before forecast evaluation. Since the minimum RV in each estimation window is strictly positive, this procedure ensures that all forecasts entering the QLIKE calculation are strictly positive. Model parameters are estimated by minimising the mean squared error (MSE) objective function:
\begin{equation}
\mathcal{L} =
\frac{1}{T}
\sum_{t=1}^{T}
\left(
RV_{t+1} - \widehat{RV}_{t+1}
\right)^2,
\end{equation}
where $RV_{t+1}$ and $\widehat{RV}_{t+1}$ denote the observed and forecasted RV, respectively. MSE is used as the optimisation criterion for estimating the NLP model, whereas out-of-sample forecast performance is evaluated using both MSE and QLIKE. Consequently, differences in performance across the two evaluation criteria may partly reflect the fact that the model is optimised directly for MSE rather than QLIKE. We therefore do not interpret the results as implying that MSE is necessarily the optimal training loss for realised volatility forecasting. To prevent overfitting, $L^2$ regularisation is applied to the CNN and fully connected output-layer parameters:
\begin{equation}
\mathcal{L}_{\text{reg}} =
\mathcal{L}
+
\lambda
\sum_{\theta \in \Theta}
\lVert \theta \rVert_2^2,
\end{equation}
where $\Theta$ denotes the set of CNN and fully connected output-layer parameters subject to $L^2$ regularisation. In implementation, the $L^2$ regularisation parameter is set to $\lambda=3$, and a dropout rate of 0.5 is applied before the output layer. Optimisation is carried out using the Adam algorithm \citep{kingma2014adam} with its default settings, using a batch size of 50 and one training epoch at each model re-estimation. All models are trained using fixed random number generator (RNG) seeds to ensure reproducibility. No early stopping is employed.

\clearpage
\subsection{LM Dictionary Results} \label{appendix_LM_Dicionary}
The LM dictionary in \citet{loughran2011liability} provides a transparent and well-established benchmark for extracting sentiment from financial text. Its relevance for the present forecasting setting is reinforced by \citet{zhang2016distillation}, who compare the LM dictionary with general-purpose lexical representations and show that text-based sentiment contains incremental information for subsequent stock volatility, trading volume, and returns. Comparing NLP-based with LM-based signals therefore serves two important purposes. First, it allows us to assess whether the added modelling flexibility and computational cost of NLP models translate into incremental predictive gains. Second, it anchors our analysis to the existing literature by evaluating performance relative to a familiar and highly interpretable textual framework.

The LM dictionary categorises words into financially meaningful sentiment classes. In this study, we focus on the standard LM sentiments: `Negative', `Positive', `Uncertainty', `Litigious', and `Constraining', as well as the modal sentiments `Modal Weak', `Modal Moderate', and `Modal Strong', which capture the strength of commitment in forward-looking statements. `Negative' and `Positive' reflect unfavourable and favourable tone, respectively, while `Uncertainty' captures ambiguity, imprecision, and lack of clarity regarding future outcomes. `Litigious' words proxy for legal, judicial, and regulatory risk, and `Constraining' words reflect language associated with limitations or restrictions on managerial actions and discretion. `Modal Weak' and `Modal Moderate' terms indicate low to intermediate levels of certainty and conditional or qualified intentions, whereas `Modal Strong' terms signal a high degree of certainty or obligation. In addition to sentiment-based measures, we include a simple news volume proxy, denoted as `News Count', defined as the daily number of stock-related news stories. This variable captures the intensity of information arrival independently of tone.

Sentiment scores are computed on a daily, stock-related basis using term frequency–inverse document frequency (tf--idf) weighting. For each stock and trading day, tf--idf--weighted counts of words belonging to a given LM sentiment are first computed at the news story level and then aggregated across all stock-related news items published during that day by simple averaging. The resulting daily sentiment measure, denoted by $S_{i,t}$, summarises the overall tone of stock-related news within the daily information set. The sentiment measures are constructed over a one-day horizon, with the start and end times exactly matching the news aggregation window described in \Cref{nlp_ml_structure_subsection}. As a result, the information content and temporal coverage of the LM-based sentiment variables are fully aligned with those of the NLP-derived signals. These sentiment variables are incorporated into the forecasting framework by extending the baseline CHAR model, which is identified as the best-performing specification in \Cref{results_section}. Building on the general HAR-family specification in \Cref{har_general_spec}, the CHAR model augmented with dictionary-based sentiment is given by
\begin{equation}
RV_{i,t+1}
= \beta_0
+ \beta_1 BPV_{i,t}
+ \beta_2 BPV^{w}_{i,t}
+ \beta_3 BPV^{m}_{i,t}
+ \gamma S_{i,t}
+ \varepsilon_{i,t+1},
\end{equation}
where $RV_{i,t+1}$ denotes next-day RV, $BPV_{i,t}$ is daily bipower variation, and $BPV^{w}_{i,t}$ and $BPV^{m}_{i,t}$ are the corresponding weekly and monthly averages, respectively. The coefficient $\gamma$ captures the incremental predictive content of LM-based sentiment for RV. Each sentiment is added individually, yielding a parsimonious extended CHAR specification. This exercise provides a useful benchmark for assessing the predictive content of simple textual variables. It should not, however, be interpreted as a fully matched comparison with the NLP ensemble in \Cref{Results_Ensemble_Subsection}, because the dictionary-based variables enter CHAR directly whereas the NLP signal is incorporated at the forecast level.

\afterpage{%
	\clearpage%
	\thispagestyle{empty}%
	\atxy{\dimexpr\paperwidth-0.45in}{.5\paperheight}{\rotatebox[origin=center]{90}{\thepage}}
\begin{landscape}
	\begin{table}
	\centering
	\begin{adjustbox}{max width=680pt}
		\begin{threeparttable}
			\centering
			\setlength{\tabcolsep}{10pt}
			\caption{\small Dictionary-Based Out-of-Sample Forecasting Performance}
			\begin{tabular}{ccc|cccccccc|c}
				\toprule
				{{\textbf{\small Full Out-of-Sample Period}}}  & {$\alpha$} & &
				Negative & Positive & Uncertainty & Litigious &
				Modal\textsubscript{\textit{weak}} &
				Modal\textsubscript{\textit{moderate}} &
				Modal\textsubscript{\textit{strong}} &
				Constraining & News Count \\
				\midrule
				MSE\tnote{a} & & Avg
				& 1.001 & 1.000 & 1.000 & 1.001 & 1.001 & 1.000 & 1.000 & 1.000 & 0.985 \\
				& & Med
				& 1.001 & 1.000 & 1.000 & 1.000 & 1.000 & 1.000 & 1.000 & 1.000 & 0.992 \\
				RC\tnote{c} & 5\% & &
				100 & 100 & 100 & 95.65 & 100 & 95.65 & 100 & 86.96 & 100 \\
				& 10\% & &
				95.65 & 95.65 & 95.65 & 82.61 & 91.30 & 95.65 & 91.30 & 86.96 & 100 \\

				[0.1cm]
                
				QLIKE\tnote{b} & & Avg
				& 0.999 & 1.000 & 1.004 & 1.003 & 0.998 & 1.001 & 1.008 & 1.006 & 0.947 \\
				& & Med
				& 1.003 & 1.001 & 1.000 & 1.002 & 1.001 & 1.001 & 1.001 & 1.001 & 0.941 \\
				RC & 5\% & &
				95.65 & 91.30 & 100 & 95.65 & 100 & 95.65 & 95.65 & 91.30 & 95.65 \\
				& 10\% &  &
				91.30 & 91.30 & 100 & 91.30 & 95.65 & 95.65 & 86.96 & 91.30 & 91.30  \\
   
				\midrule
				{\textbf{\small Normal Volatility Days}}  \\
				\midrule
				MSE & & Avg
				& 1.019 & 1.005 & 1.040 & 1.015 & 1.031 & 1.010 & 1.011 & 1.010 & 1.340 \\
				& & Med
				& 0.997 & 1.005 & 1.012 & 1.006 & 1.011 & 1.006 & 1.005 & 1.003 & 1.123 \\
				RC & 5\% & &
				73.91 & 73.91 & 47.83 & 60.87 & 56.52 & 56.52 & 60.87 & 82.61 & 43.48 \\
				& 10\% & &
				73.91 & 65.22 & 47.83 & 47.83 & 47.83 & 52.17 & 52.17 & 73.91 & 43.48 \\

				[0.1cm]
				QLIKE & & Avg
				& 0.997 & 1.002 & 1.007 & 1.003 & 1.006 & 1.004 & 1.004 & 1.005 & 0.992 \\
				& & Med
				& 0.995 & 1.001 & 1.001 & 1.002 & 1.003 & 1.003 & 1.002 & 1.003 & 0.988 \\
				RC & 5\% & &
				78.26 & 82.61 & 82.61 & 69.57 & 73.91 & 82.61 & 82.61 & 82.61 & 73.91 \\
				& 10\% &  &
				78.26 & 78.26 & 73.91 & 69.57 & 69.57 & 82.61 & 69.57 & 78.26 & 73.91  \\
                
				\midrule
				{\textbf{\small High Volatility Days}}  \\
				\midrule
				MSE & & Avg
				& 1.000 & 1.000 & 0.999 & 1.000 & 0.999 & 1.000 & 1.000 & 1.000 & 0.970 \\
				& & Med
				& 1.001 & 1.000 & 1.000 & 1.000 & 1.000 & 1.000 & 1.000 & 1.000 & 0.986 \\
				RC & 5\% & &
				91.30 & 100 & 95.65 & 91.30 & 95.65 & 91.30 & 91.30 & 100 & 100 \\
				& 10\% &   &
				91.30 & 100 & 95.65 & 86.96 & 95.65 & 91.30 & 86.96 & 100 & 100 \\
                
				[0.1cm]
				QLIKE & & Avg
				& 1.003 & 0.998 & 1.005 & 1.003 & 0.992 & 0.997 & 1.012 & 1.007 & 0.903 \\
				& & Med
				& 1.007 & 1.001 & 1.000 & 1.001 & 0.988 & 1.000 & 1.001 & 0.998 & 0.911 \\
				RC & 5\% &  &
				82.61 & 100 & 100 & 95.65 & 100 & 100 & 100 & 100 & 100  \\
				& 10\% &  &
				78.26 & 95.65 & 100 & 91.30 & 91.30 & 91.30 & 91.30 & 100 & 100 \\

				\bottomrule
			\end{tabular}
			\begin{tablenotes}[para,flushleft]
				\footnotesize
                \textit{Notes:}
                \item[a] For each ticker, MSE is expressed relative to the corresponding MSE of CHAR. The reported average (median) is the cross-sectional mean (median) of the 23 ticker-level MSE ratios.
                \item[b] For each ticker, QLIKE is expressed relative to the corresponding QLIKE of CHAR. The reported average (median) is the cross-sectional mean (median) of the 23 ticker-level QLIKE ratios.
                \item[c] RC non-rejection denotes the percentage of tickers for which White's Reality Check does not reject, at the stated significance level, the null hypothesis that no competing HAR-family model has lower expected loss than the model under evaluation. A high value therefore indicates limited statistical evidence of inferiority.
			\end{tablenotes}
			\label{har_outofsample_table}
		\end{threeparttable}
	\end{adjustbox}
	\end{table}
\end{landscape}
\clearpage%
}

\Cref{har_outofsample_table} shows that dictionary-based signals yield, at best, modest gains relative to the CHAR benchmark. Over the full out-of-sample period, most dictionary sentiments produce loss ratios that are effectively at parity, with the best-performing entries remaining very close to one. This pattern indicates that any improvements delivered by dictionary-based sentiment measures are small in magnitude. The RC rates are generally high, typically ranging from about 83--100\% depending on the sentiment measure and significance level, indicating limited evidence that a competing HAR-family model has significantly lower expected loss. By contrast, the simple stock-related `News Count' proxy stands out clearly. It reduces the full out-of-sample MSE ratio to 0.985 and the QLIKE ratio to 0.947, and it achieves 100\% RC at both the 5\% and 10\% levels in the full out-of-sample MSE comparison and above 91\% for QLIKE loss function. Together with the loss ratios below one, these results indicate that news intensity provides a comparatively strong improvement relative to the CHAR benchmark, highlighting the importance of information intensity rather than tone alone. Consequently, `News Count' provides the best-performing baseline against which the incremental predictive value of more sophisticated textual representations can be assessed.

Comparing the dictionary-based results in \Cref{har_outofsample_table} with the stock-related ensemble results in \Cref{NLP_ML_primary_experiment_table_stock_related_ensemble} provides suggestive evidence on how forecasting performance changes when richer
textual information is used. Most LM sentiments are essentially at parity with CHAR in terms of loss ratios, whereas `News Count' produces a meaningful full out-of-sample improvement, with an average MSE ratio of 0.985 and an average QLIKE ratio of 0.947. This finding shows that the intensity of firm-specific information arrival is itself an important source of predictability. The NLP-based ensemble produces further improvements in several dimensions. For example, the FinText Word2Vec (skip-gram) ensemble achieves an average MSE ratio of 0.961 and an average QLIKE ratio of 0.937. These results are consistent with the possibility that richer linguistic representations contain predictive information in addition to that captured by news intensity. However, because `News Count' and the NLP signal are incorporated through different forecasting mechanisms, the comparison does not fully isolate a semantic-content effect. Accordingly, we interpret the evidence as showing that news volume accounts for an important component of news predictability, while the richer NLP representation is associated with additional forecasting improvements.

\begin{table}
    \centering
    \begin{adjustbox}{max width=0.6\textwidth}
        \begin{threeparttable}
            \caption{Realised Utility of Dictionary-Based and News Count Models}
            \begin{tabular}{l@{\hspace{110pt}}c}
                \toprule
                Category & Realised Utility \\
                \midrule
                Negative                        & 2.7524 \\
                Positive                        & 2.7532 \\
                Uncertainty                     & 2.7485 \\
                Litigious                       & 2.7492 \\
                Modal\textsubscript{\textit{weak}}      & 2.7592 \\
                Modal\textsubscript{\textit{moderate}}  & 2.7531 \\
                Modal\textsubscript{\textit{strong}}    & 2.7406 \\
                Constraining                    & 2.7437 \\
                News Count                      & 2.8367 \\
                \bottomrule
            \end{tabular}
            \label{realised_utility_LM_categories}
            \begin{tablenotes}[para,flushleft]
                \footnotesize
                \item \textit{Notes:} This table reports realised utility values from the utility-based approach in \Cref{economic_gain_section} for dictionary-based and News Count RV forecasting models. In \Cref{utility_function}, the maximum attainable realised utility is 4\%. All values are expressed in percentage terms.
            \end{tablenotes}
        \end{threeparttable}
    \end{adjustbox}
\end{table}

The utility-based results in \Cref{realised_utility_LM_categories} reinforce and sharpen the conclusions drawn from the statistical forecasting evidence.\footnote{Details of the economic evaluation approach are provided in \Cref{economic_gain_section}.} With the exception of `News Count', all LM sentiments deliver realised utilities tightly clustered around the CHAR benchmark of 2.7540\%, ranging from 2.7406\% to 2.7592\%. These magnitudes indicate that sentiment-based dictionary signals, while occasionally marginally improving or matching CHAR in loss-based metrics, do not translate into meaningful improvements in realised utility. This finding is fully consistent with the loss ratios in \Cref{har_outofsample_table}. By contrast, `News Count' nevertheless stands out economically, achieving realised utility of 2.8367\%, compared with 2.7540\% for CHAR. The strongest stock-related NLP ensemble attains a still higher realised utility of 2.9321\%. This difference is consistent with additional economic value from richer textual information, although the two approaches incorporate textual information through different forecasting mechanisms. The utility evidence therefore reinforces the conclusion that news intensity is itself economically relevant while suggesting, rather than fully identifying, additional value from the linguistic information captured by the NLP model.

\begin{figure}
    \centering
    
	\begin{subfigure}{0.47\linewidth}
	\centering
	\begin{tikzpicture}[baseline]
    \node[label=left:\rotatebox{90}{{\footnotesize Loss}}]
    {\includegraphics[scale=0.28, trim = {0cm 1cm 0cm 1cm}]{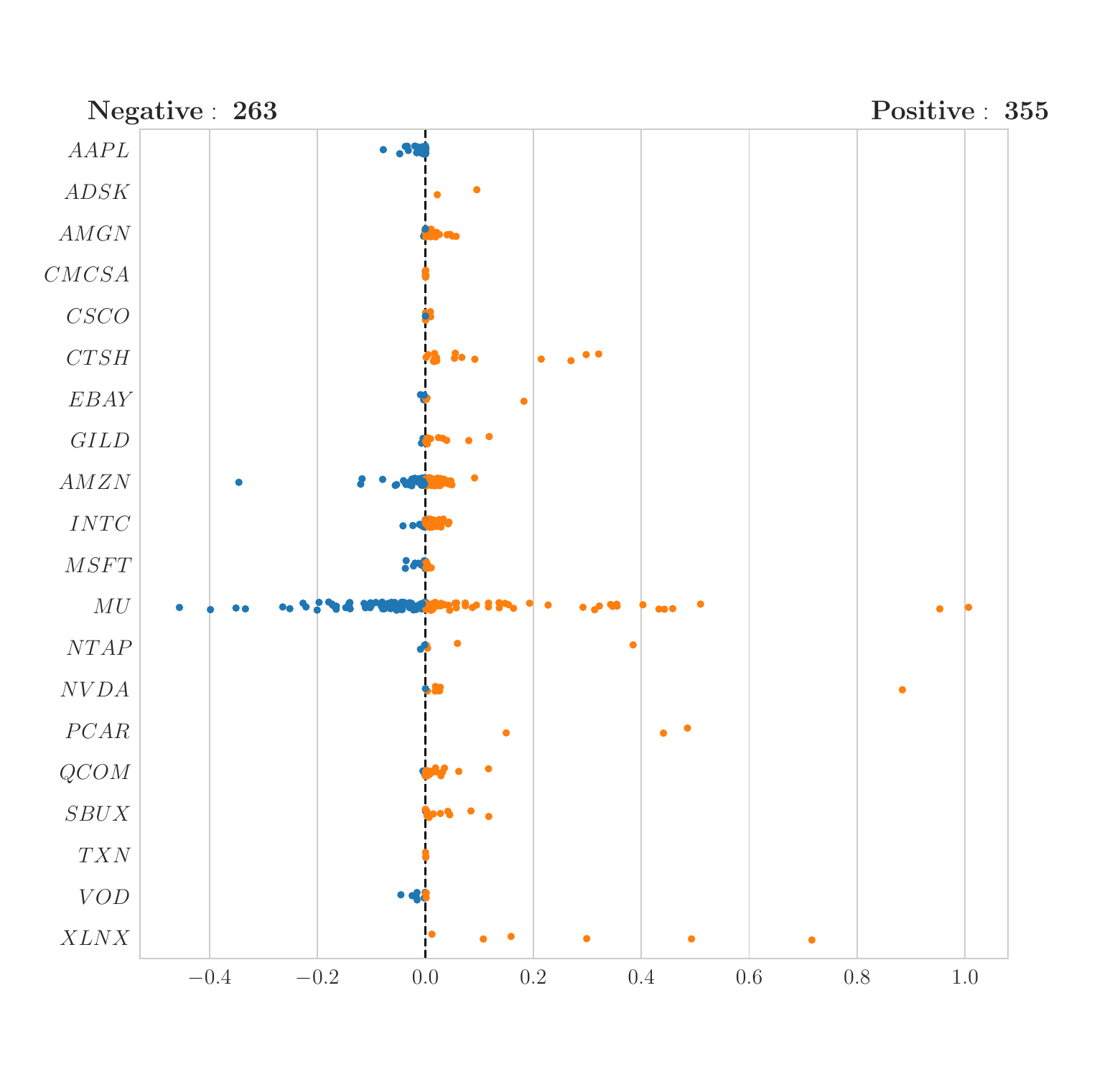}};
    \end{tikzpicture}
	\end{subfigure}	
	\begin{subfigure}{0.47\linewidth}
	\centering
	\begin{tikzpicture}[baseline]
	\node[label=left:\rotatebox{90}{{\footnotesize}}]
    {\includegraphics[scale=0.28,  trim = {0cm 1cm 0cm 1cm}]{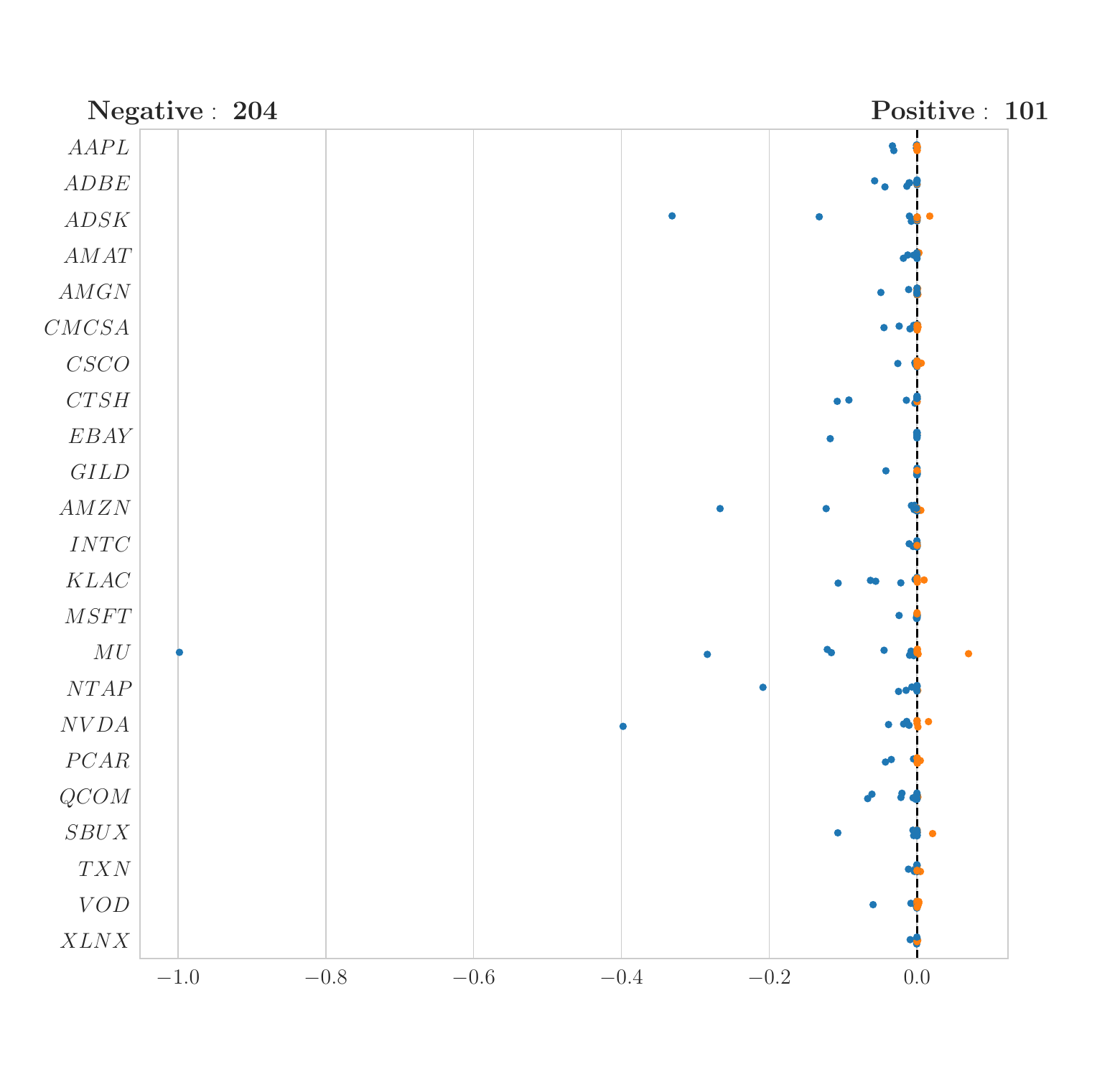}};
    \end{tikzpicture}
	\end{subfigure}
	\begin{subfigure}{0.47\linewidth}
	\centering
    \begin{tikzpicture}[baseline]
    \node[label=left:\rotatebox{90}{\footnotesize Termination}]
    {\includegraphics[scale=0.28, trim = {0cm 1cm 0cm 1cm}]{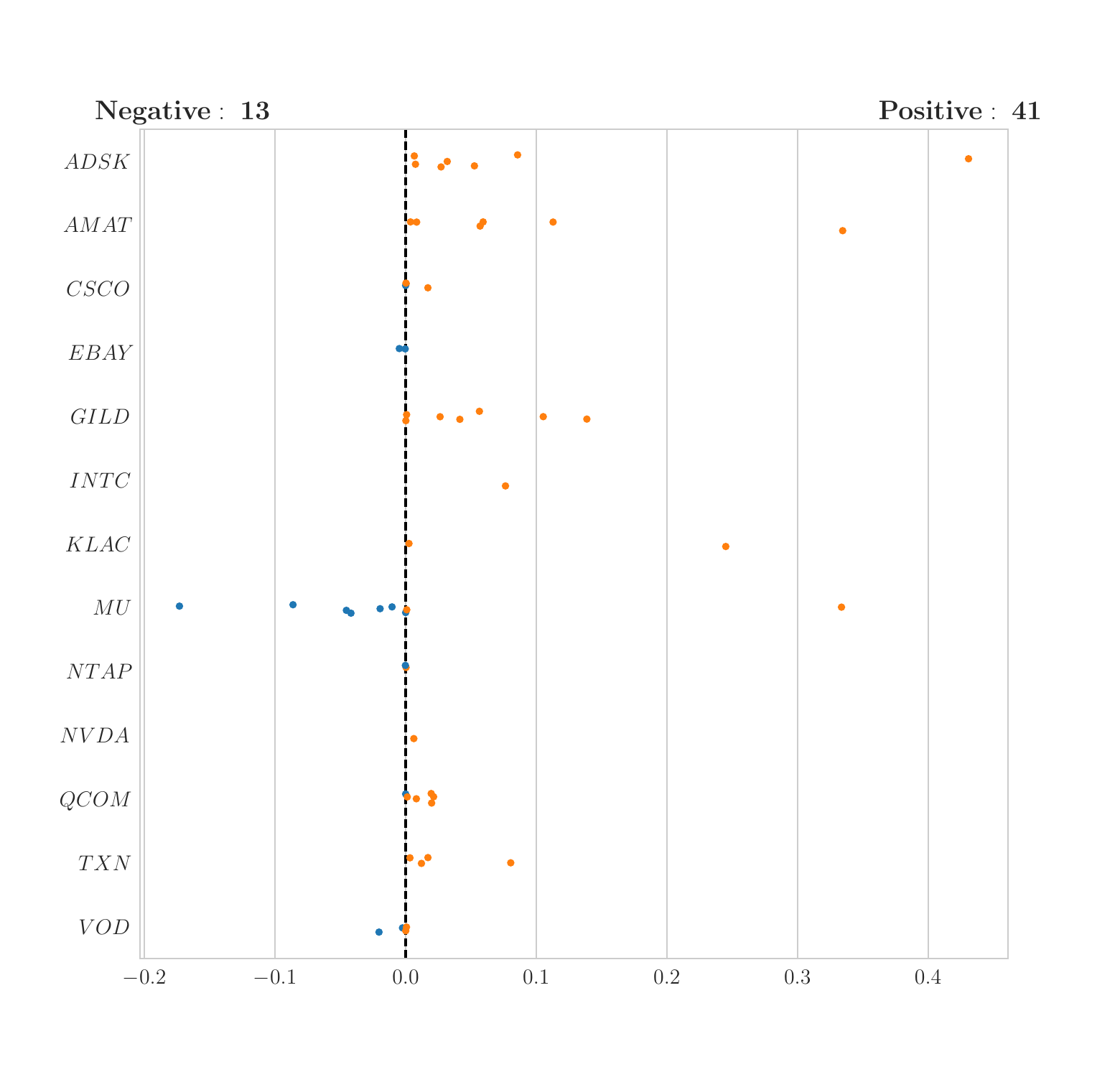}};
    \end{tikzpicture}	
	\end{subfigure}
	\begin{subfigure}{0.47\linewidth}
	\centering
    \begin{tikzpicture}[baseline]
    \node[label=left:\rotatebox{90}{\footnotesize}]
    {\includegraphics[scale=0.28, trim = {0cm 1cm 0cm 1cm}]{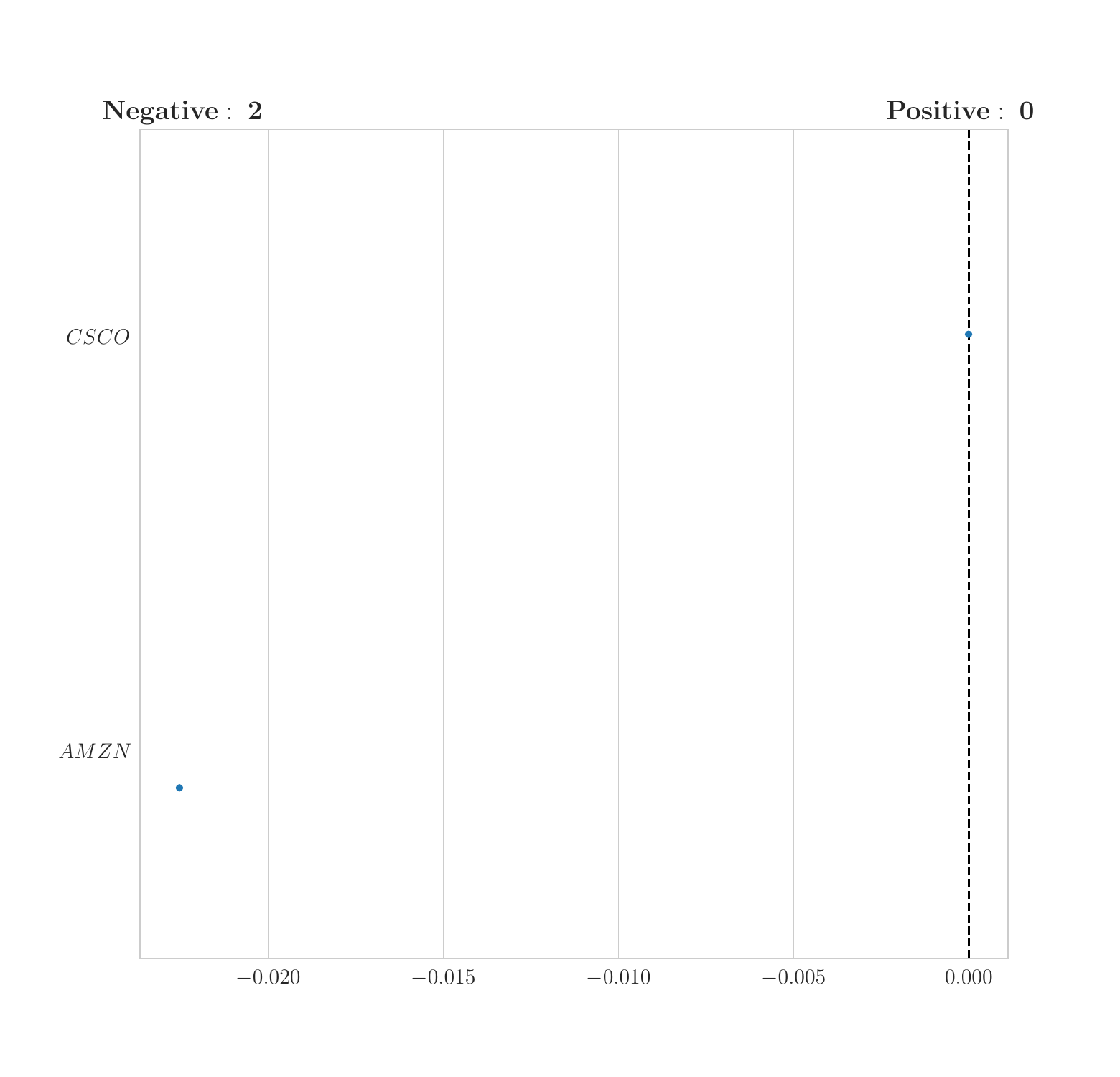}};
    \end{tikzpicture}	
	\end{subfigure}
	\begin{subfigure}{0.47\linewidth}
	\centering
    \begin{tikzpicture}[baseline]
    \node[label=below: \footnotesize Stock-Related News, label=left:\rotatebox{90}{\footnotesize Against}]
    {\includegraphics[scale=0.28, trim = {0cm 1cm 0cm 1cm}]{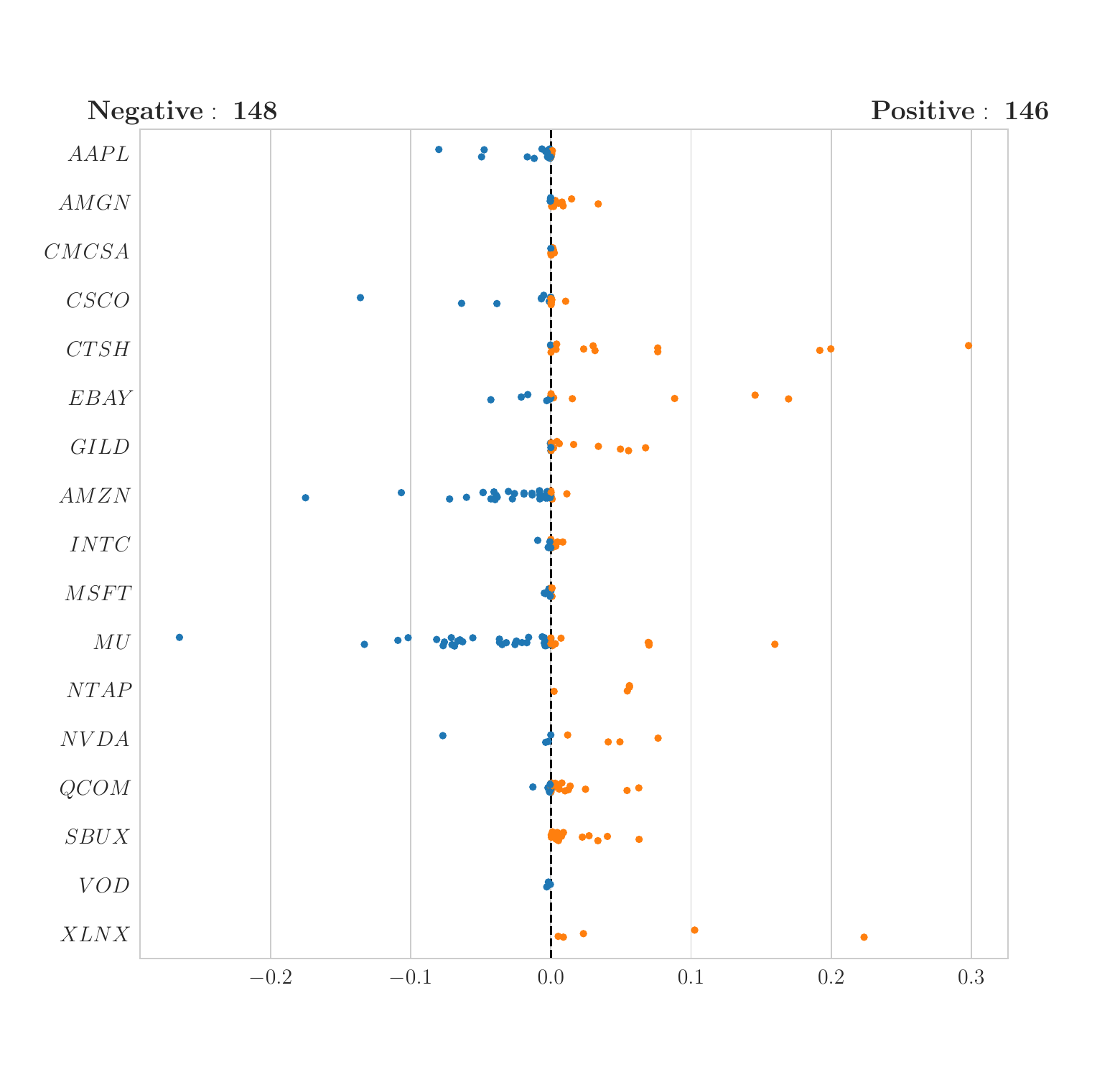}};
    \end{tikzpicture}
	\end{subfigure}
	\begin{subfigure}{0.47\linewidth}
	\centering
    \begin{tikzpicture}[baseline]
    \node[label=below: \footnotesize General Hot News, label=left:\rotatebox{90}{\footnotesize}]
    {\includegraphics[scale=0.28,  trim = {0cm 1cm 0cm 1cm}]{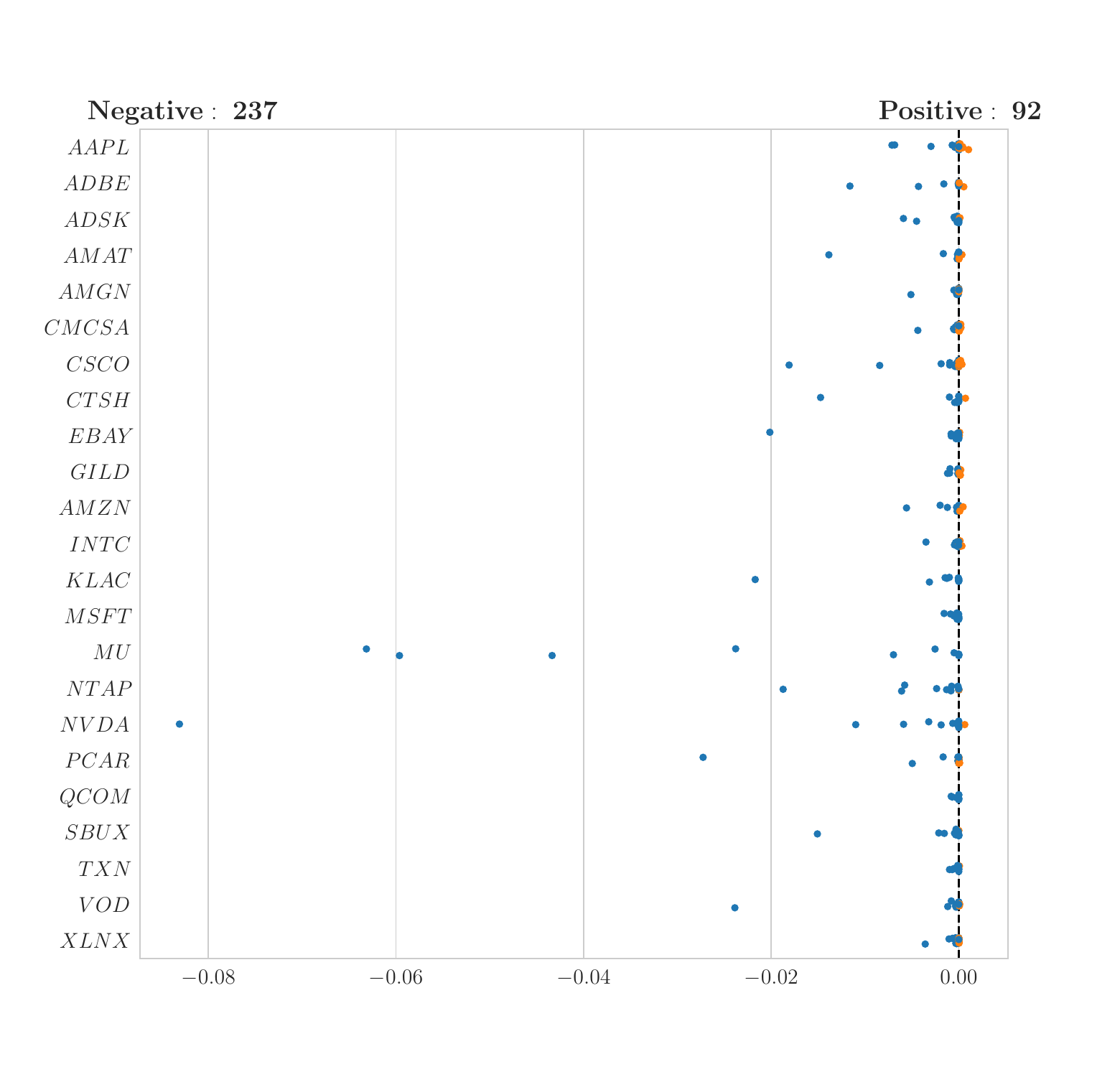}};
    \end{tikzpicture}	
	\end{subfigure}
	\caption{Shapley Values for Top Negative LM Words}
	\label{XAI_vs_LM_table}
	\begin{minipage}{1\linewidth} 
		\rule{\linewidth}{0.03em} \vspace{.15mm} \footnotesize
		{\emph{Notes:} Rows represent the most frequently occurring negative words in the LM dictionary. The left (right) column shows the Shapley values for stock-related (general hot) news published during the out-of-sample period. The x-axis displays the Shapley values, while the y-axis represents the ticker names. The vertical line indicates no impact on the RV forecast, with the left (right) side indicating a negative (positive) impact of the specified group of words. The total number of negative and positive Shapley values is displayed at the top of each figure.}
		
	\end{minipage}
	\label{Explainer_results_donald_trump}
\end{figure}

Moving to XAI in \Cref{explainable_ai_section} and extending it to the LM dictionary, \Cref{Explainer_results_donald_trump} shows the Shapley values for the three most frequent negative words in the LM dictionary, namely \textit{loss}, \textit{termination}, and \textit{against}, displayed in the top, middle, and bottom rows, respectively.\footnote{Negative sentiment is selected from the LM dictionary as it encompasses the largest number of terms compared to other sentiments. This dictionary is primarily designed for analysing financial reports, which typically convey a positive tone. Consequently, the negative sentiment category is more extensive, with a broader range of terms, to capture the nuances necessary for effective financial report analysis.} Each selected word is also grouped with its variations from the LM dictionary; thus, \{\textit{loss}, \textit{losses}\}, \{\textit{termination}, \textit{terminate}, \textit{terminates}, \textit{terminated}\}, and \{\textit{against}\} represent \textit{loss}, \textit{termination}, and \textit{against} words, respectively. The Shapley values for these words in each group are displayed individually in \Cref{Explainer_results_donald_trump}. The left (right) column shows the Shapley values for stock-related (general hot) news. A dot on the vertical zero line indicates cases where the word has no impact on the RV forecast. The left (right) side represents the negative (positive) impact of the specified word group, where a larger positive (negative) value indicates a larger increase (decrease) in the RV forecast. The y-axis displays the ticker name; if a ticker is absent from the y-axis, it implies that the specified word group either did not appear in the news stories or SHAP method did not identify words from that group as contributors to RV forecasting for that ticker. The total number of negative and positive Shapley values is indicated at the top of each figure.

\Cref{XAI_vs_LM_table} reveals important properties about using textual information to forecast RV based on a fixed LM dictionary approach versus our NLP model. First, the LM dictionary approach assigns words such as \textit{loss}, \textit{termination}, and \textit{against}, including their variations, to fixed sentiment categories and aggregates their tf--idf-weighted presence into daily sentiment measures. The contribution of each resulting sentiment measure to the RV forecast is then governed by its estimated regression coefficient. In contrast, \Cref{XAI_vs_LM_table} shows a wide variety of relationships between these top word groups and the RV forecast. For example, the `Loss' group in stock-related news increases the RV forecast in 355 appearances and decreases the RV forecast in 263 appearances (see top left subfigure). Similar patterns are observed for the other two top word groups in stock-related news and also in general hot news. Additionally, substantial variations exist across stocks. In the same subfigure, the `Loss' group has a negative impact on the RV forecast for AAPL but a predominantly positive impact on the RV forecast for CTSH. As previously noted, some of these words did not appear at all in news stories for certain stocks, or when they did, they had no discernible impact on the RV forecasts. This shows how the NLP model is able to capture more nuances in news, which ultimately translate into better statistical and economic performance compared with dictionary-based sentiments.

\clearpage
\subsection{Nonlinear modelling of HAR-Family Predictors} \label{appendix_nonlinear}
An important benchmarking consideration is whether the gains obtained by adding news to RV forecasting models reflect the incremental informational content of news, or whether similar improvements could be achieved through a more flexible specification of the benchmark models themselves. The HAR-family of models in \Cref{har_general_spec} is typically estimated under linear functional forms, despite the possibility that RV dynamics exhibit nonlinear interactions across horizons. If a nonlinear reformulation of the HAR framework were able to deliver forecasting performance comparable to, or exceeding, that of the proposed NLP models, the incremental contribution of news would be less clearly identified. This consideration motivates a direct assessment of how forecasting performance changes when the HAR-family specification is extended from a linear to a nonlinear form, while holding the information set fixed.

To address this concern, the linear HAR-family of models is generalised by allowing for a nonlinear functional form while preserving the same information set:
\begin{align}
RV_{t+1}
=
f_{\mathrm{FCNN}}\!\Big(
& RV_t,\ \overline{RV}^w_t,\ \overline{RV}^m_t,\ 
J_t,\ 
BPV_t,\ \overline{BPV}^w_t,\ \overline{BPV}^m_t,\nonumber\\
& RV_t^{+},\ RV_t^{-},\ 
RQ_t,\ \overline{RQ}^w_t,\ \overline{RQ}^m_t
\Big),
\label{nonlinear_har_twoline}
\end{align}
where all variables are defined in \Cref{review_rv_subsection}. The function \( f_{\mathrm{FCNN}}(\cdot) \) is implemented as a simple FCNN. An important advantage of this specification is that it accommodates, within a single unified framework, all predictors introduced across the HAR, SHAR, HAR-J, CHAR, ARQ, HARQ, and HARQ-F models, while allowing for nonlinear interactions among them. As a result, this specification provides a stringent benchmark for assessing whether nonlinear transformations of the full HAR-family information set can account for the forecasting gains. The nonlinear model is trained using the same forecasting design as the linear HAR benchmarks in \Cref{results_section}. In particular, training is conducted using a daily rolling-window with identical in-sample and out-of-sample periods and the same forecasting horizon, ensuring full comparability across models. To assess the role of model complexity, the FCNN is trained using different numbers of hidden units, specifically 5, 10, 15, 20, and 25. This allows for a systematic evaluation of how increasing nonlinear flexibility affects forecasting performance. Given 23 stocks, 1,604 out-of-sample trading days, and five model complexities, a total of 184{,}460 FCNN models are trained and evaluated under this specification.

The results in \Cref{HAR_FCNN_results} indicate that introducing nonlinearity into the HAR-family framework does not lead to an overall improvement in forecasting performance over the full out-of-sample period. Across both loss functions, the nonlinear HAR specification exhibits systematically higher loss ratios relative to the CHAR benchmark, together with generally weak RC results. For example, under MSE, the average loss ratios over the full out-of-sample period range between 1.200 and 1.219 across different network sizes, while the corresponding QLIKE ratios range from 4.474 to 7.402. Consistent with these results, RC values remain low, particularly under QLIKE. These findings indicate that greater model flexibility alone does not translate into superior average forecasting performance when evaluated across the full out-of-sample period. A more granular decomposition by volatility regime, however, reveals a pronounced state dependence. During normal volatility days, the results differ sharply across loss functions. Under MSE, average loss ratios fall well below unity and decline monotonically with model complexity, reaching values as low as 0.689 for the largest specification, with RC statistics exceeding 90\% for most configurations. Nevertheless, under QLIKE, with average ratios ranging from 2.873 to 5.184 and slightly higher RC values compared to the full out-of-sample results, the deterioration in performance is evident. In contrast, performance deteriorates substantially during high volatility days. In this regime, loss ratios exceed unity across all specifications, with MSE ratios averaging 1.226 and QLIKE ratios averaging 7.7838 across different model complexities, accompanied by uniformly weak RC statistics. These large forecast errors during high volatility days dominate the aggregate evaluation and account for the observed deterioration in full out-of-sample period. Overall, the results suggest that nonlinear transformations of the HAR-family predictors can improve MSE performance in normal market conditions, but these gains do not extend to QLIKE and lack robustness in periods of elevated volatility, limiting their effectiveness as standalone alternatives to the HAR-family of models.

Although nonlinear transformations of HAR-type regressors can deliver incremental MSE forecasting gains relative to linear specifications, particularly during normal volatility days, their overall performance remains systematically inferior to that of the standalone NLP models. This underperformance is especially pronounced when compared with ensemble forecasts that combine news-based signals with volatility-history benchmarks over the full out-of-sample period. A comparison with the NLP models reported in \Cref{NLP_ML_primary_experiment_table_ticker_relatedX} indicates that, while both the nonlinear HAR models and the NLP models exhibit some degradation in performance over the full out-of-sample window, the magnitude of underperformance is substantially larger for the nonlinear models. This result holds consistently across MSE and QLIKE loss ratios, as well as RC metrics. Furthermore, comparing these nonlinear specifications with the ensemble model results in \Cref{NLP_ML_primary_experiment_table_stock_related_ensemble} reveals a clear advantage of the ensemble approach. In the vast majority of cases, ensemble forecasts deliver unambiguous improvements over the full out-of-sample period, accompanied by substantially higher RC non-rejection rates than those of the nonlinear models.

\Cref{realised_utility_nonlinear_har} reports the realised utility associated with the nonlinear HAR model variants evaluated in the out-of-sample period, using the utility-based approach described in \Cref{economic_gain_section}. The results show that all nonlinear HAR specifications deliver negative realised utility, indicating a deterioration in economic performance. This stands in contrast to the higher realised utility documented in \Cref{realised_utility_models} for the NLP models and, in particular, for the ensemble models. Under the calibration in \Cref{economic_gain_section}, a negative realised utility indicates poor economic performance of the corresponding volatility-targeting strategy. It is a realised utility level rather than a utility difference relative to CHAR and therefore should not, by itself, be interpreted as evidence that average forecast errors are larger. Together with the substantially higher forecast losses reported in \Cref{HAR_FCNN_results}, however, the negative utility values confirm the weak economic performance of these nonlinear specifications. Taken together, the higher statistical forecast losses and negative realised utility indicate that these specifications are both less accurate on average and poorly aligned with the investor's objective, leading to inferior volatility-timing decisions. The observed results here suggest that the additional nonlinear structure does not enhance the decision-relevant content of volatility forecasts and instead amplifies estimation noise and overfitting, such that increased model flexibility fails to translate into higher realised utility, which, from the results in \Cref{HAR_FCNN_results}, stems from the underperformance of the models during high volatility days.

Overall, the results show that the particular nonlinear HAR specification considered in this study does not replicate the forecasting or realised-utility improvements obtained by incorporating news through the NLP models. While the nonlinear HAR specification shows improvements in MSE performance during normal volatility days, its performance deteriorates substantially during high volatility days, leading to weaker full out-of-sample forecasting performance and realised utility. These findings should not, however, be interpreted as evidence that nonlinear volatility-history models are generally ineffective, since performance may depend on the choice of architecture, regularisation, hyperparameters, and estimation procedure. Moreover, these results are consistent with research that questions the robustness of ML models for RV forecasting when they rely on broadly the same set of volatility-based predictors, suggesting that gains may be limited in the absence of genuinely new information \citep{audrino2016lassoing, branco2024forecasting, audrino2025hard}.

\vspace*{\fill}

\afterpage{%
\clearpage
\thispagestyle{empty}
\atxy{\dimexpr\paperwidth-0.45in}{.5\paperheight}{%
\rotatebox[origin=center]{90}{\thepage}}
\begin{landscape}
\begin{table}
\centering
\begin{adjustbox}{max width=1.20\textwidth}
\begin{threeparttable}
\centering
\setlength{\tabcolsep}{6pt}
\caption{FCNN Out-of-Sample Forecasting Performance}
\label{HAR_FCNN_results}
\begin{tabular}{lccc|ccccc|ccccc|ccccc}
\toprule
 &  &  &  
 & \multicolumn{5}{c|}{Full Out-of-Sample Period}
 & \multicolumn{5}{c|}{Normal Volatility Days}
 & \multicolumn{5}{c}{High Volatility Days} \\
\cmidrule(lr){5-9}\cmidrule(lr){10-14}\cmidrule(lr){15-19}
 &  & $\alpha$ &  
 & 5 & 10 & 15 & 20 & 25
 & 5 & 10 & 15 & 20 & 25
 & 5 & 10 & 15 & 20 & 25 \\
\midrule
\addlinespace[0.25cm]

MSE\tnote{a} &  &  & Avg
 & 1.215 & 1.214 & 1.219 & 1.211 & 1.200
 & 0.972 & 0.853 & 0.824 & 0.783 & 0.689
 & 1.224 & 1.228 & 1.233 & 1.226 & 1.219 \\
 &  &  & Med
 & 1.191 & 1.188 & 1.197 & 1.177 & 1.165
 & 0.919 & 0.755 & 0.776 & 0.702 & 0.662
 & 1.212 & 1.211 & 1.206 & 1.207 & 1.188 \\

RC\tnote{c} &  & 0.05 &
 & 60.87 & 60.87 & 56.52 & 60.87 & 65.22
 & 78.26 & 91.30 & 95.65 & 95.65 & 100
 & 52.17 & 43.48 & 34.78 & 34.78 & 39.13 \\
 &  & 0.10 &
 & 26.09 & 13.04 & 21.74 & 26.09 & 30.44
 & 78.26 & 91.30 & 91.30 & 95.65 & 100
 & 17.39 & 13.04 & 4.35 & 8.70 & 8.70 \\

\addlinespace[0.25cm]

QLIKE\tnote{b} &  &  & Avg
 & 5.221 & 6.152 & 7.402 & 5.141 & 4.474
 & 3.255 & 4.271 & 5.184 & 3.091 & 2.873
 & 7.376 & 8.302 & 9.870 & 7.225 & 6.146 \\
 &  &  & Med
 & 5.017 & 5.352 & 6.554 & 4.320 & 4.433
 & 3.008 & 3.492 & 3.776 & 2.106 & 2.571
 & 6.267 & 7.172 & 10.009 & 6.506 & 5.645 \\

RC &  & 0.05 &
 & 4.35 & 0.00 & 4.35 & 8.70 & 0.00
 & 8.70 & 4.35 & 4.35 & 8.70 & 21.74
 & 8.70 & 0.00 & 4.35 & 4.35 & 0.00 \\
 &  & 0.10 &
 & 0.00 & 0.00 & 0.00 & 0.00 & 0.00
 & 8.70 & 4.35 & 4.35 & 8.70 & 21.74
 & 0.00 & 0.00 & 0.00 & 0.00 & 0.00 \\

\addlinespace[0.25cm]
\bottomrule
\end{tabular}

\begin{tablenotes}[para,flushleft]
\footnotesize
\textit{Notes:} Columns report FCNN specifications with increasing numbers of hidden units (5–25), with larger values indicating greater model complexity.
\item[a] For each ticker, MSE is expressed relative to the corresponding MSE of CHAR. The reported average (median) is the cross-sectional mean (median) of the 23 ticker-level MSE ratios.
\item[b] For each ticker, QLIKE is expressed relative to the corresponding QLIKE of CHAR. The reported average (median) is the cross-sectional mean (median) of the 23 ticker-level QLIKE ratios.
\item[c] RC non-rejection denotes the percentage of tickers for which White's Reality Check does not reject, at the stated significance level, the null hypothesis that no competing HAR-family model has lower expected loss than the model under evaluation. A high value therefore indicates limited statistical evidence of inferiority.

\end{tablenotes}
\end{threeparttable}
\end{adjustbox}
\end{table}

\begin{table}
    \centering
    \begin{adjustbox}{max width=0.36\textwidth}
        \begin{threeparttable}
            \caption{Realised Utility of Nonlinear HAR Models}
            \begin{tabular}{c@{\hspace{50pt}}c}
                \toprule
                Complexity & Realised Utility \\
                \midrule
                5                & -4.4668 \\
                10               & -6.1599 \\
                15               & -8.5773 \\
                20               & -4.6595 \\
                25               & -3.3510 \\
                \bottomrule
            \end{tabular}
            \label{realised_utility_nonlinear_har}
            \begin{tablenotes}[para,flushleft]
                \footnotesize
                \item \textit{Notes:} This table reports realised utility values from the utility-based approach in \Cref{economic_gain_section} for nonlinear HAR forecasting models. Complexity column lists the number of units in FCNN, including 5, 10, 15, 20, and 25, with larger values indicating greater model complexity. In \Cref{utility_function}, the maximum attainable realised utility is 4\%. All values are expressed in percentage terms.
            \end{tablenotes}
        \end{threeparttable}
    \end{adjustbox}
\end{table}

\end{landscape}}

\vspace*{\fill}

\clearpage
\subsection{Hyperparameter Effects on NLP Model Forecasts} \label{appendix_NLP_Models_hyper}
\Cref{NLP_ML_primary_experiment_table_window} examines the sensitivity of the NLP models to the length of the input window used to aggregate stock-related news. Specifically, this table reports out-of-sample forecasting performance when headlines from the preceding one, three, five, and seven days are incorporated as model inputs, evaluated over the full sample as well as across normal and high volatility days. The purpose of this exercise is twofold. First, it assesses whether extending the information set beyond the previous day improves forecasting performance, thereby testing the temporal persistence of the predictive information in news for RV. Second, by increasing the input window, the analysis reduces the incidence of days without news coverage, thereby providing an indirect evaluation of the dedicated trainable no-news representation presented in \Cref{abstract_rep_nlp_ml}. We select the Word2Vec (skip-gram) specification from FinText, which performs among the best models in \Cref{results_NLP_subsection}, for this analysis. The 1-Day specification is equivalent to the results reported in \Cref{NLP_ML_primary_experiment_table_ticker_relatedX}.

The results indicate that extending the input window has loss-function-dependent effects. Over the full out-of-sample period, the average MSE ratio rises from 1.106--1.108 (1-Day) to 1.164--1.165 (7-Days), whereas the average QLIKE ratio decreases from 1.566--1.640 to 1.508--1.518. At the same time, RC non-rejection rates are generally strongest for the 1-Day specification (e.g., under MSE they reach 100\% at the 5\% level and 82.61\% at the 10\% level), while QLIKE RC non-rejection rates remain comparatively low. Across regimes, the effect of extending the input window differs across loss functions. Under MSE, longer windows improve performance on normal volatility days but deteriorate performance on high volatility days. Under QLIKE, longer windows generally reduce loss ratios in both regimes. The RC results provide complementary evidence on whether a HAR-family competitor has significantly lower expected loss under each window specification. Overall, the evidence is loss-function dependent: the one-day window is preferred under MSE, especially during high volatility days, whereas longer windows often improve QLIKE. Accordingly, extending the news window does not generate a systematic improvement across evaluation criteria.

\Cref{NLP_ML_primary_experiment_table_filter_size} focuses on the role of model architecture by evaluating the impact of alternative convolutional filter sizes on forecasting performance when stock-related news is used. The table contrasts results obtained from three distinct sets of filters, corresponding to shorter and longer n-gram representations of textual information, again reported for the full out-of-sample period and conditional on volatility regimes. The motivation for this analysis is to investigate whether allowing the model to capture longer linguistic patterns enhances its ability to extract volatility-relevant information from news. By holding the remaining components of the model fixed and varying only the filter sizes, this table isolates the effect of textual granularity on RV forecasting performance. Similar to the previous analysis, we select the Word2Vec (skip-gram) specification from FinText, which performs among the best models in \Cref{results_NLP_subsection}. Additionally, the $\{1,2,3\}$ filter sizes are equivalent to the results reported in \Cref{NLP_ML_primary_experiment_table_ticker_relatedX}.

The results indicate that the effect of increasing textual granularity is loss-function and regime dependent. Over the full out-of-sample period, loss ratios are generally lower under the baseline filter set $\{1,2,3\}$ for MSE but not for QLIKE, relative to extending the filters to $\{4,5,6\}$ or $\{7,8,9\}$; in contrast, RC statistics relative to the HAR-family benchmarks tend to be similar or to increase slightly with longer filter sizes. This divergence is more pronounced across volatility regimes. During normal volatility days, longer filters yield lower MSE and QLIKE ratios. However, RC values for QLIKE remain close to zero, and no material change is evident for MSE. During high volatility days, MSE ratios but not QLIKE generally increase as filter sizes lengthen, despite slight improvements in RC. Overall, the evidence is loss-function dependent: shorter filters perform better under full-sample and high-volatility MSE, whereas longer filters generally improve QLIKE and normal-volatility MSE. Accordingly, no filter set dominates across evaluation criteria.

These robustness checks do not identify a single input-window or filter configuration that dominates across evaluation criteria, and they show that forecast performance varies across loss functions and volatility regimes. They also indicate that RV forecasting performance could, in principle, be improved through a more extensive hyperparameter search. Such optimisation would require substantially greater computational resources and careful experimentation (e.g., tuning window construction, filter sets, regularisation, and learning dynamics), which is beyond the scope of this study. Moreover, moving toward heavily tuned architectures would depart from the deliberately simple and transparent modelling philosophy that underpins the HAR-family of models.

\afterpage{%
\clearpage
\thispagestyle{empty}

\atxy{\dimexpr\paperwidth-0.45in}{.5\paperheight}{%
  \rotatebox[origin=center]{90}{\thepage}
}

\begin{landscape}
\begin{table}
\centering
\begin{adjustbox}{max width={1.28\textwidth}}
\begin{threeparttable}
\centering
\setlength{\tabcolsep}{6pt}
				\caption{Out-of-Sample RV Forecasting Performance: Stock-Related News with Varying Input Windows}
				\begin{tabular}{cccc|cccc|cccc|cccc|cccc} \toprule
					{}  & {} & {} & {} &   \multicolumn{4}{c|}{1-Day} & \multicolumn{4}{c|}{3-Days} & \multicolumn{4}{c|}{5-Days} & \multicolumn{4}{c}{7-Days} \\
					{{\textbf{\small Full Out-of-Sample Period}}}  & {} & {$\alpha$} & {} & 25 & 50 & 75 & 100 & 25 & 50 & 75 & 100 & 25 & 50 & 75 & 100 & 25 & 50 & 75 & 100  \\ \midrule

                    \addlinespace[0.25cm]
					MSE\tnote{a} & &  & Avg & 1.106 & 1.108 & 1.108 & 1.107 & 1.154 & 1.153 & 1.153 & 1.153 & 1.159 & 1.159 & 1.159 & 1.159 & 1.164 & 1.165 & 1.164 & 1.164  \\
					& &  & Med & 1.101 & 1.092 & 1.096 & 1.093 & 1.122 & 1.125 & 1.123 & 1.123 & 1.124 & 1.127 & 1.125 & 1.124 & 1.124 & 1.127 & 1.125 & 1.125 \\
					
					RC\tnote{c} & & 0.05 &  & 100 & 100 & 100 & 100 & 100 & 100 & 100 & 100 & 95.65 & 95.65 & 95.65 & 95.65 & 95.65 & 91.30 & 95.65 & 91.30 \\
					& & 0.10 &  & 82.61 & 82.61 & 82.61 & 82.61 & 73.91 & 73.91 & 73.91 & 73.91 & 69.57 & 65.22 & 65.22 & 65.22 & 65.22 & 65.22 & 65.22 & 65.22 \\ \addlinespace[0.25cm]

					QLIKE\tnote{b} & &  & Avg & 1.566 & 1.640 & 1.571 & 1.571 & 1.503 & 1.504 & 1.497 & 1.494 & 1.510 & 1.508 & 1.500 & 1.498 & 1.516 & 1.518 & 1.511 & 1.508 \\
					& &  & Med & 1.570 & 1.579 & 1.563 & 1.567 & 1.440 & 1.441 & 1.430 & 1.439 & 1.453 & 1.449 & 1.438 & 1.453 & 1.475 & 1.473 & 1.460 & 1.464 \\
					
					RC & & 0.05 &  & 39.13 & 34.78 & 39.13 & 30.44 & 21.74 & 21.74 & 21.74 & 21.74 & 21.74 & 21.74 & 21.74 & 21.74 & 21.74 & 21.74 & 21.74 & 21.74 \\
					& & 0.10 &  & 21.74 & 21.74 & 21.74 & 21.74 & 17.39 & 17.39 & 17.39 & 17.39 & 13.04 & 17.39 & 17.39 & 17.39 & 8.70 & 8.70 & 8.70 & 13.04 \\ \addlinespace[0.25cm]   

					\midrule
					{\textbf{\small Normal Volatility Days}}  & {} & {} &   {} & {} & {} \\ 	
					\midrule						
					MSE & &  & Avg & 1.683 & 1.703 & 1.702 & 1.690 & 1.108 & 1.176 & 1.178 & 1.200 & 1.106 & 1.173 & 1.169 & 1.192 & 1.090 & 1.151 & 1.147 & 1.164 \\
					& &  & Med & 1.592 & 1.562 & 1.595 & 1.609 & 1.053 & 1.142 & 1.138 & 1.176 & 1.057 & 1.122 & 1.115 & 1.145 & 1.072 & 1.130 & 1.124 & 1.119 \\

					RC & & 0.05 &  & 69.57 & 69.57 & 73.91 & 69.57 & 73.91 & 60.87 & 56.52 & 56.52 & 69.57 & 56.52 & 56.52 & 56.52 & 69.57 & 65.22 & 65.22 & 60.87 \\
					& & 0.10 &  & 69.57 & 65.22 & 65.22 & 65.22 & 65.22 & 56.52 & 56.52 & 47.83 & 65.22 & 56.52 & 56.52 & 47.83 & 65.22 & 56.52 & 52.17 & 52.17 \\ \addlinespace[0.25cm]
     
					QLIKE & &  & Avg & 1.345 & 1.370 & 1.363 & 1.354 & 1.299 & 1.321 & 1.322 & 1.331 & 1.299 & 1.320 & 1.321 & 1.331 & 1.299 & 1.320 & 1.321 & 1.330 \\
					& &  & Med & 1.343 & 1.342 & 1.346 & 1.327 & 1.285 & 1.302 & 1.317 & 1.329 & 1.317 & 1.337 & 1.346 & 1.353 & 1.330 & 1.354 & 1.350 & 1.355 \\
					
					RC & & 0.05 & & 4.35 & 4.35 & 4.35 & 4.35 & 4.35 & 4.35 & 4.35 & 4.35 & 4.35 & 4.35 & 4.35 & 4.35 & 4.35 & 4.35 & 4.35 & 4.35 \\
					& & 0.10 &  & 4.35 & 4.35 & 4.35 & 4.35 & 4.35 & 4.35 & 4.35 & 4.35 & 4.35 & 4.35 & 4.35 & 4.35 & 4.35 & 4.35 & 4.35 & 4.35 \\ \addlinespace[0.25cm]	

					\midrule
					{\textbf{\small High Volatility Days}}  & {} & {} &   {} & {} & {} \\ 	
					\midrule	
                    \addlinespace[0.25cm]
					MSE & &  & Avg & 1.081 & 1.082 & 1.082 & 1.082 & 1.155 & 1.151 & 1.151 & 1.150 & 1.160 & 1.158 & 1.158 & 1.157 & 1.166 & 1.164 & 1.164 & 1.163 \\
					& &  & Med & 1.095 & 1.094 & 1.096 & 1.093 & 1.129 & 1.130 & 1.128 & 1.129 & 1.130 & 1.132 & 1.130 & 1.130 & 1.129 & 1.132 & 1.130 & 1.130 \\
					
					RC & & 0.05 &  & 95.65 & 95.65 & 95.65 & 95.65 & 91.30 & 91.30 & 91.30 & 91.30 & 91.30 & 91.30 & 91.30 & 91.30 & 91.30 & 91.30 & 91.30 & 91.30 \\
					& & 0.10 &  & 65.22 & 69.57 & 69.57 & 69.57 & 65.22 & 65.22 & 65.22 & 65.22 & 65.22 & 65.22 & 65.22 & 65.22 & 60.87 & 65.22 & 65.22 & 65.22 \\ \addlinespace[0.25cm]
     
					QLIKE & &  & Avg & 1.804 & 1.960 & 1.794 & 1.812 & 1.705 & 1.685 & 1.671 & 1.656 & 1.720 & 1.697 & 1.681 & 1.668 & 1.735 & 1.720 & 1.705 & 1.690 \\
					& &  & Med & 1.749 & 1.771 & 1.788 & 1.784 & 1.646 & 1.664 & 1.670 & 1.665 & 1.642 & 1.649 & 1.634 & 1.626 & 1.661 & 1.655 & 1.658 & 1.645 \\
					
					RC & & 0.05 &  & 60.87 & 69.57 & 65.22 & 65.22 & 60.87 & 65.22 & 65.22 & 65.22 & 56.52 & 69.57 & 65.22 & 65.22 & 47.83 & 56.52 & 56.52 & 52.17 \\
					& & 0.10 &  & 43.48 & 43.48 & 43.48 & 43.48 & 43.48 & 43.48 & 43.48 & 43.48 & 34.78 & 39.13 & 43.48 & 43.48 & 30.44 & 30.44 & 30.44 & 30.44 \\ \addlinespace[0.25cm]
                    
					\bottomrule                                
				\end{tabular}
				\begin{tablenotes} [para,flushleft]
					\footnotesize
                    \textit{Notes:} Using the FinText with Word2Vec (skip-gram) specification, identified as one of the best-performing models in \Cref{results_NLP_subsection}, the `1-Day,’ `3-Days,’ `5-Days,’ and `7-Days’ columns incorporate stock-related news from the preceding one, three, five, and seven days, respectively, as model inputs. The `1-Day’ specification corresponds to the results reported in \Cref{NLP_ML_primary_experiment_table_ticker_relatedX}.
					  \item[a] For each ticker, MSE is calculated over the relevant evaluation sample and expressed relative to the corresponding MSE of CHAR. The reported average (median) is the cross-sectional mean (median) of the 23 ticker-level MSE ratios. \item[b] For each ticker, QLIKE is calculated over the relevant evaluation sample and expressed relative to the corresponding QLIKE of CHAR. The reported average (median) is the cross-sectional mean (median) of the 23 ticker-level QLIKE ratios. \item[c] RC non-rejection denotes the percentage of tickers for which White's Reality Check does not reject, at the stated significance level, the null hypothesis that no competing HAR-family model has lower expected loss than the model under evaluation. A high value therefore indicates limited statistical evidence of inferiority.

				\end{tablenotes}
				\label{NLP_ML_primary_experiment_table_window}
			\end{threeparttable}
		\end{adjustbox}
	\end{table}
\end{landscape}
}

\afterpage{%
\clearpage
\thispagestyle{empty}

\atxy{\dimexpr\paperwidth-0.45in}{.5\paperheight}{%
  \rotatebox[origin=center]{90}{\thepage}
}

\begin{landscape}
\begin{table}
\centering
\begin{adjustbox}{max width={1.11\textwidth}}
\begin{threeparttable}
\centering
\setlength{\tabcolsep}{6pt}
				\caption{Out-of-Sample RV Forecasting Performance: Stock-Related News with Varying Filter Sizes}
				\begin{tabular}{cccc|cccc|cccc|cccc} \toprule
					{}  & {} & {} & {} &   \multicolumn{4}{c|}{$\{1, 2, 3\}$} & \multicolumn{4}{c|}{$\{4, 5, 6\}$} & \multicolumn{4}{c}{$\{7, 8, 9\}$}\\
					{{\textbf{\small Full Out-of-Sample Period}}}  & {} & {$\alpha$} & {} & 25 & 50 & 75 & 100 & 25 & 50 & 75 & 100 & 25 & 50 & 75 & 100  \\ \midrule
                    \addlinespace[0.25cm]
     
					MSE\tnote{a} & &  & Avg & 1.106 & 1.108 & 1.108 & 1.107 & 1.131 & 1.128 & 1.128 & 1.129 & 1.125 & 1.125 & 1.124 & 1.125 \\
					& &  & Med & 1.101 & 1.092 & 1.096 & 1.093 & 1.112 & 1.112 & 1.111 & 1.112 & 1.106 & 1.108 & 1.109 & 1.109 \\

					RC\tnote{c} & & 0.05 &  & 100 & 100 & 100 & 100 & 100 & 100 & 100 & 100 & 100 & 100 & 100 & 100 \\
					& & 0.10 &  & 82.61 & 82.61 & 82.61 & 82.61 & 86.96 & 86.96 & 86.96 & 86.96 & 86.96 & 86.96 & 86.96 & 86.96 \\ \addlinespace[0.25cm]

					QLIKE\tnote{b} & &  & Avg & 1.566 & 1.640 & 1.571 & 1.571 & 1.478 & 1.478 & 1.470 & 1.472 & 1.486 & 1.487 & 1.481 & 1.478  \\
					& &  & Med & 1.570 & 1.579 & 1.563 & 1.567 & 1.424 & 1.432 & 1.427 & 1.426 & 1.435 & 1.453 & 1.451 & 1.432 \\

					RC & & 0.05 &  & 39.13 & 34.78 & 39.13 & 30.44 & 43.48 & 43.48 & 43.48 & 43.48 & 43.48 & 43.48 & 43.48 & 43.48 \\
					& & 0.10 &  & 21.74 & 21.74 & 21.74 & 21.74 & 21.74 & 21.74 & 21.74 & 21.74 & 21.74 & 17.39 & 17.39 & 21.74 \\ \addlinespace[0.25cm]
          
					\midrule
					{\textbf{\small Normal Volatility Days}}  & {} & {} &   {} & {} & {} \\ 	
					\midrule		
                    \addlinespace[0.25cm]
					MSE & &  & Avg & 1.683 & 1.703 & 1.702 & 1.690 & 1.140 & 1.188 & 1.178 & 1.192 & 1.172 & 1.221 & 1.211 & 1.226 \\
					& &  & Med & 1.592 & 1.562 & 1.595 & 1.609 & 1.107 & 1.133 & 1.133 & 1.148 & 1.118 & 1.155 & 1.142 & 1.178 \\
					
					RC & & 0.05 &  & 69.57 & 69.57 & 73.91 & 69.57 & 73.91 & 60.87 & 65.22 & 60.87 & 65.22 & 65.22 & 65.22 & 65.22 \\
					& & 0.10 &  & 69.57 & 65.22 & 65.22 & 65.22 & 65.22 & 60.87 & 60.87 & 60.87 & 65.22 & 60.87 & 65.22 & 56.52 \\ \addlinespace[0.25cm]
     
					QLIKE & &  & Avg & 1.345 & 1.370 & 1.363 & 1.354 & 1.283 & 1.297 & 1.295 & 1.301 & 1.278 & 1.293 & 1.291 & 1.298 \\
					& &  & Med & 1.343 & 1.342 & 1.346 & 1.327 & 1.266 & 1.285 & 1.287 & 1.296 & 1.259 & 1.275 & 1.283 & 1.290 \\

					RC & & 0.05 & & 4.35 & 4.35 & 4.35 & 4.35 & 4.35 & 4.35 & 4.35 & 4.35 & 8.70 & 4.35 & 4.35 & 4.35 \\
					& & 0.10 &  & 4.35 & 4.35 & 4.35 & 4.35 & 4.35 & 4.35 & 4.35 & 4.35 & 8.70 & 4.35 & 4.35 & 4.35 \\ \addlinespace[0.25cm]

					\midrule
					{\textbf{\small High Volatility Days}}  & {} & {} &   {} & {} & {} \\ 	
					\midrule
                    \addlinespace[0.25cm]
					MSE & &  & Avg & 1.081 & 1.082 & 1.082 & 1.082 & 1.128 & 1.124 & 1.124 & 1.124 & 1.120 & 1.119 & 1.118 & 1.119 \\
					& &  & Med & 1.095 & 1.094 & 1.096 & 1.093 & 1.133 & 1.131 & 1.130 & 1.127 & 1.126 & 1.124 & 1.117 & 1.124 \\
					
					RC & & 0.05 &  & 95.65 & 95.65 & 95.65 & 95.65 & 95.65 & 95.65 & 95.65 & 95.65 & 95.65 & 95.65 & 95.65 & 95.65 \\
					& & 0.10 &  & 65.22 & 69.57 & 69.57 & 69.57 & 73.91 & 78.26 & 78.26 & 78.26 & 82.61 & 82.61 & 82.61 & 78.26 \\ \addlinespace[0.25cm]
     
					QLIKE & &  & Avg & 1.804 & 1.960 & 1.794 & 1.812 & 1.673 & 1.659 & 1.645 & 1.643 & 1.696 & 1.683 & 1.672 & 1.658 \\
					& &  & Med & 1.749 & 1.771 & 1.788 & 1.784 & 1.700 & 1.694 & 1.652 & 1.648 & 1.706 & 1.708 & 1.686 & 1.644 \\
					
					RC & & 0.05 &  & 60.87 & 69.57 & 65.22 & 65.22 & 73.91 & 73.91 & 73.91 & 73.91 & 73.91 & 73.91 & 73.91 & 73.91 \\
					& & 0.10 &  & 43.48 & 43.48 & 43.48 & 43.48 & 43.48 & 39.13 & 47.83 & 43.48 & 43.48 & 43.48 & 43.48 & 43.48 \\ \addlinespace[0.25cm]

					\bottomrule                                
				\end{tabular}
				\begin{tablenotes} [para,flushleft]
					\footnotesize
                        \textit{Notes:} Using the FinText with Word2Vec (skip-gram) specification, identified as one of the best-performing models in \Cref{results_NLP_subsection}, the sets $\{1, 2, 3\}$, $\{4, 5, 6\}$, and $\{7, 8, 9\}$ correspond to different filter sizes discussed in \Cref{nlp_ml_structure_subsection}. The $\{1, 2, 3\}$ group aligns with the results presented in \Cref{NLP_ML_primary_experiment_table_ticker_relatedX}. \\           
                      \item[a] For each ticker, MSE is calculated over the relevant evaluation sample and expressed relative to the corresponding MSE of CHAR. The reported average (median) is the cross-sectional mean (median) of the 23 ticker-level MSE ratios. \item[b] For each ticker, QLIKE is calculated over the relevant evaluation sample and expressed relative to the corresponding QLIKE of CHAR. The reported average (median) is the cross-sectional mean (median) of the 23 ticker-level QLIKE ratios. \item[c] RC non-rejection denotes the percentage of tickers for which White's Reality Check does not reject, at the stated significance level, the null hypothesis that no competing HAR-family model has lower expected loss than the model under evaluation. A high value therefore indicates limited statistical evidence of inferiority.
                    \end{tablenotes}
				\label{NLP_ML_primary_experiment_table_filter_size}
			\end{threeparttable}
		\end{adjustbox}
	\end{table}
\end{landscape}
}

\clearpage

\subsection{DeepLIFT (Deep Learning Important FeaTures)}
\label{appendix_deeplift}
DeepLIFT \citep{shrikumar2017learning} explains a prediction by comparing the model output at the observed input to the output at a reference (baseline) input, and then attributing the difference to the inputs via additive contribution scores. In our setting, we apply DeepLIFT to the fully connected output block and attribute changes in the RV forecast to changes in the pooled convolutional features $\mathbf{z}_t$, where each component of $\mathbf{z}_t$ is a pooled convolutional activation and hence corresponds to an $n$-gram detector. The model produces a nonnegative RV forecast through the final ReLU:
\begin{equation}
\widehat{RV}_{t+1}=\mathrm{ReLU}(u_{t+1}), \qquad
u_{t+1}=\mathbf{w}^{\top}\mathbf{z}_t+b.
\end{equation}
Following \citet{shrikumar2017learning}, we take the DeepLIFT target to be the pre-activation $t:=u_{t+1}$, and interpret attributions as explaining how inputs shift $u_{t+1}$, which then maps into $\widehat{RV}_{t+1}$ via the ReLU. Given a reference input producing $(z_t^{0},t^{0})$, define differences from the reference as
\begin{equation}
\Delta z_{t,k}=z_{t,k}-z_{t,k}^{0}, \qquad \Delta t=t-t^{0}.
\end{equation}
DeepLIFT assigns contribution scores $C_{\Delta z_{t,k}\rightarrow \Delta t}$ such that the total change in the target is exactly decomposed into additive input contributions:
\begin{equation}
\sum_{k} C_{\Delta z_{t,k}\rightarrow \Delta t}=\Delta t.
\end{equation}
Computationally, these contributions are obtained efficiently using a single backward pass with finite-difference-style propagation rules. In Deep SHAP, DeepLIFT-style attributions are averaged over a background set of reference inputs to approximate SHAP values \citep{lundberg2017unified}.

\clearpage

\bibliography{bib_list}

\end{document}